\documentclass{article}
\usepackage[utf8]{inputenc}
\usepackage[a4paper,left=3cm,right=3cm,top=3cm,bottom=3cm]{geometry}
\usepackage{amsmath}
\usepackage{enumitem}
\usepackage{tikz}
\usepackage{cite}

\usepackage{floatrow}
\usepackage{caption}

\usepackage{amssymb}
\usepackage{amsbsy}
\usepackage{fixmath}
\usepackage{graphicx}
\usepackage{hyperref}
\usepackage{booktabs}
\usepackage{xspace}

\newcommand{\be}{\begin{equation}}
\newcommand{\ee}{\end{equation}}

\newcommand{\GeV}{\ensuremath{\,\mathrm{GeV}}\xspace}

\newcommand{\gluonTMD}{gluon\ TMD}

\newcommand{\calF}{{\cal F}}

\newcommand{\eg}{{\it e.g.}}
\newcommand{\ie}{{\it i.e.}}
\newcommand{\cf}{{\it c.f.}}

\newcommand{\that}{{\hat t}}
\newcommand{\shat}{{\hat s}}
\newcommand{\uhat}{{\hat u}}
\newcommand{\tlt}{\bar{t}}
\newcommand{\tls}{\bar{s}}
\newcommand{\tlu}{\bar{u}}

\definecolor{red}{rgb}{1,0,0}

\def\bea{\begin{eqnarray}}
\def\eea{\end{eqnarray}}
\hypersetup{
  colorlinks   = true,    % Colours links instead of ugly boxes
  urlcolor     = blue,    % Colour for external hyperlinks
  linkcolor    = blue,    % Colour of internal links
  citecolor    = black     % Colour of citations
}

\newcommand{\KaTie}{{\sc Ka\hspace{-0.2ex}Tie}}
\newcommand{\Pythia}{{\sc Pythia}}

\title{\mbox{Searching for saturation in forward dijet production at the LHC}}
\author{
A. van Hameren$\,\,^a$, H. Kakkad$\,\,^b$, P. Kotko$\,\,^b$, \\
K. Kutak$\,\,^a$, S. Sapeta$\,\,^a$ \\\\
$^a$ {\it Institute of Nuclear Physics, Polish Academy of Sciences} \\
     {\it  Radzikowskiego 152, 31-342 Krak\'ow, Poland } \\ \\
$^b$ {\it AGH University Of Krakow, }\\
     {\it Faculty of Physics and Applied Computer Science,} \\ 
    {\it al. Mickiewicza 30, 30-059 Krak\'ow, Poland} \\ \\
 }

\date{}

\begin{document}
%=======================================================================
\maketitle

\begin{abstract}
   We review recent results for forward jests at the LHC and EIC as obtained within small-x Improved Transverse Momentum Dependent factorization (ITMD). 
   In addition to elementary overview of various approaches to perturbative QCD at high energy, including High Energy Factorization, Color Glass Condensate and  ITMD, we describe the Monte Carlo implementation and discuss the existing and unpublished phenomenological results for forward dijets.
\end{abstract}

%-----------------------------------------------------------------------
\section{Introduction}
\label{sec:Intro}
Quantum Chromodynamics (QCD) is a well established theory that describes interactions of quarks and gluons. However, it still has its challenges. In the high energy domain, one of the long standing problems is finding clear experimental signals of gluon saturation, which is a signature of quasi equilibrium between gluon splitting and gluon fusion in dense nuclear systems. Gluon saturation has been predicted from QCD long time ago \cite{Gribov:1984tu,Mueller:1985wy} and has been extensively studied using various approaches, most recently the Color Glass Condensate (CGC) -- \ie\ effective theory obtained within QCD (for a reviews see \cite{Iancu:2003xm,Jalilian-Marian:2005ccm,Gelis:2010nm,Albacete:2014fwa,Blaizot:2016qgz} and the textbook \cite{Kovchegov:2012mbw}). 

While there is no doubt that gluon distributions must saturate at some point due to the  unitarity constraints on the cross section, and there are strong indications of saturation in the data \cite{Golec-Biernat:1998zce,Albacete:2010pg,Dusling:2012cg,Kutak:2012rf,Stasto:2018rci,vanHameren:2019ysa,Mantysaari:2019nnt,Ducloue:2019jmy,ArroyoGarcia:2019cfl,STAR:2021fgw,Beuf:2020dxl,Morreale:2021pnn}, a complete consensus on reaching it is still to be achieved. This is mainly due to the demanding kinematics. It requires the final states  to be measured in the forward region of the  detectors and dealing with nuclear targets, which posses additional challenges due to the rich collision environment (see \eg\ \cite{Cacciari:2010te}). Moreover, theoretical predictions in high energy QCD are at present not as precise as those requiring ordinary collinear factorization.

Among various final states that can be measured in the context of saturation searches, the system consisting of two identified jets in the forward region (and everything else) plays a special role \cite{Marquet:2007vb,vanHameren:2014lna,Kutak:2014wga}.
In addition to the possibility of studying correlations, it has an important advantage over the single-inclusive jet production. Namely, the jet transverse momenta can be quite large, of the order of twenty GeV or so, and still be sensitive to the saturation effects in the back-to-back region. In single inclusive jet production, the jet transverse momentum is demanded to be of the  order  of the saturation scale $Q_s$ and determines the average transverse momenta of gluons being ``saturated''. The latter is of the order of a few GeV, for sufficiently high collision energy and sufficiently forward. Jets with larger transverse momenta are easier to reconstruct and are ``cleaner''. Moreover, on theoretical grounds, the description of harder dijets becomes possible in terms of ``generalized'' transverse momentum dependent (TMD) factorization, which makes it easier to implement in Monte Carlo generators than the full CGC.

 Dijet yields at relatively forward rapidity have already been measured at the Large Hadron Collider (LHC) by ATLAS collaboration for both proton-proton and proton-lead collisions \cite{ATLAS:2019jgo}. Since there was no cross sections measurement, the conclusions regarding saturation signals do not give yet a definite answer to whether saturation has been observed or not. Inclusive single forward jet energy spectra for proton-lead collisions measured at CMS CASTOR detector \cite{CMS:2018yhi} does not provide convincing proof either.
 Further research and analysis are necessary to gain a better understanding of gluon saturation and its effects on nuclear systems. In particular, more measurements of both proton-proton and proton-lead collisions in the same kinematics are needed. In addition to the ATLAS forward detector, the ALICE collaboration plans to build a more forward detector FoCal \cite{ALICECollaboration:2719928} that hopefully will shed more light on the saturation phenomenon. The forward jet physics at LHC is complementary to the physics of the Electron Ion Collider (EIC) \cite{Accardi:2012qut,AbdulKhalek:2021gbh} one of the primarily goals of which is to study gluon saturation physics.

In this work, we review the theoretical framework suitable for a description of forward dijet production in hadro and lepto-production in the full azimuthal angle range. The formalism lies in the intersection of the CGC theory and more traditional factorization approach, utilizing TMD gluon distributions. Since the formalism accounts for power corrections it has been dubbed as Improved TMD Factorization (ITMD) \cite{Kotko:2015ura}. A review of some essential aspects of high energy QCD and the ITMD framework itself is contained in Section~\ref{sec:dijets_HEQCD}.
We shall also review Monte Carlo implementation of the ITMD framework \KaTie\ (Section~\ref{sec:katie}), construction of the  TMD gluon distributions, as well as recent phenomenological predictions for azimuthal dijet correlations, both for proton-proton and proton-lead collisions, in the kinematics of the forward calorimeters of ATLAS detector and planned FoCal of ALICE (Section~\ref{sec:Results}). We also include new unpublished earlier computations of the rapidity distributions for different kinematic cuts.

%========================================================
\section{Forward dijet production in high energy QCD}
\label{sec:dijets_HEQCD}

%--------------------------------------------------------
\subsection{Kinematics}
\label{subsec:Kinematics}

\begin{figure}
  \begin{center}
    \includegraphics[width=0.45\textwidth]{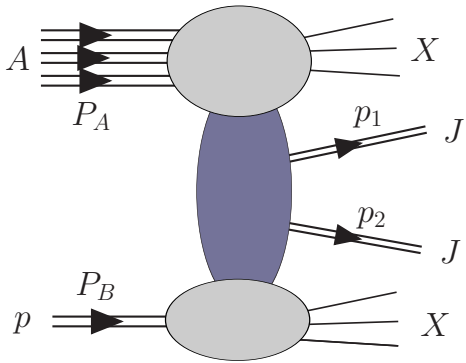}
  \end{center}
  \caption{\small 
  The momentum assignment in the inclusive dijet production in p+A collision. The multiple double lines represent the nuclear target $A$. $J$ are the jets with momenta $p_1$ and $p_2$. $X$ denotes arbitrary final states.}
  \label{fig:dijets-pA}
\end{figure}

The process in question is the inclusive production of two jet, that is, there are at least two jets with transverse momenta above a certain threshold, and both are produced in the forward rapidity interval in the center of mass frame:
\begin{equation}
    A(P_A)+p(P_B) \rightarrow J(p_1) + J(p_2)+X \,,
    \label{eq:process1}
\end{equation}
where $A$ is either the proton state $p$ or a nucleus having the four momentum $P_A$ that is hit by the proton state with four momentum $P_B$ (see Fig.~\ref{fig:dijets-pA}). The produced jets have four momenta $p_1$ and $p_2$ and are defined by a proper jet algorithm. $X$ corresponds to other particles produced in the process; there are no kinematic cuts imposed on those additional states. 
We shall also discuss the complementary DIS process 
\begin{equation}
    A(P_A)+e^{-} (P_B) \rightarrow J(p_1) + J(p_2)+e^{-} (P'_B)+X \,,
    \label{eq:processDIS}
\end{equation}
which will be studied at EIC and will be essential in testing the theoretical formalism and constraining the nonperturbative input.

In what follows, we define the $P_A$ and $P_B$ four momenta as being the two light cone vectors, defining the``plus'' and ``minus'' light cone components:
\begin{equation}
    P_A^{\mu}=(E_A,0,0,-E_A)=E_A n_+^{\mu}\,, \qquad P_B^{\mu}=(E_B,0,0,E_B)=E_B n_-^{\mu} \,,
\end{equation}
with the center of mass energy $s=2P_A\cdot P_B$. Using these, the Sudakov decomposition of any four vector reads:
\begin{equation}
    k^{\mu}=k^+n_-^{\mu} + k^-n_+^{\mu} + k_T^{\mu} \,,
\end{equation}
where
\begin{equation}
    k_T \cdot P_A = k_T \cdot P_B = 0\,, \qquad k_T\cdot k_T = - |\vec{k}_T|^2 \,.
\end{equation}

The central assumption in our investigations is that the nucleus (or a proton) is probed at small longitudinal momentum transfers compared to the hadron longitudinal momentum. That is, we define longitudinal momentum fractions
\begin{equation}
    x_A=\frac{k_1\cdot P_B}{P_A\cdot P_B}\,, \qquad  x_B=\frac{k_2\cdot P_A}{P_B\cdot P_A}\,,
\end{equation}
where $k_1$ is the momentum transferred to the target $A$, whereas $k_2$ is the momentum transferred to the proton target. We assume
\begin{equation}
    x_A \ll x_B \,.
\end{equation}
These fractions can be also expressed in terms of the final state rapidities $y_i$ and transverse momenta
\begin{equation}
    x_A = \frac{1}{\sqrt{s}} \left( p_{T1} e^{-y_1} + p_{T2} e^{-y_2} \right)\,, \qquad 
    x_B = \frac{1}{\sqrt{s}} \left( p_{T1} e^{y_1} + p_{T2} e^{y_2} \right)\,, 
\end{equation}
Therefore, restricting the final states to a large rapidity window and keeping the jet transverse momenta relatively low guarantees the smallness of $x_A$.

In phenomenological applications, we shall follow the realistic setup of the LHC experiments that measured (or plan to measure) forward dijets in proton-proton and proton-nucleus collisions: ATLAS and ALICE with their planned forward calorimeter FoCal. 
We shall discuss the kinematic cuts more precisely in Section~\ref{sec:Results} devoted to phenomenology.

%---------------------------------------------
\subsection{QCD at high energy}

One of the biggest achievements in theoretical developments of perturbative QCD are hard factorization theorems (see \cite{Collins:2011zzd} for a review). These include the collinear factorization and the TMD factorization, both types proved to all orders in coupling constant $\alpha_s$ and the leading power in the hard scale $Q^2$, for few sufficiently inclusive processes. In practice, not only the truncation of perturbation series in $\alpha_s$ is necessary (at present at very low orders), but also the limit $Q^2 \to \infty$ is rarely adequate. Indeed, when comparing with experimental data, for example for large transverse momenta dijet production \cite{ATLAS:2011kzm,CMS:2011hzb}, both a normalization factor (so-called $K$-factor) is needed, as well as a resummation (parton shower) with the addition of semi-nonperturbative effects (like multiple interactions).

In the collinear factorization or TMD factorization,  one encounters powers of logarithms of both $x_A$ and $x_B$ at every perturbative order. Therefore, at very large energies, small transverse momenta, forward production, or a mixture of all of these conditions, the perturbative expansion becomes less reliable. Although within the collinear factorization, it is in principle possible to resum the logarithms of small $x$ (see \cite{Bonvini:2022xio} and references therein), this does not address the two important issues. First, the evolution of the PDFs is always linear, whereas at high energy QCD, one must include the nonlinear effects that will tame the growth of the cross section as required by the unitarity of the scattering matrix \cite{Froissart:1961ux}. Second, the collinear factorization neglects higher twist effects (that contribute to the power corrections). For a discussions of higher twists in inclusive DIS and Drell-Yan see \eg~\cite{Bartels:2009tu,Motyka:2012ty,Motyka:2014jpa,Motyka:2014lya,Brzeminski:2016lwh,Motyka:2017xgk}.

Below, we briefly recall the existing approaches that attempt to address the above issues (for an extensive and pedagogical review of high energy QCD see \cite{Kovchegov:2012mbw}). We focus on the central ideas and main references, skipping the technical details.

\subsubsection*{Pomeron and The Reggeon Field Theory}

Historically, Pomeron -- the Regge trajectory with intercept greater than one and vacuum quantum numbers, was introduced in order to explain the growth of the total hadronic cross section with energy. The nature of this quasi-particle is essentially non-perturbative. Together with its parity-odd partner, the Odderon, they constitute a possible effective theory of strong interactions with color singlet degrees of freedom. Typically, their interactions are described by the Euclidean field theory called the Regge field theory \cite{Abarbanel:1975me} (see also \cite{Bartels:2015gou,Bartels:2015gou} and \cite{Kovner:2020jfb,Kovner:2020exf}). In its most general form, it includes the multiple Pomeron and Odderon interaction, but in the context of unitarity, one typically studies a truncation to the triple Pomeron vertex.

In perturbative QCD the parity even color singlet state exchanged in the $t$ channel can be made out of two gluons. This leads to the perturbative Pomeron of Balitsky, Fadin, Kuraev and Lipatov (BFKL) \cite{Balitsky:1978ic,Kuraev:1976ge,Kuraev:1977fs,Fadin:1975cb} (for a pedagogical review see eg.~\cite{Forshaw:1997dc}). The corresponding energy evolution equation of the Pomeron Green function gives the power like behavior of the cross section with energy. The triple Pomeron vertex needed for nonlinear taming of  the growth can also be constructed in QCD \cite{Bartels:1994jj,Bartels:1993ih,Bartels:2007dm}. 
The difference in the Pomeron intercept derived in perturbative QCD and the one needed to describe for example elastic hadron-hadron scattering leads to a distinction between the hard and the soft Pomeron.

There are numerous applications of the Pomeron calculus to particle production; in particular, there exist several Monte Carlo event generators: PHOJET and DPMJET \cite{Engel:1994vs,Roesler:2000he}, EPOS \cite{Werner:2005jf,Werner:2023zvo}, SIBYLL \cite{Fletcher:1994bd,Ahn:2009wx}, QGSJET \cite{Kalmykov:1997te,Ostapchenko:2010vb}.

\subsubsection*{Lipatov effective action}

The effective building blocks of the diagrams in the BFKL approach are the ``reggeized'' gluons $R$ and effective interaction  vertices $RR\rightarrow g$. The former is characterized by the eikonal coupling to other vertices (due to the high energy approximation) and the ``reggeization'' factor $(-s)^{\omega(t)}$ appearing due to radiative corrections, where $\omega(t)$ is the perturbative Regge trajectory. 
The $RRg$ effective vertices are separated in rapidity (the quasi-multi-Regge kinematics). 

Stripping off the reggeization factors, the building blocks can be formalized into an effective gauge invariant action \cite{Lipatov:1995pn,Antonov:2004hh} that contains both the QCD degrees of freedom, and the reggeized gluon fields $A_+$, $A_{-}$, that, roughly, by virtue of equations of motions are straight semi-infinite Wilson lines along two light-cone directions $P_A$ and $P_B$. The vertices are naturally generalized into $RRg\dots g$ vertices, as well as multiple reggeon vertices. For a review see \cite{Hentschinski:2020rfx} (chapter 11 in \cite{Bartels:2021zvx}).

Using Lipatov's high energy effective action beyond tree level is rather cumbersome due to double counting between gluon field and composite reggeons, but several results have been obtained (see for example \cite{Chachamis:2012gh,Chachamis:2012mw,Chachamis:2012cc,Chachamis:2013hma,Hentschinski:2014lma}). 
The important feature of that action is that it includes all the necessary ingredients to unitarize the cross section. In particular in \cite{Bondarenko:2017ern,Hentschinski:2018rrf,Bondarenko:2017vfc,Bondarenko:2018eid} a relation to the Color Glass Condensate was established and investigated. 

\subsubsection*{High Energy Operator Product Expansion}

Consider a projectile, a photon for simplicity, scattering off a hadron in a frame where it is moving slowly. The photon fluctuates into a pair of quarks well before it hits the target and -- at high energies -- interacts with the target eikonally. In the considered high energy kinematics, the target is made of gluons, that are treated as background field, which shrinks to a shock wave. The quark-antiquark dipole interacts with that shockwave becoming a straight infinite Wilson lines of the background field. 
Since the scattering cross section is related to the hadronic matrix element of the time ordered product of quark current, effectively, the above picture provides a means of decomposing the hadronic tensor into products of Wilson lines and the so-called impact factors. This is the essence of the high energy operator product expansion \cite{Balitsky:1995ub}.
More precisely
\begin{equation}
    \mathcal{T}\left\{J^{\mu}(x)J^{\nu}(y)\right\} = 
    \int\! d^2z_{1T}d^2z_{2T} \,
    \mathcal{I}^{\mu\nu}_2(x,y;\vec{z}_{1T},\vec{z}_{2T})
     \mathrm{Tr}\, U(\vec{x}_T) U^{\dagger}(\vec{y}_T) + \dots \,,
\end{equation}
where
\begin{equation}
    U(\vec{x}_T)=\mathcal{P} \exp\left\{ ig \int\! dw^+\, \hat{A}^-(x_T+w^+) \right\} \,
\end{equation}
is the straight infinite Wilson line along the plus light cone direction and fixed transverse position, whereas $\mathcal{I}^{\mu\nu}_2$ is the leading order impact factor. The dots represent the subleading corrections.
Similar to the treatment of the ordinary factorization, where the hadronic matrix element is traded to the partonic one when deriving the evolution equation, here it is traded to the matrix element of the background gluon field
\begin{equation}
     \left< P_A \right| \dots \left| P_A \right> \rightarrow \left< \dots \right> \,.
\end{equation}
The distinction between the background field and the projectile is related to the rapidity; formally the matrix elements depend on the rapidity cutoff. This dependence is explicitly introduced when regulating the rapidity divergence appearing when considering perturbative correction. However, a perturbative gluon interacting with the shock wave generates an adjoint Wilson line that gives a product of four fundamental Wilson lines. The evolution of this four-point correlator generates in turn further higher point correlators. Therefore the resulting evolution equation is not closed and consists of a tower of products of an increasing number of Wilson lines, known as the Balitsky hierarchy. However, at the large $N_c$ limit, the hierarchy is terminated and one obtains a single  evolution equation for the product of two Wilson lines, the so-called \emph{dipole} amplitude
\begin{equation}
   N(\vec{x}_T,\vec{y}_T;\eta) = 1 -\frac{1}{N_c} \left< \mathrm{Tr}\, U(\vec{x}_T)U^{\dagger}(\vec{y}_T) \right> \,,
\end{equation}
where $\eta$ is the implicit rapidity cutoff. The evolution equation \cite{Balitsky:1995ub} reads
\begin{multline}
    \frac{\partial}{\partial \eta} N(\vec{x}_T,\vec{y}_T;\eta) =\frac{\alpha_s N_c}{2\pi} \int\! d^2z_T\,  \frac{(\vec{x}_T-\vec{y}_T)^2}{(\vec{x}_T-\vec{z}_T)^2(\vec{y}_T-\vec{z}_T)^2} \,  \Bigg\{ N(\vec{x}_T,\vec{z}_T;\eta) + N(\vec{y}_T,\vec{z}_T;\eta)\\ - N(\vec{x}_T,\vec{y}_T;\eta) - N(\vec{x}_T,\vec{z}_T;\eta)N(\vec{z}_T,\vec{y}_T;\eta) \Bigg\} \,.
    \label{eq:BK_equation}
\end{multline}
It was also obtained independently in  \cite{Kovchegov:1999yj}  using the Mueller dipole approach \cite{Mueller:1993rr} and is therefore called the Balitsky-Kovchegov (BK) equation.

Typically one introduces new variables, the dipole size
\begin{equation}
    \vec{r}_T=\frac{\vec{x}_T-\vec{y}_T}{2} \,,
\end{equation}
and the impact parameter
\begin{equation}
    \vec{b}_T=\frac{\vec{x}_T+\vec{y}_T}{2} \,.
\end{equation}
The usual treatment of that equation assumes an infinite target and thus neglects the impact parameter dependence \cite{Kovchegov:1999ua,Levin:1999mw,Levin:2000mv,Levin:2001cv,Braun:2000wr,Armesto:2001fa,Lublinsky:2001yi,Lublinsky:2001bc,Golec-Biernat:2001dqn}, but solutions with impact parameter are also known, see 
\cite{Golec-Biernat:2003naj} and the following works, including \cite{Gotsman:2004ra,Ikeda:2004zp,Berger:2010sh,Rezaeian:2013tka,Bendova:2019psy,Cepila:2018faq}.

Let us finally mention the huge progress in controlling next-to-eikonal corrections, see \cite{Altinoluk:2020oyd,Chirilli:2021lif,Kovchegov:2021iyc,Altinoluk:2021lvu,Agostini:2022ctk,Altinoluk:2022jkk,Altinoluk:2023qfr}, as well as the development of the next-to-leading order BK equation \cite{Balitsky:2007feb,Kovner:2013ona,Kovner:2014lca,Lappi:2015fma,Lappi:2016fmu,Ducloue:2019ezk}.

\subsubsection*{Color Glass Condensate}

%____
\begin{figure}
\begin{center}
    \includegraphics[width=8cm]{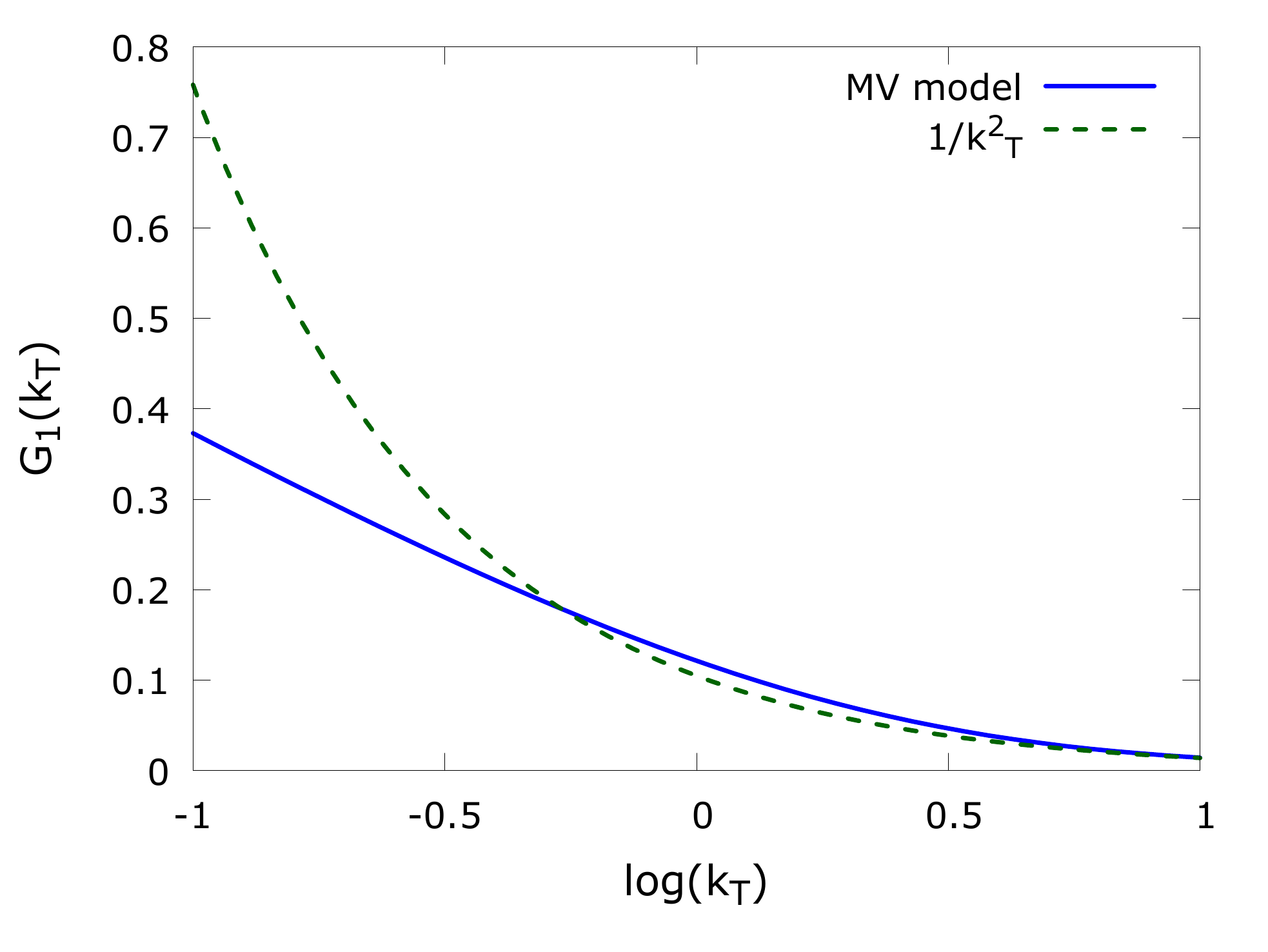}
\end{center}
\caption{
\small 
Gluon saturation phenomenon in the MV model. The perturbative behaviour $\sim 1/k_T^2$ is tamed at $k_T\sim Q_s$, with $Q_s$ increasing with energy.
}\label{fig:WW_MV}
\end{figure}
%___

The High Energy Operator Product expansion does provide the evolution equation but does not allow for the calculation of the non-perturbative correlators itself.

On the other hand, the effective theory of high energy QCD -- the Color Glass Condensate (CGC) is a convenient framework to approach both the evolution and averages of Wilson lines in the background color field (for a review see for example \cite{Gelis:2010nm}).

It is convenient to review now  a famous model of large nucleus -- the McLerran-Venugopalan (MV) model \cite{McLerran:1993ni,McLerran:1993ka}. Assume we view a large nucleus with a large number of nucleons in a frame where it is moving with a very large velocity. In that boosted frame, large $x$ partons are localized in one of the light-cone directions and can be considered as color sources  for the classical Yang-Mills fields. These sources are some, apriori unknown, distributions in the transverse plane $\rho_a(x_T)$ and are considered independent within the large nucleus because there are large numbers of nucleons inside. One can solve the Yang-Mills equation for such configuration to obtain the gluon ``wee'' fields. In order to compute observables, one needs to average over the color sources. For example, a basic quantity to consider is the particle number density with the generic definition
\begin{equation}
    \frac{dn}{d^3\mathbf{k}} \sim 
%    = \frac{2k^+}{(2\pi)^3}\, 
    \left<P_A\right| \tilde{A}^i_b\left(y^+,\mathbf{k}\right) \tilde{A}^i_b\left(y^+,-\mathbf{k}\right) \left|P_A\right> \, ,
    \label{eq:WW_MV1}
\end{equation}
where $\tilde{A}^i_b$ are partially Fourier transformed gauge fields,  $\mathbf{k}=(k^+,\vec{k}_T)$ is a three-vector conjugate to $\mathbf{y}=(y^-,\vec{y}_T)$. We have assumed here that the large boost is in the light cone ``plus'' direction.
In the MV model, the hadronic matrix element of the gluon field correlator is traded to 
\begin{equation}
    \left<A^i_b\left(\mathbf{y}\right) A^i_b\left(\mathbf{z}\right)\right>_{x} = \int \left[d\rho\right] 
    \mathcal{W}_{x}[\rho] A^i_b[\rho]\left(\mathbf{y}\right) A^i_b[\rho]\left(\mathbf{z}\right) \, ,
    \label{eq:WW_MV2}
\end{equation}
where the averaging is over the color sources in the functional sense. $\mathcal{W}_{x}$ is the functional weight for random color sources, which in the MV model is a Gaussian.
Direct computation of the above gluon distribution leads to the so-called Weizsacker-Williams (WW) gluon distribution \cite{Kovchegov:1996ty,Jalilian-Marian:1996mkd}.
Interestingly, despite it being the most basic quantity in the MV model it is not the easiest one to compute nor can it be probed in the simplest scattering processes. Rather, as we discuss later, the quantities that appear for example in inclusive DIS are correlators of Wilson lines.
The WW gluon distribution can be approximated as
\begin{equation}
    G_{1}\left(k_{T}\right) \sim \int d^{2}r_{T}\,\,\frac{e^{-i\vec{k}_{T}\cdot\vec{r}_{T}}}{|\vec{r}_{T}|}\left[1-\exp\left(-\frac{1}{4}\alpha_{s}N_{c}\mu_{A}\,|\vec{r}_{T}|^{2}\log\frac{1}{r_{T}\Lambda_{\mathrm{QCD}}}\right)\right] \,,
    \label{eq:WW_MV3}
\end{equation}
where $\mu_A$ is the average color charge per unit transverse area per color.
The saturation phenomenon is visible in the above distribution as follows. If the two-point correlator \eqref{eq:WW_MV1} was an ordinary perturbative correlator it would behave like $\sim 1/k_T^2$; this behavior leads to the power-like growth of the gluon distribution. It turns out that \eqref{eq:WW_MV3} behaves like $\sim 1/k_T^2$ in the perturbative domain (large $k_T$), but at small $k_T$ it behaves like $\frac{1}{\alpha_{s}}\log\frac{Q_{s}^{2}}{k_{T}^{2}}$, where $Q_s$ is the scale at which the suppression happens, called the \emph{saturation scale} (see Fig.~\ref{fig:WW_MV}). 
This dynamical scale depends on $x$ (through the evolution) and increases with decreasing $x$.

The subscript $x$ in the functional weight $\mathcal{W}_{x}$ denotes a longitudinal cutoff between the sources and the ``wee'' partons. Decreasing the cutoff produces new color sources and thus the Gaussian distribution gets distorted.
The CGC high energy effective theory predicts the
evolution equation of $\mathcal{W}_x$ in $x$. It is the so-called Jalilian-Marian-Iancu-McLerran-Weigert-Leonidov-Kovner (JIMWLK) equation \cite{Jalilian-Marian:1997qno,Kovner:2000pt,Kovner:1999bj,Weigert:2000gi,Iancu:2000hn,Ferreiro:2001qy} and has the following general form
\begin{equation}
    \frac{d\mathcal{W}_x[\rho]}{d\ln x} = -H_{\mathrm{JIMWLK}}\left[\rho,\frac{\delta}{\delta\rho}\right] \mathcal{W}_x[\rho] \,,
    \label{eq:JIMWLK}
\end{equation}
where $H_{\mathrm{JIMWL}}$ is the so-called JIMWLK Hamiltonian (we skip its exact form here). It gives a nonlinear evolution of $\mathcal{W}_x$ and in turn also a nonlinear evolution of any gauge field-dependent quantity due to the averaging procedure in $\rho$. These equations are consistent with the Balitsky hierarchy mentioned earlier, therefore it is often called B-JIMWLK equation.

%-----------------------------------------------------
\subsection{Small-$x$ Improved TMD (ITMD) factorization}

In this section, we shall review the formalism that is suitable for phenomenological studies of forward jet production processes at moderate transverse momenta. The formalism lies at the intersection of the two aspects of QCD that in the past were rarely overlapping and, essentially, were exercised by distinct communities. One is the high energy scattering reviewed in the previous Section and the other is the hard factorization theorems, in particular the TMD factorization (see eg. \cite{Collins:2011zzd} for a review).

%-----------------------------------------------------
\subsubsection*{Dijet production in dilute-dense collisions}

We start with a CGC description of a scattering of a proton or a photon off a nuclear target to produce two partons.
The basic assumption here is that, in the case of the proton projectile, it is dilute, that is  the partons extracted from the projectile are moderate $x$ partons. With that assumption, we can treat the scattering of color dipoles, such as $\gamma\rightarrow q\bar{q}$, $q\rightarrow qg $, etc., off a color field of a nucleus. This is the essence of the dilute-dense (or hybrid) approach which, as explained in Section~\ref{subsec:Kinematics}, is kinematically well suited to study forward jets \cite{Dumitru:2005gt}.

%____
\begin{figure}
\begin{center}
    \includegraphics[width=6cm]{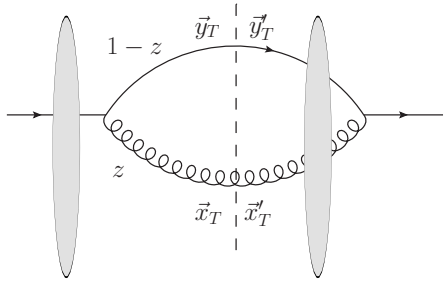}
\end{center}
\caption{
\small 
Scattering of a color dipole $q\rightarrow qg$ off a nuclear target in CGC. The shaded ellipses symbolize the color field of a nucleus. 
This particular diagram corresponds to a contribution, where in the amplitude the incoming quark scatters off the color field of the nucleus before the splitting, while in the amplitude conjugate both gluon and a quark scatter after splitting.
Particles scattering off the color field of the nucleus gain infinite straight Wilson line along the light cone, in their respective representations.
}\label{fig:CGC1}
\end{figure}
%___

Consider, as an example, the scattering of a $q\rightarrow qg$ color system off a nuclear target (Fig.~\ref{fig:CGC1}).
In the shock wave approximation, the interaction with the color field of the nucleus can take place either before the quark splits or after. For example, in Fig.~\ref{fig:CGC1} the scattering takes place before the splitting in the amplitude and after the splitting in the amplitude conjugate. Each colored particle gains a nonabelian phase when scattering, given by the Wilson line
\begin{equation}
    U_R\left(\vec{x}_T\right) = \mathcal{P} \exp\left\{ ig \int_{-\infty}^{+\infty}\! dx^- A^+_a(x^-,\vec{x}_T)T_R^a\right\} \, ,
    \label{eq:CGCWilsonLine}
\end{equation}
where the generator $T_R^a$ is in the fundamental representation ($R=F$) for a quark or in the adjoint representation ($R=A$) for a gluon.
The cross section for the scattering of that system can be written as \cite{Marquet:2007vb}
\begin{multline}
    \frac{d\sigma_{qA\rightarrow qg}}{d^2p_{1T}d^2p_{2T}dy_1dy_2} \sim 
    \int \frac{d^2x_T}{(2\pi)^2} \frac{d^2x'_T}{(2\pi)^2} \frac{d^2y_T}{(2\pi)^2} \frac{d^2y'_T}{(2\pi)^2} 
    e^{-i\vec{p}_{1T}\cdot \left(\vec{x}_T-\vec{x}'_T\right)} e^{-i\vec{p}_{2T}\cdot \left(\vec{y}_T-\vec{y}'_T\right)} \\
    \psi^*_q\left(z,\vec{x}'_T-\vec{y}'_T\right)\psi_q\left(z,\vec{x}_T-\vec{y}_T\right) 
    \Bigg\{S_x^{(4)}\left(\vec{y}_T,\vec{x}_T,\vec{y}'_T,\vec{x}'_T\right) \\
    -S_x^{(3)}\left(\vec{y}_T,\vec{x}_T,(1-z)\vec{y}'_T+z\vec{x}'_T\right) 
    -S_x^{(3)}\left((1-z)\vec{y}_T+z\vec{x}_T,\vec{y}'_T,\vec{x}'_T\right) \\
    -S_x^{(2)}\left((1-z)\vec{y}_T+z\vec{x}_T,(1-z)\vec{y}'_T+z\vec{x}'_T\right) \Bigg\} \, ,
    \label{eq:CGCdijets1}
\end{multline}
where $\psi_q(z,\vec{x}_T)$ is the wave function of the dipole, with $z$ being a fraction of the incoming quark momentum taken by a gluon. This wave function can be computed in perturbative QCD.
The quantities in the curly bracket correspond to the color averages of the Wilson lines:
\begin{gather}
    S_x^{(2)}\left(\vec{x}_T,\vec{y}_T\right) = \frac{1}{N_c} 
    \left< \mathrm{Tr}\, U_F\left(\vec{x}_T\right)U_F^{\dagger}\left(\vec{y}_T\right) \right>_x \\
    S_x^{(3)}\left(\vec{x}_T,\vec{y}_T,\vec{z}_T\right) = \frac{1}{C_FN_c} 
    \left< \mathrm{Tr}\left\{ U_F^{\dagger}\left(\vec{z}_T\right)t^a U_F\left(\vec{x}_T\right)t^b\right\}
    U_A^{ab}\left(\vec{y}_T\right) \right>_x \\
    S_x^{(4)}\left(\vec{x}_T,\vec{y}_T,\vec{x}'_T,\vec{y}'_T\right) = \frac{1}{C_FN_c} 
    \left< \mathrm{Tr}\left\{ U_F\left(\vec{x}_T\right)U_F^{\dagger}\left(\vec{x}'_T\right)t^bt^a\right\}
    \left\{U_A\left(\vec{y}_T\right)U_A^{\dagger}\left(\vec{y}'_T\right)\right\}^{ab} \right>_x \, .
\end{gather}
The above correlators depend on $x$ through the B-JIMWLK equations. In practice, one often uses the Gaussian approximation \cite{Fujii:2006ab}, which states that the color source distribution stays Gaussian throughout the evolution. This allows us to compute the evolution of correlators in closed form.

In order to compute the differential cross section for dijet production in proton-nucleus scattering in the above setup, one needs to find the remaining color dipole contributions \cite{Marquet:2007vb,Iancu:2013dta} and convolute the parton-nucleus cross section with the ordinary collinear PDFs.

The CGC formulation of the dilute-dense scattering provides the high-energy description of the jet production at both small transverse momenta $p_T$ of jets and moderate (at very large $p_T$ one should switch the approach from leading energy to leading power in the hard scale). 
The drawback of the above approach, in addition to being hard to generalize and implement in a Monte Carlo code, is that the perturbative information is contained not only in the color dipole wave function but also in the correlators. As we discuss later, in certain limits this information can be extracted and combined with the dipole wave function to obtain the hard matrix elements, similar to those known from hard factorization.

At the end, let us mention a huge  progress in computing higher orders in particle production within the CGC framework. The NLO results for single inclusive jets are available already for some time \cite{Chirilli:2011km,Chirilli:2012jd,Stasto:2013cha}, also with resolutions of the ``negative cross section'' problem \cite{Ducloue:2016shw,Iancu:2016vyg,Liu:2019iml,Liu:2022ijp}. Further, there are NLO computations for inclusive DIS \cite{Ducloue:2017ftk,Beuf:2020dxl}, also with heavy quarks \cite{Beuf:2022ndu,Hanninen:2022gje}, dijets in DIS \cite{Caucal:2021ent,Caucal:2022ulg,Caucal:2023nci}, vector meson production \cite{Mantysaari:2022bsp,Mantysaari:2022kdm}, dijet and dihadron hadroproduction \cite{Iancu:2020mos,Iancu:2022gpw,Taels:2022tza}. Higher multiplicity LO computations in CGC include dijet and photon production \cite{Altinoluk:2018byz}, trijets \cite{Iancu:2018hwa} and trijets in photoproduction \cite{Altinoluk:2020qet}.
%-----------------------------------------------------
\subsubsection*{High Energy (or $k_T$) factorization}

Usually, the high energy factorization (HEF) (or $k_T$-factorization) refers to a description of particle production at asymptotically high energies in terms of ``unintegrated'' PDFs that undergo the BFKL\cite{Collins:1991ty,Catani:1990eg} or CCFM \cite{ Ciafaloni:1987ur,Catani:1989sg} evolution. More precisely, within HEF the cross section for a multi-jet production in a collision of two hadrons can be written as
\begin{multline}
d\sigma_{AB}\left(P_A,P_B;p^{(J)}_1,\dots,p^{(J)}_n\right) =  \\ \int\! dx_Adx_B\, \int\! d^2k_{1T}d^2k_{2T} \, 
    \mathcal{F}_{g/A}\left(x_A,|\vec{k}_{1T}|\right) \mathcal{F}_{g/B}\left(x_B,|\vec{k}_{2T}|\right) \\
\, d\hat{\sigma}_{RR}\left(x_AP_A+k_{1T},x_B P_B+k_{2T};p^{(J)}_1,\dots,p^{(J)}_n;\mu\right) 
\label{eq:HEFhadronicXsec}
\, , \end{multline}
where $\mathcal{F}_{g/H}$ are unintegrated gluon distributions defined as
\begin{equation}
    \mathcal{F}_{g/H}\left(x,|\vec{k}_T|\right)=\int \frac{d^2p_T}{2\pi} \, \frac{\Phi_H\left(\vec{p}_T\right)}{|\vec{p}_T|^{2}}\mathcal{G}\left(\vec{p}_T,\vec{k}_T;x\right) \,,
    \label{eq:UnintegratedGlue}
\end{equation}
where $\Phi_H$ is the non-perturbative impact factor of a hadron  and $\mathcal{G}$ is the BFKL Green's function. 
At leading order, the partonic cross section in \eqref{eq:HEFhadronicXsec} reads
\begin{multline}
d\hat{\sigma}_{RR}^{(0)}\left(k_1,k_2;p_1^{(J)},\dots,p_n^{(J)}\right) \\ 
= \frac{1}{2x_Ax_Bs} 
\left|\overline{ V}_{RR\rightarrow P\dots P} \left(k_1,k_2;p_1,\dots,p_n\right)\right|^2   
 F\left(\{p_i\},\{p_i^{(J)}\}\right)  d\Gamma_n\left(\{p_i\}\right) \, ,
 \label{eq:HEFpartonicXsec}
\end{multline}
where $V_{RR\rightarrow P\dots P}$ (the bar denotes the usual spin/color summation and averaging) is the suitably normalized tree-level $RR\rightarrow P\dots P$ vertex to produce $n$ partons (jets), with $R$ being  reggeon states understood as fields from the high energy Lipatov's action (\ie\ without the ``reggeization'' factors), while $P$ being on-shell quarks or gluons. Above, $d\Gamma_n$ is ordinary on-shell phase space and $F$ is a jet function that implements the cuts. The momenta of reggeons are
\begin{equation}
    k_1^{\mu}=x_A P_A^{\mu} + k_{1T}^{\mu}\, \qquad k_2^{\mu}=x_B P_B^{\mu} + k_{2T}^{\mu} \, .
    \label{eq:QMRK}
\end{equation}
Notice that they are off-shell $k_{1,2}^2=-|\vec{k}_{T1,2}|^2$ and lack one of the light-cone components, due to the high energy approximations.

\begin{figure}
\begin{center}
    \includegraphics[width=8.5cm]{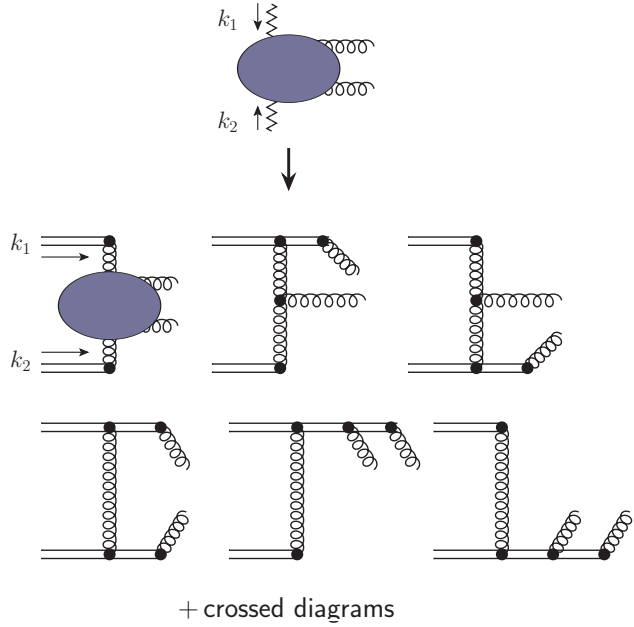}
\end{center}
\caption{
\small 
Off-shell gauge invariant amplitude (upper blob -- the zigzag lines represent off-shell gauge invariant gluons) can be constructed by promoting a single off-shell gluon coupled eikonally to a straight infinite Wilson line along corresponding hadron momenta $P_A$ or $P_B$. Here we show diagrams for the production of two gluons at tree level. The double line represents the momentum space Wilson line along $P_A$ (for the top line) and $P_B$ (for the bottom line). Gluons couple to these Wilson lines via $igt^a p_{A(B)}^\mu$ and the double-line propagators have form  $i/(k\cdot p_{A(B)}+i\epsilon)$. Only planar diagrams are shown. The blob represents all possible connections of gluons via standard vertices.
}\label{fig:HEF_offshell_amp1}
\end{figure}

In the original work on HEF in heavy quark production \cite{Catani:1990eg} the partonic cross section \eqref{eq:HEFpartonicXsec} was expressed in terms of off-shell amplitude to produce heavy quark pair, with initial gluons having momenta \eqref{eq:QMRK} and projected onto $P_A$ and $P_B$ (eikonal coupling). In that simple case, it turns out that the amplitude calculated in terms of ordinary diagrams is gauge invariant, despite the initial gluons being off-shell. In practical applications, it proves to be very convenient to stick to that logic also for more complicated processes, instead of using Lipatov's effective action.
Indeed, it is possible to define off-shell gauge invariant amplitude for arbitrary processes. In addition to ordinary Feynman diagrams (with off-shell space-like gluons), one needs ``gauge restoring'' contributions. Those additional diagrams can be reconstructed for example by the ``embedding'' method \cite{vanHameren:2012if}, where an off-shell process is embedded in an on-shell process with special kinematics of the auxiliary quarks or gluons. This method was automatized for arbitrary Standard Model tree level process in the \KaTie\ Monte Carlo discussed in Section~\ref{Katie}\footnote{The first Monte Carlo implementation of the High Energy Factorization was achieved within the CASCADE framework \cite{CASCADE:2021bxe}}. Also, other methods suitable for automatization were also developed \cite{vanHameren:2012uj,Kotko:2014aba,vanHameren:2014iua,vanHameren:2015bba}. In Fig.~\ref{fig:HEF_offshell_amp1} we illustrate one of the methods.

%---
%---

In the context of forward jets, one can apply similar considerations to the hybrid approach explained before. Namely, when, say, $x_B$ is moderate and not small, one should extract partons from collinear PDF rather than unintegrated PDF undergoing BFKL evolution. This was discussed in \cite{Deak:2009xt} and the corresponding HEF formula reads
\begin{multline}
    d\sigma_{AB}^{\mathrm{forward}}\left(P_A,P_B;p^{(J)}_1,\dots,p^{(J)}_n\right) =  \\ \sum_{a} \int\! dx_Adx_B\, \int\! d^2k_{T} \,\, 
    f_{a/B}\left(x_B,\mu\right) \mathcal{F}_{g/A}\left(x_A,|\vec{k}_{T}|;\mu\right) \\
\, d\hat{\sigma}_{ag^*}\left(x_B P_B,x_A P_A+k_{T};p^{(J)}_1,\dots,p^{(J)}_n;\mu\right) 
\label{eq:HybridHEFhadronicXsec}
\, ,
\end{multline}
where we have explicitly denoted the fact that the partonic cross section is constructed from amplitude with one off-shell (gauge invariant) gluon. Above $f_{a/B}$ is the collinear PDFs for parton $a$ (quark or gluon) in hadron $B$.

Let us now refocus our attention to the dijet case. It turns out that the $k_T$-factorization formula with proper gauge invariant off-shell amplitudes can be retrieved from the CGC expressions discussed before \cite{Iancu:2013dta,Kotko:2015ura}.
In the dilute limit, which corresponds to $|\vec{k}_T|\gg Q_s$, one can neglect the multiple scattering off the target. This means that the triple and quadrupole operators $S^{(3)}$, $S^{(4)}$ can be expressed only in terms of the dipole $S^{(2)}$. In that limit, one obtains for the $q\rightarrow qg$ dipole scattering 
\begin{multline}
    \frac{d\sigma_{pA\rightarrow qg}}{d^2p_{1T}d^2p_{2T}dy_1dy_2} 
    = x_Bf_{q/B}(x_B,\mu^2)\mathcal{F}_{g/A}(x_A,|\vec{k}_T|) \\
     \frac{\alpha_s}{2\pi} \frac{(1-z)(1+(1-z)^2)}{|\vec{p}_{1T}|^2|\vec{p}_{2T}|^2} \left[ 1 + \frac{(1-z)^2 |\vec{p}_{T1}|^2}{\big|\vec{P}_T\big|^2} - \frac{1}{N_c^2}\frac{z^2 |\vec{p}_{T2}|^2}{\big|\vec{P}_T\big|^2}  \right] \,,
\end{multline}
where 
\begin{equation}
    \vec{P}_T = (1-z)\vec{p}_{1T} + z \vec{p}_{2T} 
\end{equation}
and the unintegrated gluon distribution is related to the average of the weak field limit of the CGC dipole operator as follows
\begin{equation}
    \mathcal{F}_{g/A}(x,|\vec{k}_T|)= \frac{N_c}{\alpha_s (2\pi)^3} \int\! d^2y_T d^2z_T\, e^{-i \vec{k}_T\cdot (\vec{y}_T-\vec{z}_T)}\nabla^2_{\vec{y}_T-\vec{z}_T} \left[1-S^{(2)\, \mathrm{weak}}_x(\vec{y}_T,\vec{z}_T) \right]\,.
\end{equation}
The hard factor obtained above turns out to be exactly the off-shell gauge invariant amplitude.

%-----------------------------------------------------
\subsubsection*{TMD factorization}

Formally, the TMD factorization is the leading power (in the hard scale $\mu^2$) factorization of a cross section into TMD dependent PDFs and hard factors, that to leading power are on-shell (for a review see \cite{Collins:2011zzd} and \cite{Angeles-Martinez:2015sea,Boussarie:2023izj}). This factorization does not resum the large small-$x$ logarithms, therefore becomes unreliable at very high energies. Moreover, formally it does not hold for processes we are interested in. However, the formalism provides sturdy theoretical definitions of the TMD gluon distributions that, as it turns out, can be matched to the CGC correlators.

\begin{figure}
\begin{center}
\parbox{8.5cm}{\includegraphics[width=8.5cm]{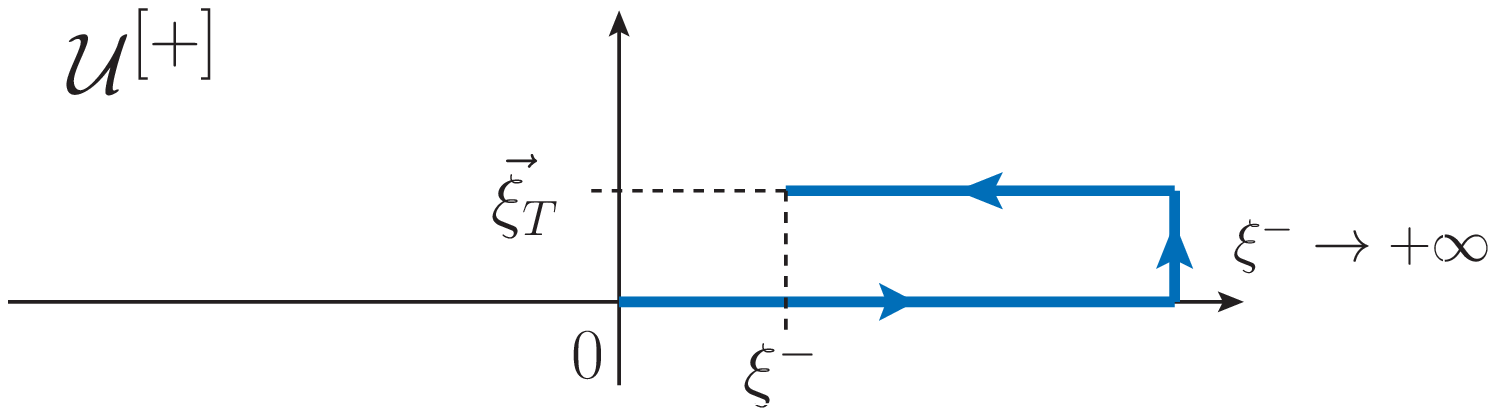}\vspace{0.8cm}\\
\includegraphics[width=8.5cm]{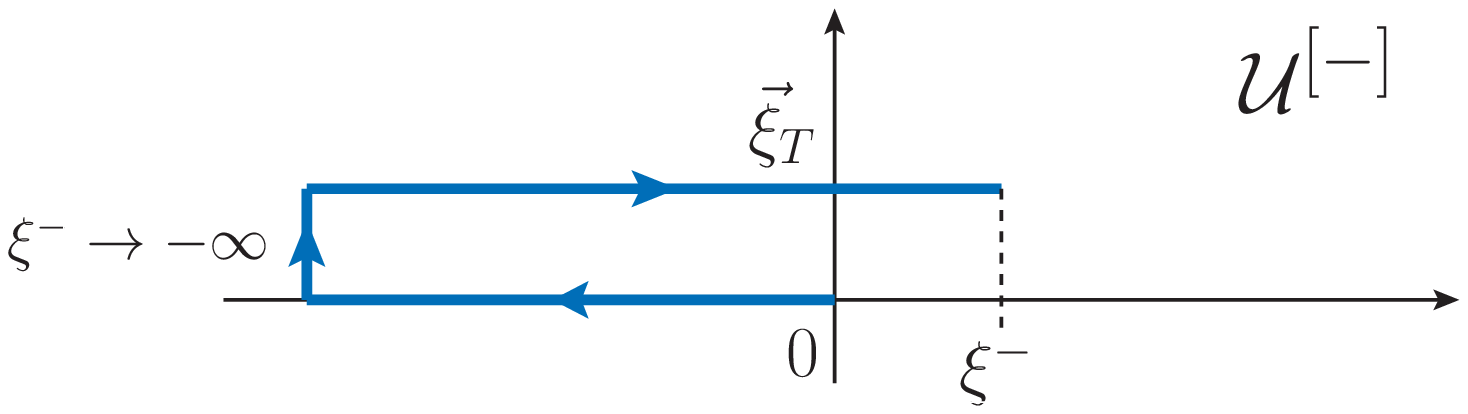}
}
\end{center}
{\caption{
\small
The shape of the ``staple-like'' gauge links $\mathcal{U}^{[+]}$ and $\mathcal{U}^{[-]}$. The horizontal axis represents the light-cone minus direction, the vertical axis symbolizes the transverse displacement. The transverse pieces are placed at $+\infty$ and $-\infty$, respectively.
}\label{fig:Staples}}
\end{figure}

In TMD factorization, gluon distribution is given by the Fourier transform of the bilocal matrix element of the gluon field strength tensor
\begin{equation}
\mathcal{F}_{C_1C_2}\big(x,|\vec{k}_T|\big) = 2\int\frac{d\xi^{-}d^{2}{\xi_T}}{(2\pi)^{3}P_A^{+}}\, e^{\,ix P_A^{+}\xi^{-}-i\vec{k}_{T}\cdot\vec{\xi}_{T}} 
\left\langle P_A\right|
\mathrm{Tr} \hat{F}^{j+}\big(\xi^-,\vec{\xi}_T,0\big)\mathcal{U}_{C_1}
\hat{F}^{j+}\big(0\big)\mathcal{U}_{C_2}
\left|P_A\right\rangle \,,
\label{eq:TMD_generic}
\end{equation}
where $\hat{F}^{j+}=F^{j+}_at^a$ is the gluon field strength tensor; the two operators are displaced in both the light cone and transverse direction (unlike in the collinear PDF, where the displacement is only along the light cone). The bilocal operator would not be gauge invariant, therefore the proper general definition requires gauge links $\mathcal{U}_{C_1}$ and $\mathcal{U}_{C_2}$ (here everything is in the fundamental representation) that connect the two space-time points. There is also a possibility of the double-trace over the fundamental representation (see below). We consider here only the unpolarized case, therefore the transverse index $j$ is summed over. 
The TMD gluon distributions are connected to a partonic process by virtue of factorization. Since we consider gauge theory, there are multiple soft and collinear gluons that can be connected to various places in the diagrams. The definition given above corresponds to the bare operator. In perturbation theory, it contains UV and rapidity divergences that, for some processes, can be removed order-by-order by the  operator renormalization and by (part of) the soft factor accumulating soft gluons. This gives the hard scale and rapidity evolution of the TMD PDFs. The collinear gluons (collinear to the target hadron) can be resummed into the Wilson lines $\mathcal{U}_{C_1}$ and $\mathcal{U}_{C_2}$. As can be easily understood, the form of these Wilson lines depends on the actual hard process (its color flow). In \cite{Bomhof:2006dp} a general procedure of determining these Wilson lines via resummation of collinear gluons was given. It turns out, that they can become quite complicated for colored partonic processes, see for example \cite{Bury:2018kvg}. 
In Table~\ref{tab:TMDs} we collect the operators relevant for the dijet production. The notation $\mathcal{F}_{gg}^{(i)}$ and $\mathcal{F}_{gg}^{(i)}$ corresponds to TMD PDFs that appear for incoming gluons (the ``gg'' subscript) or quark-gluon system (the ``qg'' subscript), and various color flows (the $(i)$ superscript). 
The Wilson lines $\mathcal{U}^{[\pm]}$ are defined as
\begin{equation}
    \mathcal{U}^{[-]} =\! \left[\big(\xi^-,\vec{\xi}_T,0\big),\big({-\infty},\vec{\xi}_T,0\big)\right]\!
    \left[\big({-\infty},\vec{\xi}_T,0\big),\big({-\infty},\vec{0}_T,0\big)\right]\!
    \left[\big({-\infty},\vec{0}_T,0\big),\big(0,\vec{0}_T,0\big)\right] \, ,
    \label{eq:U-}
\end{equation}
which is the past-pointing staple-like gauge link and
\begin{equation}
    \mathcal{U}^{[+]} =\! \left[\big(\xi^-,\vec{\xi}_T,0\big),\big({+\infty},\vec{\xi}_T,0\big)\right]\!
    \left[\big({+\infty},\vec{\xi}_T,0\big),\big({+\infty},\vec{0}_T,0\big)\right]\!
    \left[\big({+\infty},\vec{0}_T,0\big),\big(0,\vec{0}_T,0\big)\right] \, 
    \label{eq:U+}
\end{equation}
is the future pointing ``staple'', see Fig.~\ref{fig:Staples}.
The square brackets above $[x,y]$ is a standard notation for segments of straight gauge link between the points $x$ and $y$.
Out of two staples it is possible to make a Wilson loop
\begin{equation}
\mathcal{U}^{\left[\square\right]}=\mathcal{U}^{\left[-\right]\dagger}\mathcal{U}^{\left[+\right]}\,.\label{eq:WilsonLoopDef}
\end{equation}

%---
%---

The relation of the field theoretical definitions of TMD PDFs and small-$x$ QCD has been a subject of intense work, see for example \cite{Xiao:2010sp,Dominguez:2011wm,Metz:2011wb,Akcakaya:2012si,Balitsky:2015qba,Balitsky:2016dgz,Kovchegov:2015zha,Kovchegov:2015pbl,Kovchegov:2017lsr,Kovchegov:2018znm,Boussarie:2020fpb,Boussarie:2021wkn}.
In the context of forward dijet production, we are interested in the small-$x$ limit of the TMD gluon distributions. This is achieved by literally taking the limit $x_A\to0$ in the definitions. On the other hand, one can consider the leading power limit of the CGC expressions \eqref{eq:CGCdijets1}. This leads to the identification of the TMD gluon distributions and the leading terms in the gradient expansions of the CGC correlators \cite{Dominguez:2011wm}. For example
\begin{equation}
    \mathcal{F}_{qg}^{(1)}=\frac{1}{\pi \alpha_s} \int\!\frac{d^2x_T d^2y_T}{2\pi^3} e^{-i \vec{k}_T\cdot (\vec{x}_T-\vec{y}_T)} \left< \mathrm{Tr}\, \partial_j U_F\big(\vec{y}_T\big) \partial_j U_F^{\dagger}\big(\vec{x}_T\big) \right>_x \,,
\end{equation}
\begin{multline}
    \mathcal{F}_{qg}^{(2)}=-\frac{1}{\pi \alpha_s N_c} \int\!\frac{d^2x_T d^2y_T}{2\pi^3} e^{-i \vec{k}_T\cdot (\vec{x}_T-\vec{y}_T)} \Bigg< \mathrm{Tr}\left\{ \partial_j U_F\big(\vec{x}_T\big)U_F^{\dagger}\big(\vec{y}_T\big) \partial_j U_F^{\dagger}\big(\vec{y}_T\big)U_F^{\dagger}\big(\vec{x}_T\big)\right\} \\ \times 
    \mathrm{Tr}\left\{ U_F\big(\vec{y}_T\big)U_F^{\dagger}\big(\vec{x}_T\big)\right\}
    \Bigg>_x \,,
\end{multline}
that is, the $\mathcal{F}_{qg}^{(1)}$ is identified with the expansion of the dipole operator. For the complete list of similar relations for other TMDs see \cite{Dominguez:2011wm,Marquet:2016cgx}.
Ultimately, in the leading power limit, the CGC formula  for the $qg\rightarrow qg $ contribution can be written as 
\begin{equation}
\frac{d\sigma_{pA\rightarrow qg+X}}{d^{2}P_{T}d^{2}k_{T}dy_{1}dy_{2}}=\frac{1}{(x_A x_B s)^{2}}
 x_{B}f_{q/B}\big(x_{B}, \mu^2\big)\sum_{i=1}^{2}\mathcal{F}_{qg}^{(i)}\big(x_A,|\vec{k}_T|\big)\mathcal{H}_{qg\rightarrow qg}^{(i)} \, ,
\label{eq:TMDfact_qg}
\end{equation}
where $s$ is the hadronic center of mass energy and  $\mathcal{H}_{qg\rightarrow qg}^{(i)}$ are on-shell hard factors corresponding to the two independent color flows. They read
\begin{equation}
    \mathcal{H}_{qg\rightarrow qg}^{(1)} = \alpha_s^2\big(\hat{u}^2 +\hat{s}^2\big)\left( -\frac{\hat{u}}{2\hat{s}\hat{t}^2} + \frac{1}{2N_c^2} \frac{1}{\hat{s}\hat{u}} \right) \,,
\end{equation}
\begin{equation}
    \mathcal{H}_{qg\rightarrow qg}^{(2)} = \alpha_s^2\big(\hat{u}^2 +\hat{s}^2\big)\left( -\frac{\hat{s}}{2\hat{u}\hat{t}^2}  \right) \,,
\end{equation}
where $\hat{s}$, $\hat{t}$, $\hat{u}$ are the Mandelstam variables.

In collinear LO factorization, both TMD gluon distributions would be replaced by the collinear gluon PDFs and the above hard factors would be added. It is easy to check that the sum corresponds to the known on-shell hard factor for $qg\rightarrow qg$ process. The list of all hard factors for all subprocesses beyond the large $N_c$ limit is given in \cite{Kotko:2014aba}.

In the end, let us mention progress in establishing the CGC-leading power TMD factorization correspondence. In \cite{Bury:2018kvg} the Authors explicitly obtained the structure of all TMD operators corresponding to five and six colored parton processes. In \cite{Altinoluk:2020qet} the TMD factorization limit was studied for three jet production in CGC. Finally, in \cite{Taels:2022tza} the TMD factorization was studied in the context of dijet production in photoproduction at NLO.

%---
\begin{table}
\begin{centering}
\begin{tabular}{c|c c c c c c}
\hline\hline 
\begin{minipage}[c][1cm]{1cm}
$\displaystyle \mathcal{F}_{C_1C_2}$
\end{minipage} &
$\displaystyle \mathcal{F}^{(1)}_{gg}$ &
$\displaystyle  ^\mathrm{(*)} \mathcal{F}^{(2)}_{gg}$ &
$\displaystyle \mathcal{F}^{(3)}_{gg}$ &
$\displaystyle \mathcal{F}^{(4)}_{gg}$ &
$\displaystyle \mathcal{F}^{(5)}_{gg}$ &
$\displaystyle \mathcal{F}^{(6)}_{gg}$
\tabularnewline
\hline
\begin{minipage}[c][1.3cm]{0.7cm}
$\displaystyle \mathcal{U}_{C_1} $ 
\end{minipage} & 
$\displaystyle \frac{\mathrm{Tr}\mathcal{U}^{[\square]\dagger}}{N_c}\mathcal{U}^{[-]\dagger}$ &
$\displaystyle \mathcal{U}^{[\square]\dagger} $&
$\displaystyle \mathcal{U}^{[+]\dagger} $ &
$\displaystyle \mathcal{U}^{[-]\dagger}$ &
$\displaystyle \mathcal{U}^{[\square]\dagger}\mathcal{U}^{[+]\dagger} $&
$\displaystyle \frac{\mathrm{Tr}\mathcal{U}^{[\square]\dagger}}{N_c}\mathcal{U}^{[+]\dagger}$ 
\tabularnewline
\hline
\begin{minipage}[c][1.3cm]{0.7cm}
$\displaystyle \mathcal{U}_{C_2} $
\end{minipage} &
$\displaystyle \mathcal{U}^{[+]}$ &
$\displaystyle \mathcal{U}^{[\square]}$ &
$\displaystyle \mathcal{U}^{[+]}$ &
$\displaystyle \mathcal{U}^{[-]} $&
$\displaystyle \mathcal{U}^{[\square]}\mathcal{U}^{[+]}$ &
$\displaystyle \frac{\mathrm{Tr}\mathcal{U}^{[\square]}}{N_c}\mathcal{U}^{[+]} $ 
\tabularnewline
\hline\hline
\end{tabular}
%\end{doublespace}

\vspace{0.8cm}
%\begin{doublespace}
\begin{tabular}{c|cc}
\hline\hline
\begin{minipage}[c][1cm]{1cm}
$\displaystyle \mathcal{F}_{C_1C_2}$
\end{minipage} &
$\displaystyle \mathcal{F}^{(1)}_{qg}$ &
$\displaystyle \mathcal{F}^{(2)}_{qg}$ 
\tabularnewline
\hline
\begin{minipage}[c][1.3cm]{0.7cm}
$\displaystyle \mathcal{U}_{C_1} $ 
\end{minipage} & 
$\displaystyle \mathcal{U}^{[-]\dagger}$ &
$\displaystyle \mathcal{U}^{[+]\dagger}$ 
\tabularnewline
\hline
\begin{minipage}[c][1.3cm]{0.7cm}
$\displaystyle \mathcal{U}_{C_2} $
\end{minipage} &
$\displaystyle \mathcal{U}^{[+]}$ &
$\displaystyle \frac{\mathrm{Tr}\mathcal{U}^{[\square]}}{N_c}\mathcal{U}^{[+]} $
\tabularnewline
\hline\hline
\end{tabular}
%\end{doublespace}
\end{centering}
%\end{doublespace}

\vspace{0.5cm}
\caption{
\small
Gauge links $\mathcal{U}_{C_1}$ and $\mathcal{U}_{C_2}$ in terms of the ``staple-like'' Wilson lines contributing to TMD gluon distributions that are coupled to independent color flows of  $gg\rightarrow gg$ and $gg\rightarrow q\bar{q}$ (upper table) and $qg\rightarrow qg$ (lower table) processes. The operators in the column marked with a star $*$ should be traced with the gluon field strength tensor independently.
\label{tab:TMDs}}
\end{table}
%---

%-----------------------------------------------------
\subsubsection*{ITMD and resummation of kinematic twists}
\label{sec:ITMDsec}
It is important to stress, that the TMD factorization in dilute-dense collisions, despite being leading power, does take into account gluon saturation. First, notice that the leading power means here $k_T \ll P_T$. The saturation scale is not neglected and is of the order of $k_T$. Second, the leading twist TMD gluon distribution is considered in the strict high energy limit and they still can undergo nonlinear evolution, even at leading twist (for an interesting discussion of two types of saturation see \cite{Altinoluk:2019wyu}). 
To summarize, the dilute-dense TMD factorization is suitable when the transverse momenta of dijets are rather large and we are interested in the back-to-back dijet region. For very large transverse momenta of the jets, the highest scale is not given by the energy but the hard scale and one should switch the framework to the collinear factorization.

The main limitation of the dilute-dense TMD factorization is that it works only in the back-to-back region. In dijet studies, especially at small $x$, dijet azimuthal correlations are the most important observables and can be measured over a wide range of the azimuthal angle. Therefore, it is essential to account not only for jet correlations but also decorrelations.
A related issue is that of the transverse momentum conservation. In the TMD factorization, the transverse momentum $k_T$ of the gluon scattered (or extracted) from the target nucleus enters only TMD gluon distributions, and not the hard process. This complicates for example Monte Carlo realization of such an approach.

It turns out, that, in practical terms, it is actually quite easy to improve the dilute-dense TMD factorization. The hard factors have to take into account power corrections due to the transverse momentum of the incoming gluon. This is uniquely realized in the high energy limit by promoting the on-shell hard factors to the gauge invariant off-shell ones, following the rules described in the part devoted to HEF. This procedure was described in detail in  \cite{Kotko:2014aba}, where also off-shell gauge invariant hard factors are calculated for all channels and color flows. A proof and proper interpretation within the CGC was later given in \cite{Altinoluk:2019fui}. First, a formal distinction between kinematic and genuine twist is made. The genuine twist counting can be simply understood as counting gluon operators in the TMD matrix element (but not the Wilson lines), for example, an operator with two strength field tensors corresponds to twist two and two-body contribution. The kinematic twists are power corrections in $k_T$ to the $n$-body process and come from the hard matrix element. In \cite{Altinoluk:2019fui} a resummation of the kinematic twists was performed for the two-body contributions, showing that the resulting hard factors indeed correspond to the off-shell gauge invariant hard factors.

Before we review the dilute-dense improved TMD (ITMD) factorization framework, let us first comment, that when determining the TMD gluon distribution functions it is important to work with gauge invariant subset of diagrams and not the individual ones. The systematic way of doing that in small-$x$ physics was suggested in \cite{Kotko:2014aba}, where a decomposition of an amplitude into color-ordered amplitudes  was used. The color-ordered amplitudes \cite{Mangano:1990by} are gauge invariant, but contain only planar diagrams and correspond to the different ordering of the external partons. Each color-ordered amplitude comes with a color structure, for which the TMD operator can be determined. It is then usually given as some combination of the basic operators build from $\mathcal{U}^{[\pm]}$ Wilson lines, but corresponds to the gauge invariant hard factor. The ITMD framework given below was formulated with that feature in mind.

\begin{figure}
  \begin{center}
    \includegraphics[width=0.45\textwidth]{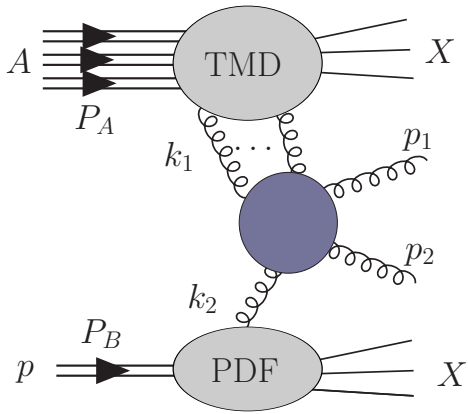}
  \end{center}
  \caption{\small 
  The ITMD factorization (pure gluon channel) of the p+A dijet cross section into TMD PDF (upper blob), the collinear PDF (lower blob) and the off-shell hard part (center blow). Since the exchanged momentum between the upper blob and the center is off-shell, multiple eikonal gluon exchanges are required to maintain the gauge invariance. In ITMD factorization, these gluons do not increase the genuine twist of the TMD operator. The TMD distribution is given by a certain linear combination of the operators listed in Table~\ref{tab:TMDs}.}
  \label{fig:factorization}
\end{figure}

The ITMD factorization formula, accounting for all channels, reads
\begin{equation}
\frac{d\sigma_{pA\rightarrow JJ+X}}{d^{2}P_{T}d^{2}k_{T}dy_{1}dy_{2}}=\frac{\alpha_{s}^{2}}{(x_A x_B s)^{2}}
\sum_{a,c,d} x_{B}f_{a/B}\big(x_{B}, \mu^2\big)\sum_{i=1}^{2}K_{ag^*\to cd}^{(i)}\big(\vec{k}_T\big)\Phi_{ag\rightarrow cd}^{(i)}\big(x_A,|\vec{k}_T|\big)\, \frac{1}{1+\delta_{cd}}\ ,
\label{eq:ITMD_factorization}
\end{equation}
where $K_{ag^*\to cd}^{(i)}\big(\vec{k}_T\big)$ are off-shell gauge invariant hard factors and $\Phi_{ag\rightarrow cd}^{(i)}\big(x_A,|\vec{k}_T|\big)$ are the corresponding TMD gluon distributions. We collect them in Table~\ref{tab:ITMDs}.
They are expressed in terms of the ordinary Mandelstam variables, as well as ``modified'' Mandelstam variables. The former read explicitly
\begin{subequations}
  \label{eq:mandelstam}
  \begin{align}
    \shat & = (k_1+k_2)^2 = (p_1 + p_2)^2=\frac{|\vec{P}_T|^2}{z(1-z)}\,, \\
    \that & = (p_2-k_1)^2 = (p_1 - k_2)^2=-\frac{|\vec{p}_{2T}|^2}{1-z}\,, \\
    \uhat & = (p_1-k_1)^2 = (p_2 - k_2)^2=-\frac{|\vec{p}_{1T}|^2}{z}\,,
  \end{align}
\end{subequations}
where the incoming momenta are
\begin{equation}
    k_1=x_A P_A + k_T,\qquad k_2 = x_B P_B \,
\end{equation}
and
\begin{equation}
z=\frac{p_1^+}{p_1^+ + p_2^+} \quad\quad \text{and}
\quad\quad \vec{P}_T=(1-z)\vec{p}_{1T}-z\vec{p}_{2T}\ .
\label{eq:zdef}
\end{equation} They sum up to $\shat + \that + \uhat = -|\vec{k}_T|^2$.
The ``modified'' Mandelstam variables take into account only the longitudinal component of the off-shell initial state $k_1$ and read
\begin{subequations}
  \begin{align}
  \tls & \ = \ (x_A P_{A} +k_2)^2 \ =
         \   \ \frac{|\vec{P}_T|^2}{z(1-z)}+|\vec{k}_T|^2=x_Ax_Bs\,,\\
  \tlt & \ = \ (p_2-x_A P_{A})^2=-z\tls\,, \\
  \tlu & \ = \ (p_1-x_A P_{A})^2=-(1-z)\tls\,,
  \end{align}
  \label{eq:gen-mandelstam}
\end {subequations}
which are related via the equation
\begin{equation}
  \tls + \tlt + \tlu = 0\,.
\end{equation}

The ITMD factorization was also investigated for other processes then dijets in proton-nucleus collisions. In \cite{Altinoluk:2021ygv}
heavy quark pair production was studied and the problem of longitudinal gluons was discussed in depth. In \cite{Boussarie:2021ybe} the effect of the genuine twists vs kinematic twists was studied in detail for dijets in DIS. Finally, neglecting the longitudinal gluon contribution, the ITMD framework was formulated and applied to the trijet production \cite{Bury:2018kvg,Bury:2020ndc}.

%---
\begin{table}
%\begin{doublespace}
\begin{centering}
\begin{tabular}{c|c|c}
\hline\hline 
\begin{minipage}[c][0.7cm]{0.5cm}
     $i$
 \end{minipage}
 & 1 & 2 \tabularnewline
\hline 
${\displaystyle \Phi_{gg^{*}\to gg}^{(i)}}$  & %
\begin{minipage}[c][2.3cm]{5cm}%
\vspace{-0.8cm}
\begin{multline*}
\frac{1}{2N_{c}^{2}}\big(N_{c}^{2}\mathcal{F}_{gg}^{\left(1\right)}-2\mathcal{F}_{gg}^{\left(3\right)}\\
+\mathcal{F}_{gg}^{\left(4\right)}+\mathcal{F}_{gg}^{\left(5\right)}+N_{c}^{2}\mathcal{F}_{gg}^{\left(6\right)}\big)
\end{multline*}
\end{minipage}  & %
\begin{minipage}[c][2.3cm]{5cm}%
\vspace{-0.8cm}
\begin{multline*}
\frac{1}{N_{c}^{2}}\big(N_{c}^{2}\mathcal{F}_{gg}^{\left(2\right)}-2\mathcal{F}_{gg}^{\left(3\right)}\\
+\mathcal{F}_{gg}^{\left(4\right)}+\mathcal{F}_{gg}^{\left(5\right)}+N_{c}^{2}\mathcal{F}_{gg}^{\left(6\right)}\big)
\end{multline*}
\end{minipage} \tabularnewline
\hline 
${\displaystyle \Phi_{gg^{*}\to q\overline{q}}^{(i)}}$  & %
\begin{minipage}[c][1.5cm]{5cm}%
\vspace{-0.3cm}
\begin{equation*}
\frac{1}{N_{c}^{2}-1}\left(N_{c}^{2}\mathcal{F}_{gg}^{\left(1\right)}-\mathcal{F}_{gg}^{\left(3\right)}\right)
\end{equation*}
\end{minipage}  & %
\begin{minipage}[c][1.5cm]{5cm}%
\vspace{-0.3cm}
\begin{equation*}
-N_{c}^{2}\mathcal{F}_{gg}^{\left(2\right)}+\mathcal{F}_{gg}^{\left(3\right)}
\end{equation*}
\end{minipage} \tabularnewline
\hline 
${\displaystyle \Phi_{qg^{*}\to qg}^{(i)}}$  & %
\begin{minipage}[c][1.5cm]{5cm}%
\vspace{-0.3cm}
\begin{equation*}
\mathcal{F}_{qg}^{\left(1\right)}
\end{equation*}
\end{minipage}  & %
\begin{minipage}[c][1.5cm]{5cm}%
\vspace{-0.3cm}
\begin{equation*}
\frac{1}{N_{c}^{2}-1}\left(-\mathcal{F}_{qg}^{\left(1\right)}+N_{c}^{2}\mathcal{F}_{qg}^{\left(2\right)}\right)
\end{equation*}
\end{minipage} \tabularnewline
\hline\hline 
\end{tabular}

\vspace{1cm}

\begin{tabular}{c|c|c}
\hline\hline  
 \begin{minipage}[c][0.7cm]{0.5cm}
     $i$
 \end{minipage}
 &  1 &   2  \\
\hline 
$\displaystyle K_{gg^*\to gg}^{(i)}$ & 
\begin{minipage}[c][2.0cm]{5cm}
\vspace{-0.3cm}
\begin{equation*}
\frac{N_{c}}{C_F}\,\frac{\left(\overline{s}^{4}+\overline{t}^{4}+\overline{u}^{4}\right)\left(\overline{u}\hat{u}+\overline{t}\hat{t}\right)}{\tlt\that\tlu\uhat\tls\shat}
\end{equation*}
\end{minipage}
& 
\begin{minipage}[c][2.0cm]{5.5cm}
\vspace{-0.3cm}
\begin{equation*}
-\frac{N_{c}}{2C_F}\,\frac{\left(\overline{s}^{4}+\overline{t}^{4}+\overline{u}^{4}\right)\left(\overline{u}\hat{u}+\overline{t}\hat{t}-\overline{s}\hat{s}\right)}{\tlt\that\tlu\uhat\tls\shat}
\end{equation*}
\end{minipage}
\\
\hline 
$\displaystyle K_{gg^*\to q\overline{q}}^{(i)}$ &
\begin{minipage}[c][2.0cm]{5cm}
\vspace{-0.3cm}
\begin{equation*}
 \frac{1}{2N_{c}}\,\frac{\left(\overline{t}^{2}+\overline{u}^{2}\right)\left(\overline{u}\hat{u}+\overline{t}\hat{t}\right)}{\overline{s}\hat{s}\hat{t}\hat{u}}
\end{equation*}
\end{minipage} 
 & 
 \begin{minipage}[c][2.0cm]{5cm}
\vspace{-0.3cm}
\begin{equation*}
  \frac{1}{4N_{c}^2 C_F}\,\frac{\left(\overline{t}^{2}+\overline{u}^{2}\right)\left(\overline{u}\hat{u}+\overline{t}\hat{t}-\overline{s}\hat{s}\right)}{\overline{s}\hat{s}\hat{t}\hat{u}}
\end{equation*}
\end{minipage}   
\\
\hline 
$\displaystyle K_{qg^*\to qg}^{(i)}$ &
 \begin{minipage}[c][2.0cm]{5cm}
\vspace{-0.3cm}
\begin{equation*}
 -\frac{\overline{u}\left(\overline{s}^{2}+\overline{u}^{2}\right)}{2\overline{t}\hat{t}\hat{s}}\left(1+\frac{\overline{s}\hat{s}-\overline{t}\hat{t}}{N_{c}^{2}\ \overline{u}\hat{u}}\right)
 \end{equation*}
\end{minipage}  
 & 
  \begin{minipage}[c][2.0cm]{5cm}
\vspace{-0.3cm}
\begin{equation*}
  -\frac{C_F}{N_c}\,\frac{\overline{s}\left(\overline{s}^{2}+\overline{u}^{2}\right)}{\overline{t}\hat{t}\hat{u}}
   \end{equation*}
\end{minipage} 
\\
\hline\hline 
\end{tabular}
\end{centering}
%\end{doublespace}
\vspace{0.5cm}
\caption{
\small
The top table lists the combinations of the TMD gluon distributions listed in Table~\ref{tab:TMDs} that correspond to gauge invariant off-shell hard factors of the ITMD factorization formula. The bottom table gives explicit formulae for the LO off-shell gauge invariant hard factors of the ITMD factorization. The $\hat{s}$, $\hat{t}$ and $\hat{u}$ are ordinary Mandelstam variables, whereas $\bar{s}$, $\bar{t}$, $\bar{u}$ are invariants where instead of the incoming off-shell gluon momentum its longitudinal component is used.
\label{tab:ITMDs}}
\end{table}
%---

%=====================================================
%=====================================================
\section{\label{Katie}ITMD with the Monte Carlo tool \KaTie}
\label{sec:katie}
\newcommand{\xVar}{\omega}
\newcommand{\dxVar}{d^{\hspace{0.1ex}3n}\hspace{-0.2ex}\omega}
\newcommand{\Vol}{V}
\newcommand{\xInst}{x_{\mathrm{ist}}}
\newcommand{\xFnst}{x_{\mathrm{fst}}}
\newcommand{\Npoints}{N}
\newcommand{\Ffunc}{F}
\newcommand{\Gfunc}{G}
\newcommand{\Ord}{\mathcal{O}}
\newcommand{\bin}{\mathrm{b}}
\KaTie~\cite{vanHameren:2016kkz} is a parton-level Monte Carlo event generator that can deal with space-like initial-state partons for arbitrary tree-level processes within the Standard Model.
This means that it can be provided with $k_T$-dependent PDFs and that it will automatically calculate the necessary matrix elements with space-like initial-state partons to create parton-level event files.
%
%\cite{Kanaki:2000ey,Krauss:2001iv,Mangano:2002ea,Moretti:2001zz,Kilian:2007gr,Maltoni:2002qb}
%
These event files can be chosen to be in the LHEF-format~\cite{Alwall:2006yp}.
The $k_T$-dependent PDFs can be provided via TMDlib~\cite{Abdulov:2021ivr}, or via independent grid files, which however must be in a specific format.
Furthermore, \KaTie\ can perform calculations within ITMD factorization, and automatically calculates the necessary gauge invariant matrix elements corresponding to the color structures associated with the several gluon distributions that appear in this factorization scheme.
While \KaTie\ primarily creates event files, it also provide tools to create histograms of differential distributions.

\KaTie\ can be downloaded from \url{ https://bitbucket.org/hameren/katie/downloads/}, and a description of use is provided in the manual there.
Here, we provide some background on the Monte Carlo method it employs, elucidating the procedures to be followed when using \KaTie.

In hybrid factorization like ITMD factorization, the cross section can be written as a $3n$-dimensional integral
%
%%%%%%%%%%%%%%%%%%%%%%%%%%%%%%%%%%%%%%%%
\begin{equation}
\sigma = \int \dxVar\,\Ffunc(\xVar)
~,
\end{equation}
%%%%%%%%%%%%%%%%%%%%%%%%%%%%%%%%%%%%%%%%
%
where $n$ is the number of final-state momenta.
There are $4$ initial-state variables, and $3n-4$ final-state variables.
For hadro-production of dijets at tree-level the number $n=2$.
The function $F$ includes the collinear PDF, the TMDS, the flux factor, and the hard matrix elements.
Given a probability density $\Gfunc(\xVar)$ in the integration space (or {\em phase space}) that is non-zero whenever $\Ffunc(\xVar)$ is non-zero, we can write
%
%
%%%%%%%%%%%%%%%%%%%%%%%%%%%%%%%%%%%%%%%%
\begin{equation}
\sigma = \int \dxVar\,\Gfunc(\xVar)\,\frac{\Ffunc(\xVar)}{\Gfunc(\xVar)}
~.
\end{equation}
%%%%%%%%%%%%%%%%%%%%%%%%%%%%%%%%%%%%%%%%
%
The Monte Carlo method is based on the Central Limit Theorem, which dictates that if $\{\xVar_1,\xVar_2,$ $\ldots,$ $\xVar_{\Npoints}\}$ is a sequence of random phase space points, or {\em events}, independently drawn from density $G(\xVar)$, then
%
%%%%%%%%%%%%%%%%%%%%%%%%%%%%%%%%%%%%%%%%
\begin{equation}
\frac{1}{\Npoints}\sum_{i=1}^{\Npoints}\frac{\Ffunc(\xVar_i)}{\Gfunc(\xVar_i)}
= \sigma 
+ \Ord\bigg(\frac{1}{\sqrt{\Npoints}}\bigg)
~.
\end{equation}
%%%%%%%%%%%%%%%%%%%%%%%%%%%%%%%%%%%%%%%%
%
Clearly, the approximation converges to the correct result faster if the fluctuation over the terms in the sum is smaller, with the optimum when $\Gfunc(\xVar)=\Ffunc(\xVar)/\sigma$.
Successfully applying the Monte Carlo method means that one solves the problem of bringing $\Gfunc(\xVar)$ to this optimum to satisfactory degree.
Reaching the optimum implies having solved the integration problem and eliminates the need for the Monte Carlo procedure.
Satisfaction means that the result can be obtained within an acceptable time, and in practice means finding a compromise between the number of terms needed, and the complexity of the algorithm to produce the sequence of events. 

An advantage of the Monte Carlo method is that, given a ``satisfactory'' sequence {\em generator} to calculate the cross section $\sigma$, it can also be used to estimate differential distributions.
Let $\varphi(\xVar)$ be an observable, and $\bin_{j}(\varphi)=\theta(\varphi-\varphi_j)\theta(\varphi_{j+1}-\varphi)$ be a bin for this observable between values $[\varphi_j,\varphi_{j+1}]$, then
%
%%%%%%%%%%%%%%%%%%%%%%%%%%%%%%%%%%%%%%%%
\begin{equation}
\frac{1}{\Npoints}\sum_{i=1}^{\Npoints}\frac{\Ffunc(\xVar_i)}{\Gfunc(\xVar_i)}
\,\bin_j(\varphi(\xVar_i))
=
\int_{\varphi_j}^{\varphi_{j+1}}\hspace{-3.5ex}d\varphi\,\frac{d\sigma}{d\varphi}
+ \Ord\bigg(\frac{1}{\sqrt{\Npoints}}\bigg)
~.
\end{equation}
%%%%%%%%%%%%%%%%%%%%%%%%%%%%%%%%%%%%%%%%
%
The idea of an {\em event file} is to store events $\xVar_i$ and their {\em weight} $W_i=\Ffunc(\xVar_i)/\Gfunc(\xVar_i)$ to produce arbitrary distributions without having to redo the generation of events.

In the context of the foregoing, \KaTie\ performs roughly speaking two tasks: it calculates the hard matrix elements, automatically and efficiently, as part of the evaluation of $\Ffunc(\xVar)$, and, provided with PDFs and TMDs, it creates reasonably-sized event files.
In order to perform the first task, \KaTie\ employs the Dyson-Schwinger approach to calculate tree-level helicity amplitudes numercially, as first proposed in~\cite{Caravaglios:1995cd} and also utilized in other tree-level programs, notably {\sc Alpgen}~\cite{Mangano:2002ea}.
The feature of space-like initial-state momenta is dealt with following the auxiliary-parton method outlined in~\cite{vanHameren:2012if}.
The second task is achieved through the application of many well-known optimization techniques, a few of which are worth to mention in order to understand the procedures in \KaTie's operation.

One is {\em adaptive importance sampling}, which optimizes adaptable probability densities iteratively, using the information of events generated so far in each iterative step.
While being very effective, it can cause bias in those events, and they must not be used for the actual event file.
Consequently, there must be a separate optimization stage before the generation of the event file starts.
This happens for each partonic subprocess separately.
Once this has been performed, the information of the optimized densities is stored, and is from then on used to generate an arbitrary number of event files.

Secondly, \KaTie\ employs {\em rejection}.
Phase space cuts are required to mimic detector acceptance and to avoid singularities in tree-level matrix elements.
These typically are formulated in terms of variables that are non-trivial functions of the variables in $\xVar$ that are actually generated, and cannot be implemented as exact integration bounds.
Instead, the integrand $\Ffunc(\xVar)$ is imagined to vanish outside those phase space cuts, and a bigger enveloping space is generated.
This leads to many events with $W_i=0$, which are however not stored but are included in the eventual normalization of the weights.
In order to achieve the unbiased normalization of the event weights in case several event files are generated, so called {\em raw} files are stored instead of actual event files, containing non-normalized weights and more statistical information.
Event files in the LHF format can be extracted from these.

The third method that needs to be mentioned is {\em unweighting}.
This is a statistical procedure to reduce the fluctuations of the weights $W_i$ while keeping the event file sound.
In practice this procedure ``weeds out'' low-weight events and reweighs remaining ones.
The cross section estimate is not affected (is also not getting better) but it allows to reduce the required size of the event file.
While it causes many events to get the same constant weight, there may occur events with higher weights.

The phase space is only $3n$-dimensional, but it is much more convenient to write the events in terms of the $4n$ variables being the parton-level initial-state and final-state momenta.
We will still denote events with the symbol $\xVar$.
While all components of $\Ffunc(\xVar_i)$, like the matrix element, the PDFs, the strong coupling etc., can be re-evaluated using these, it is more convenient to also store their values besides $W_i$.
This is useful if one wants to employ the {\em reweighting} procedure.
For example, let an event file be created within hybrid $k_T$-factorization with a TMD $\mathcal{F}(\xVar)$ (we simply imply that this function selects the appropriate variables from $\xVar$. Realize that this in practice includes the final-state ones due to the factorization scale dependence).
Suppose a user has their own TMD $\mathcal{F}'$ but only in a numerical form that cannot (yet) be married with \KaTie\ or TMDlib.
The value of $\mathcal{F}_i=\mathcal{F}(\xVar_i)$ is stored for each event, and the event file can be transformed to be valid for $\mathcal{F}'$ by multiplying
\begin{equation}
W_i \to W_i\,\frac{\mathcal{F}'(\xVar_i)}{\mathcal{F}_i}
\end{equation}
for each event.
It will increase the fluctuation of the weighs and reduce the quality of the event file, but most likely not in a drastic way.

As mentioned earlier, the use of \KaTie\ is described in the manual, but below we present the complete single input card that allows to create event files for dijets within ITMD.
{\small
\begin{verbatim}
# Only QCD interaction, QED and Weak are on by default
switch = withQCD   Yes
switch = withQED   No
switch = withWeak  No
# List of processes
process = g g -> g g  , factor = 1
process = g g -> u u~ , factor = Nf
process = u g -> g u  , factor = 1 
process = d g -> g d  , factor = 1 
process = s g -> g s  , factor = 1 
process = c g -> g c  , factor = 1 
process = b g -> g b  , factor = 1
# In the process definition B A -> 1 2..., put initial state A off-shell
offshell = 0 1
itmdf = yes
# Collinear pdf set
lhaSet = CT10nlo
# Example directory and filenames below
tmdTableDir = /home/user/projects/TMDs/
tmdpdf = qg1  myGridFile-Fqg1.dat
tmdpdf = qg2  myGridFile-Fqg2.dat
tmdpdf = gg1  myGridFile-Fgg1.dat
tmdpdf = gg2  myGridFile-Fgg2.dat
tmdpdf = gg6  myGridFile-Fgg6.dat
Nflavors = 5
# Number of non-zero weight events to be spent on optimization
Noptim = 100,000
# Summing squared helicity amplitudes is more efficient in number of
# events (less fluctuation), but slower in time than sampling.
helicity = sum
# Center-of-mass (CM) energy, events will be in CM frame.
Ecm = 8160
# Jet definition, 1 and 2 refer to the final-state momenta
cut = {deltaR|1,2|} > 0.4
# {pT|1|} is the pT of final-state number 1
# {pT|1|1,2} is the hardest of the pTs of final-state number 1 and 2
cut = {pT|1|1,2} > 28
cut = {pT|1|1,2} < 35
cut = {pT|2|1,2} > 28
cut = {pT|2|1,2} < 35
# In the process definition B A -> 1 2...,
# initial-state A has equal-sign energy and z-component,
# while initial-state B has opposite-sign.
# The off-shell initial-state A has lower x than B,
# so the final state is boosted towards negative rapidity.
cut = {rapidity|1|} > -4.0
cut = {rapidity|2|} > -4.0
cut = {rapidity|1|} < -2.7
cut = {rapidity|2|} < -2.7
# Renormalization/factorization scale
scale = ({pT|1|}+{pT|2|})/2
\end{verbatim}
}

%=====================================================
%=====================================================
\section{Phenomenology of forward dijets}
\label{sec:Results}

%-----------------------------------------------------
\subsection{TMD gluon distributions}
 From the BK equation  discussed in the earlier sections one can obtain TMD dipole gluon density. 
 In this section we will use $k^2=|\vec{k}_T|^2$ as argument of gluon density. This is the standard notation used in discussion of angular averaged distributions.\\
Let's define the following Fourier transforms \cite{Kimber:2001nm,Kutak:2003bd}
\begin{equation}
{\cal F}(x,k^2)=\frac{N_c}{\alpha_s (2\pi)^3}\int d^2b\int d^2re^{ik\cdot r}\nabla^2_{r}\,N(r,b,x),\,\,\,
\Phi(x,k^2)=\frac{1}{2\pi}\int d^2b \int\frac{d^2 r}{r^2}e^{i k \cdot r}N(r,b,x)
\label{eq:dipolampl}
\end{equation}
as well as explicit relation between the two functions 
${\cal F}(x,k^2)$ and $\Phi(x,k^2)$.

\begin{equation}
{\cal F}(x,k^2)=\frac{N_c}{4\alpha_s \pi^2}k^2\nabla^2_{k}\Phi(x,k^2),\,\,\,\,
\Phi(x,k^2)=\frac{\alpha_s\pi^2}{N_c}\int_{k^2}^{\infty}\frac{dl^2}{l^2}\ln\frac{l^2}{k^2}{\cal F}(x,l^2).
\end{equation}
Since we are working in the  large target approximation as well as in the forward limit \ie\ we neglect any momentum transfer therefore we introduce the following normalization conditions
for integration over the impact parameter
\begin{equation}
\int d^2{\bf b}\,S(b)=1,\,\,\,\int d^2{\bf b}\,S^2(b)=\frac{1}{\pi R^2}
\end{equation}
where $S(b)$ is the profile  function $ S(b)=\theta(R-b)/\pi R^2$.\\
The momentum space equation for the $\Phi(x,k^2)$ assumes the form:
\begin{equation}
\begin{split}
\Phi(x,k^2)= \Phi_0(x,k^2)+\overline\alpha_s\int_{x/x_0}^1\frac{dz}{z}
\int_0^{\infty}\frac{dl^2}{l^2}
\bigg[\frac{l^2\Phi(x/z,l^2)- k^2\Phi(x/z,k^2)}{|k^2-l^2|}+ \frac{
k^2\Phi(x/z,k^2)}{\sqrt{(4l^4+k^4)}}\bigg]\\
-\frac{\overline\alpha_s}{\pi R^2}\int_{x/x_0}^1\frac{dz}{z}\Phi^2(x/z,k^2).
\label{eq:wwglue}
\end{split}
\end{equation}
where $\overline\alpha=N_c\alpha_s/\pi$.\\
The the equation for ${\cal F}(x,k^2)$
is obtained from eq. (\ref{eq:wwglue}) by inserting in the nonlinear part of it relation expressing $\Phi (x,k^2)$  in terms of ${\cal F}(x,k^2)$ and acting on the whole equation with the operator that transforms $\Phi(x,k^2)$ to  ${\cal F}(x,k^2)$ see \cite{Kutak:2003bd} for the details  of this transformations. In the end one obtains \cite{Kutak:2003bd,Nikolaev:2006za,Bartels:2007dm}
\begin{equation}
\begin{split}
 {\cal F}(x,k^2)={\cal F}_0(x,k^2)+\overline\alpha_s\int_{x/x_0}^1\frac{dz}{z}\int_0^{\infty}\frac{dl^2}{l^2}
\bigg[\frac{l^2{\cal F}(x/z,l^2)- k^2{\cal F}(x/z,k^2)}{|k^2-l^2|}+ \frac{
k^2{\cal F}(x/z,k^2)}{\sqrt{(4l^4+k^4)}}\bigg]\\
-\frac{2\alpha_s^2\pi}{N_c R^2} \int_{x/x_0}^1\frac{dz}{z}\Bigg\{
\bigg[\int_{k^2}^{\infty}\frac{dl^2}{l^2}{\cal F}(x/z,l^2)\bigg]^2 
+\;{\cal F}(x/z,k^2)\int_{k^2}^{\infty}\frac{dl^2}{l^2}\ln\left(\frac{l^2}{k^2}\right){\cal F}(x/z,l^2)
\Bigg\}.
\label{eq:faneq1}
\end{split}
\end{equation}
The linear part   of the equation above is the well  known BFKL kernel while  the  nonlinear part  is the triple pomeron vertex.
The triple pomeron vertex  has such property that it is dominated by the anticollinear pole. We see  that as one evaluates the gluon at lower and lower values of $k^2$ the integration in the nonlinear term is over larger domain giving larger contribution and suppressing the  gluon density. 
One can also see that in the collinear  limit the nonlinear term completely drops and one  is left with linear equation.
\begin{equation}
{\cal F}(x,k^2) = {\cal F}^{(0)}(x,k^2)
+ \overline{\alpha}_s \int_{x/x_0}^1 \frac{dz}{z} \int^{k^2}_{k_{0}^2} {dk'^2} 
\frac{{\cal F}(\frac{x}{z},k'^2)}{k^2}
\; .
    \label{eq:virtual_kms_F}
\end{equation}

\subsection{BK equation within Kwiecinski Martin Sta\'sto model}
The (\ref{eq:faneq1}) is a leading order BK equation for dipole gluon density. 
In order to make it applicable to phenomenology it has been extended to account for higher order corrections following Kwiecinski-Martin-Stasto (KMS) prescription \cite{Kwiecinski:1997ee} that was originally applied to the BFKL equation.
Those include
\begin{itemize}
\item kinematical constraints which enforce that the virtuality of exchanged gluon is dominated by its transverse component $k^2=|\vec{k}_T|^2$. This constrain suppresses anticollinear pole and therefore suppresses the diffusion into infrared
\item nonsingular at small $z$ pieces of the splitting function
\item running coupling constant
\item sea  quark singlet contribution  to match the DGLAP limit at large $z$
\end{itemize}
With these corrections the equation assumes the form \cite{Kutak:2003bd}

\begin{multline}
\mathcal{F}\left(x,k^{2}\right)=\mathcal{F}_{0}\left(x,k^{2}\right)\\
+\frac{\alpha_{s}(k^2)N_{c}}{\pi}\int_{x/x_0}^{1}\frac{dz}{z}\int_{k_{0}^{2}}^{\infty}\frac{dl^{2}}{l^{2}}\left\{ \frac{l^{2}\mathcal{F}\left(\frac{x}{z},l^{2}\right)\theta\left(\frac{k^{2}}{z}-l^{2}\right)-k^{2}\mathcal{F}\left(\frac{x}{z},k^{2}\right)}{\left|l^{2}-k^{2}\right|}+\frac{k^{2}\mathcal{F}\left(\frac{x}{z},k^{2}\right)}{\sqrt{4l^{4}+k^{4}}}\right\} \\
+\frac{\alpha_{s}(k^2)}{2\pi k^{2}}\int_{x/x_0}^{1}dz\,\left\{ \left(P_{gg}\left(z\right)-\frac{2N_{c}}{z}\right)\int_{k_{0}^{2}}^{k^{2}}dl^{2}\mathcal{F}\left(\frac{x}{z},l^{2}\right)+zP_{gq}\left(z\right)\Sigma\left(\frac{x}{z},k^{2}\right)\right\} \\
-\frac{2\pi\alpha_{s}^{2}(k^2)}{N_cR^{2}}\left\{ \left[\int_{k^{2}}^{\infty}\frac{dl^{2}}{l^{2}}\mathcal{F}\left(x,l^{2}\right)\right]^{2}+\mathcal{F}\left(x,k^{2}\right)\int_{k^{2}}^{\infty}\frac{dl^{2}}{l^{2}}\,\ln\left(\frac{l^{2}}{k^{2}}\right)\mathcal{F}\left(x,l^{2}\right)\right\} \,,\label{eq:BKKMS}
\end{multline}
where $\Sigma(x,k^2)$ is a sea quark distribution
obeying essentially DGLAP equation in unintegrated form (further details can be found in \cite{Kutak:2003bd}).
Please  note that in the equation above lower cut in $k^2=k_0^2$ was introduced.
The origin of the cut is related to the method of solving of the equation and in principle  can be set to arbitrarily small value. 

In Ref.~\cite{Kutak:2012rf}, the following initial condition for Eq.~(\ref{eq:BKKMS}) was fitted to the $F_2$  proton
structure data from HERA \cite{Aaron:2009aa}
\begin{equation}
    {\cal F}_0(x,k^2)=\frac{\alpha_s(k^2)}{2\pi k^2}\int_x^1 dz\,
    P_{gg}(z)\,\frac{x}{z}\,g\left(\frac{x}{z},k_0^2\right)\,,
\end{equation}
with
\begin{equation}
    xg(x)=N(1-x)^{\beta}(1-xD)\,.
\end{equation}
For $k^2 \leq 1\, \text{GeV}^2$, the gluon distribution was taken as
$\mathcal{F} (x,k^2) = k^2
\mathcal{F}(x,1)$, which is motivated by the shape obtained
from the solution of the LO BK equation in the saturation regime
\cite{Sergey:2008wk}.

The fitting procedure gave the following numerical values for the parameters:
$N=0.94$, $\beta=18.6$, $D=-82.1$ and $R=2.40 \GeV^{-1}$. 
The overall quality of the fit was good,
with $\chi^2/\text{ndof} = 1.73$. We shall refer to this gluon distribution as the
Kutak-Sapeta or KS gluon.

For completeness, the fit of linearized version of Eq.~(\ref{eq:BKKMS}), \ie\ with
the last term dropped, was performed as well, and the following parameters were
obtained: $N=0.004$, $\beta=26.7$, $D=-51102$ and $\chi^2/\text{ndof} = 3.86$.
The presence of the parameter $R$, characterizing the target, allows one to
obtain dipole gluon distribution of nuclei.  In order to do that, one uses
relation $R_A=d\, A^{1/3} R $, which in the end gives the enhancement by $A^{1/3}$ of
the nonlinear term for gluon density normalized to the number of nucleons
\cite{Kutak:2012rf}. The parameter $d$ is a phenomenological factor that was was varied between $d=0.5$ and $d=1.0$. In the following computations we used the ``least saturation'' scenario with $d=0.5$.

%-----------------------------------------------------
\subsection{The Sudakov Resummation}
An important class of perturbative corrections to scattering process, appearing when the emitted partons are both collinear and soft, is resummed in terms of the Sudakov form factor \cite{Mueller:2013wwa,Mueller:2015ael,Mueller:2016xoc}.
It appears due to not exact cancellation of virtual corrections and real corrections, as a consequence of certain exclusivity of the final state.
In the  processes considered here the largest effect of the Sudakov is expected to affect the cross section when the jets are  in nearly back-to-back configuration. The large logarithm comes since the transverse momenta of jets can be sizable while the  imbalance $k_T$ of incoming space-like parton is small. The effect of Sudakov form factor is to suppress the back-to-back  configuration and enhance the moderate angle part of  the distribution, leading to so-called broadening.  
Within the small-$x$ formalism the Sudakov form factor can be factorized, in the coordinate space, from the hard process \cite{Mueller:2013wwa,Mueller:2015ael,Mueller:2016xoc}. The complete cross-section accounting for the Sudakov  form factor reads:  

\begin{multline}
\frac{d\sigma^{\mathrm{pA}\rightarrow j_1j_2+X}}{d^{2}P_{T}d^{2}k_{T}dy_{1}dy_{2}}
=
\sum_{a,c,d} x_{\mathrm{p}} 
\sum_{i=1}^{2}\mathcal{K}_{ag^*\to cd}^{\left(i\right)}\left(P_T,k_T;\mu\right) \\
\times\int db_T b_T J_0(b_T k_T) 
f_{a/\mathrm{p}}\left(x_{\mathrm{p}},\mu_b\right) 
\widetilde{\Phi}_{ag\rightarrow cd}^{\left(i\right)}\left(x_{\mathrm{A}},b_T\right)
e^{-S^{ag\to cd}(\mu,b_T)} \,,
\label{eq:itmd_Sud}
\end{multline}
where $\widetilde{\Phi}_{ag\rightarrow cd}^{\left(i\right)}$ is the Fourier transform of the TMD gluon distributions and 
$S^{ag\to cd}$ are the Sudakov factors written for each each channel
\begin{equation}
  \label{eq:sudpnp}
  S^{ab\to cd} (\mu,b_T) =\sum_{i=a,b,c,d} S_p^{i} (\mu,b_T) +
\sum_{i=a, c, d}S^{i}_{np} (\mu,b_T), 
\end{equation} 
where $S_p^{i}$ and $S_{np}^{i}$ are the perturbative and
non-perturbative contributions. 
The perturbative Sudakov factors, including double and single logarithms, are given by~\cite{Mueller:2012uf,Mueller:2013wwa}
\begin{gather}
S_p^{qg\to qg} (\mu, b_T) = \int_{\mu_b^2}^{\mu^2} \frac{dq_T^2}{q_T^2} \left[
2 (C_F + C_A) \frac{\alpha_s}{2\pi} \ln \left( \frac{\mu^2}{q_T^2} \right)
- \left(\frac{3}{2}C_F +  C_A \beta_0 \right) \frac{\alpha_s}{\pi}
  \right],
  \label{eq:sudpertqg}
  \\
S_p^{gg\to gg} (\mu, b_T) =  \int_{\mu_b^2}^{\mu^2} \frac{dq_T^2}{q_T^2} \left[
  4 C_A \frac{\alpha_s}{2\pi} \ln \left( \frac{\mu^2}{q_T^2} \right)
- 3 C_A \beta_0 \frac{\alpha_s}{\pi} \right]\,,
  \label{eq:sudpertgg}
\end{gather}
where 
$\beta_0 = (11-2n_f/N_c)/12$. 
The $g g \to q \bar q$ channel is negligible in  kinematic domain considered here.
In the above the scale $\mu_b$ is the inverse of the impact parameter:
\begin{equation}
\mu_b=2e^{-\gamma_E}/b_*    
\end{equation}
with
\begin{equation}
\label{eq:bstar-def}
 b_* = b_T/\sqrt{1+b_T^2/b_{\rm max}^2}\,.
\end{equation}
Given this selection, the scale $\mu_b$ becomes constant at the point of high $b_T$, where its value is determined to be $2 e^{-\gamma_E}/b_{\max}$, which is considerably greater than $\Lambda_\text{QCD}$. Following  Ref.~\cite{Marquet:2019ltn}, we will adopt $b_{\max} = 0.5, \text{GeV}^{-1}$.\\
For completeness we comment briefly on DGLAP based prescriptions for incorporating the Sudakov form factor.
The method relies on constructing the Sudakov form factor from DGLAP splitting function and using it to  
reshuffle events according to relation between hard scale and transverse momentum of the gluon. In such constructions one chooses some inclusive quantity to be unmodified while allowing for modification of unintegrated quantity.
Two such methods were presented in \cite{vanHameren:2014ala,Kutak:2014wga}. In the former the total cross section was preserved while in the later the integrated gluon density was unmodified. 

%-----------------------------------------------------
\subsection{Kutak-Sapeta (KS) gluon distribution}

\begin{figure}[p]
  \begin{center}
    \includegraphics[width=0.99\textwidth]{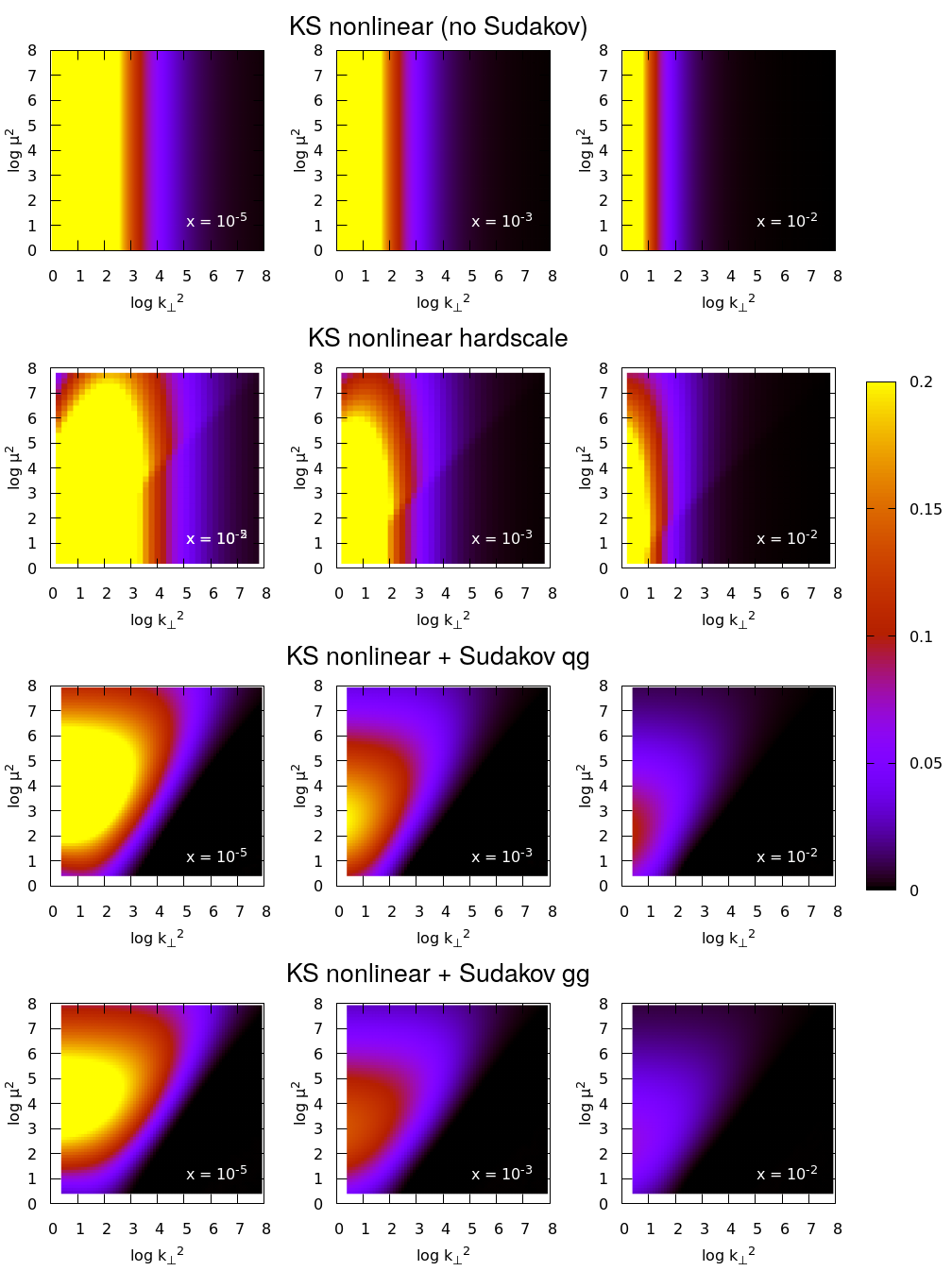}
  \end{center}
  \caption{
  KS gluon distribution, without and with the Sudakov
  form factors. The second row corresponds to the simple model-Sudakov given in
  Eq.~(\ref{eq:survival}), while the third and the fourth rows show results
  obtained with the Sudakov factors derived from QCD and given in
  Eqs.~(\ref{eq:sudpnp})--(\ref{eq:sudpertgg}).
  }
  \label{fig:gluon-map}
\end{figure}

We shall now discuss the KS gluon, introduced above, in variants with and
without the Sudakov resummation. The parameters were fixed by the original
fit~\cite{Kutak:2012rf} with no Sudakov factors and the gluon was later used
without any modifications. Hence, combining it with the Sudakov does not
introduce new parameters. This is true because the perturbative
part~(\ref{eq:sudpertqg}) and (\ref{eq:sudpertgg}) is parameter-free while the
non-perturbative terms are universal in the kinematic domain of our
study~\cite{Su:2014wpa}. 

We introduce the Sudakov effects into the KS gluon distribution following the
formalism described above. In addition, for reference, we use two methods
employed in our earlier studies~\cite{vanHameren:2014ala, Kutak:2014wga}.  Those
calculations used the Sudakov form factor, understood as the DGLAP evolution
kernel, that has been applied on the top of the \gluonTMD, together with
constrains such as unitarity. Those methods should therefore be considered as
models, in contrast to the proper resummation of Sudakov logarithms described
in the preceding section.
Nevertheless, the approaches used in Refs.~\cite{vanHameren:2014ala,
Kutak:2014wga} were phenomenologically successful (see
also~\cite{vanHameren:2019ysa}), and it is therefore useful and interesting to
compare the predictions of those simplistic models with the proper way
of including the Sudakov effects into the small-$x$ gluon.  

The reference models are:
\begin{itemize}
  %\justifying
  %\setlength\itemsep{1.2em}
  \item
  Model 1:
  The survival probability model~\cite{vanHameren:2014ala}, where the Sudakov
  factor of the form \cite{Watt:2003mx}
  \begin{equation}
    T_s(\mu_{F}^2,k_{T}^2)=
    \exp\left(-\int_{k_{T}^2}^{\mu_F^2}\frac{dk_{T}^{\prime
    2}}{k_{T}^{\prime 2}}\frac{\alpha_s(k^{\prime 2}_{T})}
    {2\pi}\sum_{a^\prime}\int_0^{1-\Delta}dz^{\prime}P_{a^\prime
    a}(z^\prime)\right)\,,
    \label{eq:survival}
  \end{equation}
  is imposed at the level of the cross section.  This procedure
  corresponds to performing a DGLAP-type evolution from the scale $\mu_0\sim
  k_T$ to $\mu$, decoupled from the small-$x$ evolution. 

  \item
  Model 2:
  The model with a hard scale introduced in Ref.~\cite{Kutak:2014wga}. The
  Sudakov form factor of the same form as in Eq.~(\ref{eq:survival}) is imposed
  on top of the KS gluon distribution in such a way that, after integration of
  the resulting hard scale dependent \gluonTMD, one obtains the same result as
  by integrating the KS gluon distribution.
\end{itemize}

In Fig.~\ref{fig:gluon-map}, we show the KS gluon distributions, with and without
Sudakov form factors, as functions of the transverse momentum $k_T$ and the
hard scale $\mu$. Three columns correspond to three different $x$ values. The
first row shows the original KS gluon distribution, which, as expected, does not
depend on the value of $\mu$. In the second row, we show the KS hardscale gluon
distribution of Ref.~\cite{Kutak:2014wga} (the other
model~\cite{vanHameren:2014ala} does not allow one to plot gluon distribution,
as it applies Sudakov effects at the cross section level via a reweighting
procedure).
Here, the dependence on $\mu$ is non-trivial and we see that the gluon develops
a maximum in that variable. As shown in the figure, this maximum is rather
broad.  In the third and the fourth row of Fig.~\ref{fig:gluon-map}, we present
the KS gluon distribution with the Sudakov form factor from
Eqs.~(\ref{eq:sudpnp})--(\ref{eq:sudpertgg}).  As explained earlier, this gluon
exists in two versions, one for the $qg$ and the other for the $gg$ channel. The
dependence on $k_T$ and $\mu$ is qualitatively similar between the these
gluons and the naive KS hardscale gluon distribution.  In the former case,
however, the peak is significantly narrower in $\mu$ as compared to the naive
model of Ref.~\cite{Kutak:2014wga}. It is interesting to note that the  $qg$
gluon is broader than the $gg$ gluon. This can be understood by comparing the
colour factors in the Sudakov functions~(\ref{eq:sudpertqg}) and~(\ref{eq:sudpertgg}).  
Since the colour factor for the $gg$ channel is larger than
for the $qg$, the Sudakov suppression is stronger along the $\mu$ direction in
the former case.

We have as well computed linear versions of the KS gluon distributions with the
Sudakov, using the KS linear gluon distribution of Ref.~\cite{Kutak:2012rf}.
The gluons discussed in this section are available publicly as part of the
{\tt KS package} and can be downloaded from 
\url{http://nz42.ifj.edu.pl/~sapeta/KSgluon-2.0.tar.gz}.

%--------------------------------------------------------
\subsection{ITMD gluons distributions}
\label{subsec:ITMD_distrib}

In the limit of small $x$ and large number of colors (Gaussian approximation), the
gluons listed in the table (\ref{tab:TMDs}) can be expressed in terms of dipole
gluon density as \cite{Dominguez:2011wm} 
\begin{eqnarray}
  {\cal F}_{qg}^{(1)}\big(x,|\vec{k}_T|\big)
  & = & xG^{(2)}\big(x,|\vec{q}_T|\big)\,,
  \label{GeneralFqg1}
  \\
  {\cal F}_{qg}^{(2)}\big(x,|\vec{k}_T|\big)
  & = &
  \int d^2q_T\,xG^{(1)}\big(x,|\vec{q}_T|\big) F\big(x, |\vec{k}_T-\vec{q}_T|\big)\,,
  \\
  {\cal F}_{gg}^{(1)}\big(x,|\vec{k}_T|\big)
  & = &
  \int d^2q_T\,xG^{(2)}\big(x,|\vec{q}_T|\big) F\big(x, |\vec{k}_T-\vec{q}_T|\big)\,,
  \\
  {\cal F}_{gg}^{(2)}\big(x,|\vec{k}_T|\big)
  & = &
  -\int d^2q_T\frac{(\vec{k}_T-\vec{q}_T)\cdot \vec{q}_T}{|\vec{q}_T|^2}
  xG^{(2)}\big(x,|\vec{q}_T|\big) F\big(x,|\vec{k}_T-\vec{q}_T|\big)\,,
  \\
  {\cal F}_{gg}^{(6)}\big(x,|\vec{k}_T|\big)
  & = &
  \int d^2q_Td^2q_T'\,x\,G^{(1)}\big(x,|\vec{q}_T|\big) F\big(x,|\vec{q}\,'_{\!T}|\big) F\big(x,|\vec{k}_T-\vec{q}_T-\vec{q}\,'_{\!T}|\big)\ .
  \label{GeneralFgg6}
\end{eqnarray}
where

\begin{equation} 
x G^{(2)}\big(x, |\vec{k}_T|\big) = \frac{N_c\ |\vec{k}_T|^2\ S_\perp}{2\pi^2 \alpha_s} F\big(x,|\vec{k}_T|\big) \ ,
\label{eq:dipolegluon}
\end{equation}
and $F\big(x,|\vec{k}_T|\big)$ is a Fourier transform of the fundamental dipole
\begin{equation} 
F\big(x,|\vec{k}_T|\big) = \int \frac{d^2{\vec{r}}}{(2\pi)^2} e^{-i{\vec{k}_T}\cdot{\vec{r}}} \left< \text{Tr} \left [ U({\vec{r}}\,)U^\dagger({0})\right]\right>_x/N_c ,
\label{eq:FTinv}
\end{equation}

The gluons listed above form a set from which one can construct the gluon densities $\Phi^{(i)}$ entering the ITMD factorization formula (\ref{eq:ITMD_factorization}) as obtained in \cite{Kotko:2015ura}:
\bea 
\Phi_{qg\to qg}^{(1)} = \mathcal{F}_{qg}^{(1)}~~~~~&,&~~~~~\Phi_{qg\to qg}^{(2)} \approx \mathcal{F}_{qg}^{(2)} \\
\Phi_{gg\to q\bar{q}}^{(1)} \approx \mathcal{F}_{gg}^{(1)}~~~~~&,&~~~~~\Phi_{gg\to q\bar{q}}^{(2)} \approx - N_c^2\mathcal{F}_{gg}^{(2)} 
\label{eq:phigg2}
\\
\Phi_{gg\to gg}^{(1)} \approx 
\frac{1}{2}\left(\mathcal{F}_{gg}^{(1)}+\mathcal{F}_{gg}^{(6)}\right)~~~~~&,&~~~~~
\Phi_{gg\to gg}^{(2)} \approx \mathcal{F}_{gg}^{(2)}+\mathcal{F}_{gg}^{(6)}\,.
\eea

%-----------------------------------------------------------------------------
\subsection{ITMD distributions from KS gluon}

\begin{figure}[t]
\begin{center}
\includegraphics[width=0.47\textwidth]{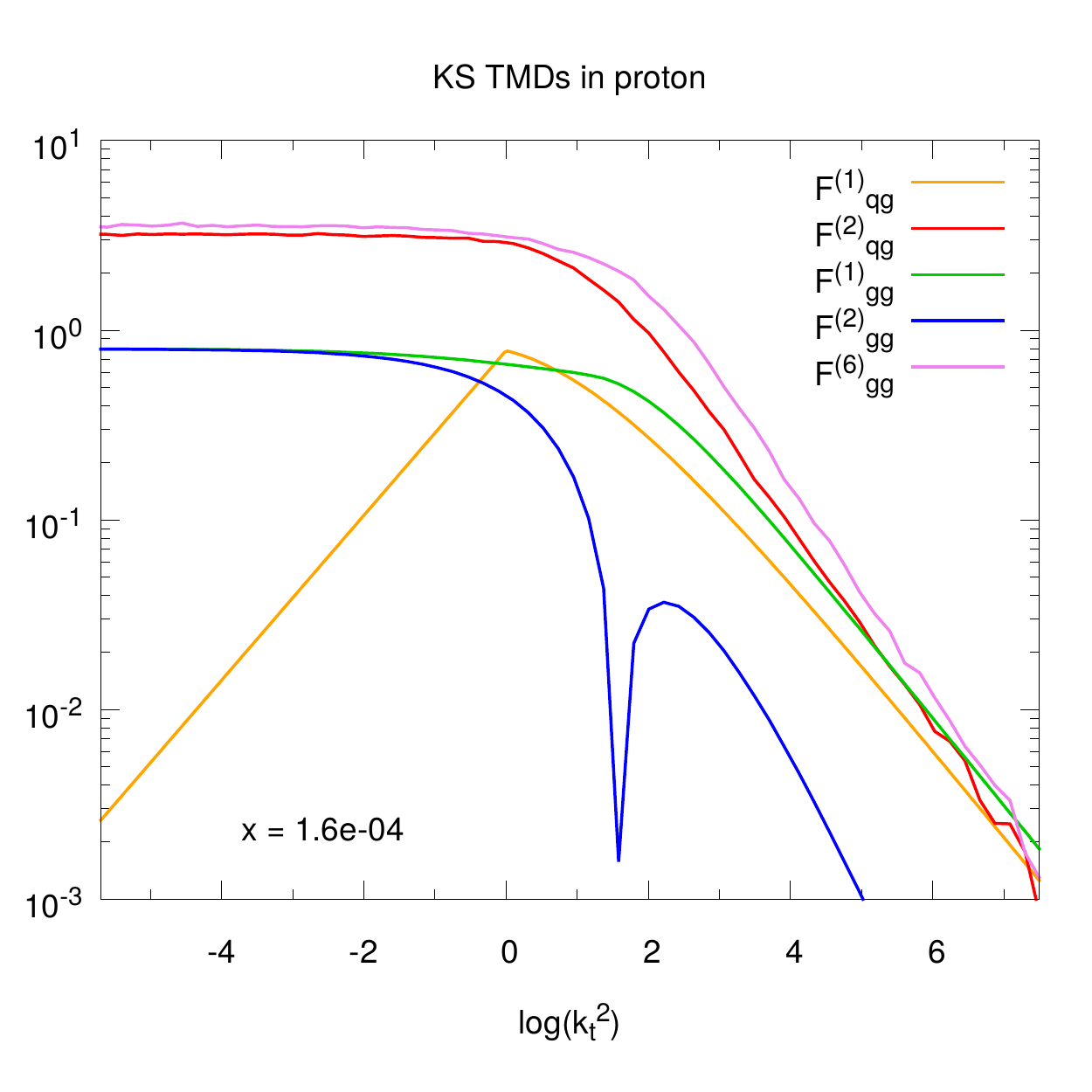}
\hfill
\includegraphics[width=0.47\textwidth]{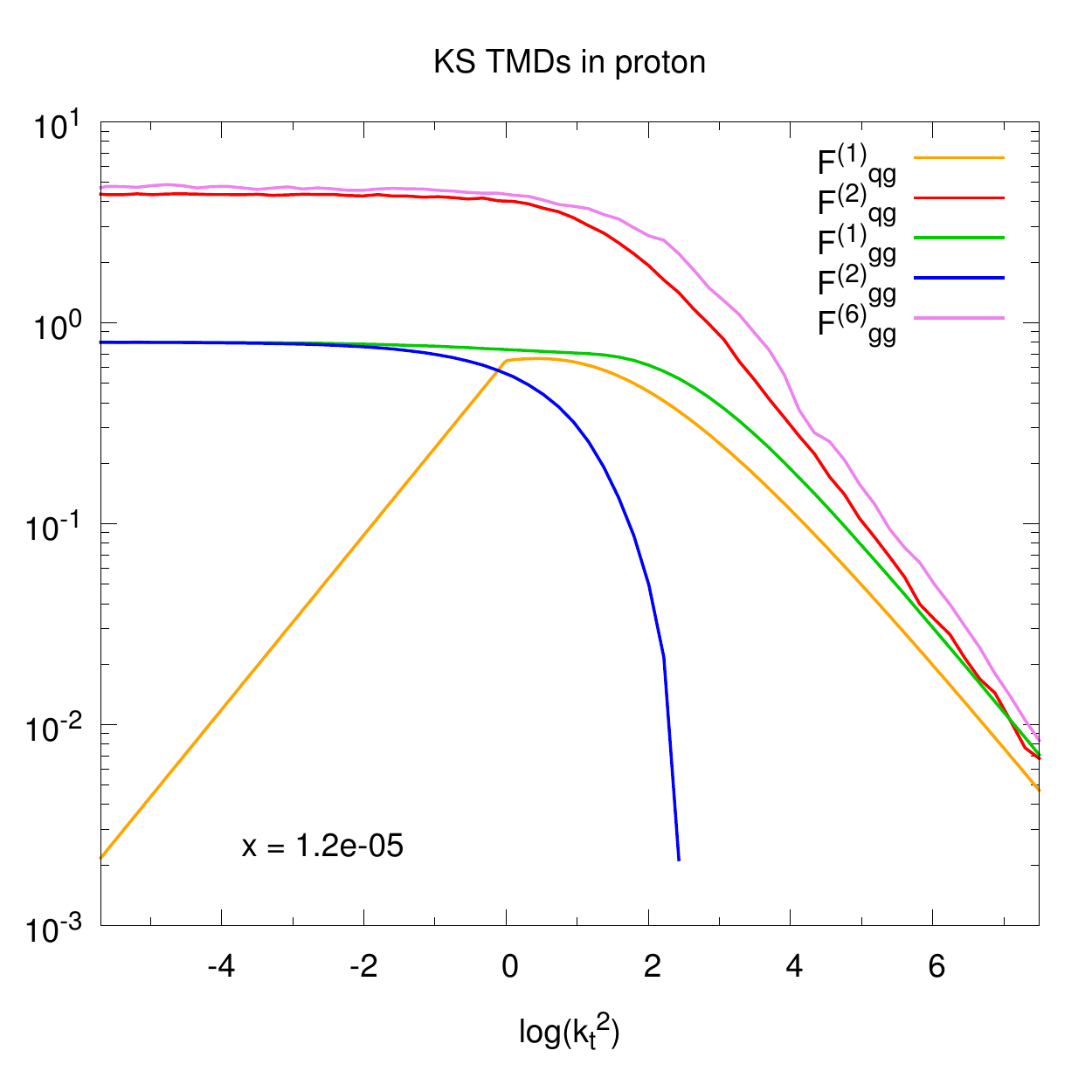}
 \end{center}
\caption{The KS gluon TMDs as a function of $\log (|\vec{k}_T|^2/\mbox{GeV}^2)$ at $x=1.6\ 10^{-4}$ for the proton (left) and the lead nucleus (right).
Since ${\cal F}_{gg}^{(2)}$ goes negative, its absolute value is shown on the figures.}
\label{fig:KSgluons}
\end{figure}

All the gluon TMDs (\ref{GeneralFqg1})-(\ref{GeneralFgg6}) can be calculated
from a single $xG^{(2)}\big(x,|\vec{k}_T|\big)$ distrubution in the above Gaussian
approximation. However, because
the KS gluon provides directly an impact-parameter-integrated distribution,
it is not straightforward to identify $S_\perp$ and obtain $F\big(x,|\vec{k}_T|\big)$ from
Eq.~(\ref{eq:dipolegluon}). To address this issue, we aplied the following
procedure~\cite{vanHameren:2016ftb}.
We first computed the dipole cross section
$\sigma_{\rm dipole}(x,r\!=\!|{\bf{r}}|)=2\int d^2b\ N_F(x, {\bf{r}})$ from
$xG^{(2)}\big(x,|\vec{k}_T|\big)$ by inverse Fourier transformation of Eq.~(\ref{eq:FTinv}),
and then defined $S_\perp$ as its value at large $r$, \ie\ when it saturates (since in that limit $N_F\to1$):
\begin{equation}
\frac{1}{2} \sigma_\text{dipole}(x,r\!=\!\infty)= S_\perp(x)=
\lim_{r \to \infty}
\frac{4\pi^3}{N_c}\alpha_s\int\frac{d|\vec{k}_T|}{|\vec{k}_T|}\Big[1-J_0\big(r|\vec{k}_T|\big)\Big] xG^{(2)}\big(x,|\vec{k}_T|\big)\ .
\end{equation}
We can now obtain $F\big(x,|\vec{k}_T|\big)$ and calculate all the needed gluon TMDs. Their
behavior as a function of $k_t=|\vec{k}_T|$ is plotted in Fig.~\ref{fig:KSgluons}, both for
the proton and the lead nucleus. The small mismatch between their high-$k_t$
behavior, expected due to the initial condition for the $x$ evolution, can be
observed.

Similarly, we computed~\cite{Al-Mashad:2022zbq} the ITMD distributions from the
KS gluon with the proper QCD Sudakov, shown in Fig.~\ref{fig:gluon-map}.

Within the Gaussian approximation, one can also derive the following formula for
the WW gluon density \cite{vanHameren:2016ftb}
\begin{equation}
  \calF^{(3)}_{gg} (x,
  |\vec{k}_T|)=\frac{2\pi^2\alpha_s}{N_c|\vec{k}_T|^2S_\perp}\frac{1}{2}\int_{|\vec{k}_T|^2}{d^2k_T^{\prime}}\ln\frac{|\vec{k}_T^{\prime
  }|^2}{|\vec{k}_T|^2}
  \int\frac{d^2q_T}{|\vec{q}_T|^2}{\cal F}_{qg}^{(1)}\big(x,|\vec{q}_T|\big){\cal
  F}_{qg}^{(1)}\big(x,|\vec{k}_T'-\vec{q}_T|\big)\,,
  \label{eq:WWfromDip}
\end{equation}
where ${\cal F}_{qg}^{(1)}$ is the dipole gluon density and $S_\perp$ is the
target's transverse area. 

\begin{figure}[t]
  \begin{center}
    \includegraphics[width=0.8\textwidth]{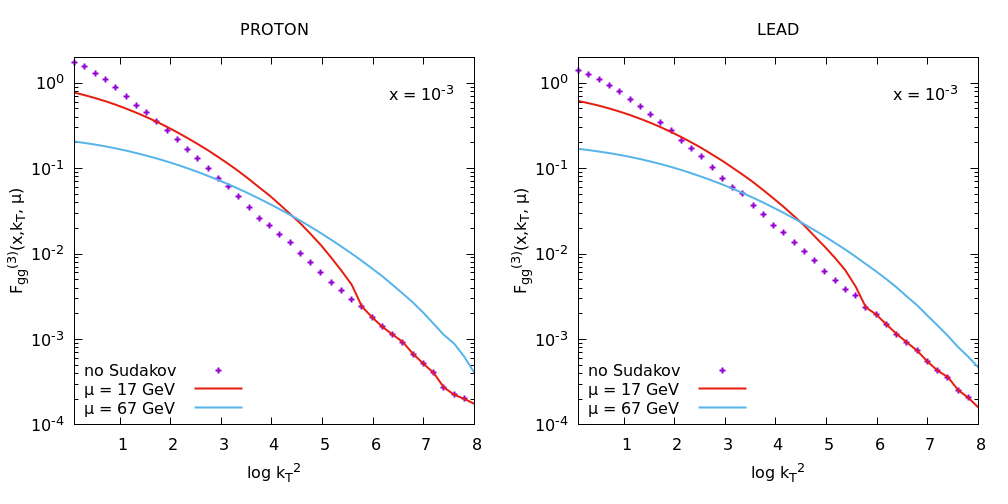}
  \end{center}
  \caption{
  The WW gluon density in the proton (left) and lead
  (right),  with and without Sudakov resummation, as a function of the
  transverse momentum of the gluon, for various values of the hard scale.
  }
  \label{fig:Fgg3}
\end{figure}

Using the procedure described above, we computed also the WW gluon, which we
show in Fig.~\ref{fig:Fgg3} in proton~(left)
and lead~(right), with and without Sudakov form factors, as functions of the
transverse momentum $|\vec{k}_T|$ and the hard scale $\mu$, for one particular $x =
10^{-3}$. (The gluon density is available from the TMDlib
\cite{Abdulov:2021ivr}.) 

First of all, let us notice that the WW gluon distribution has no maximum,
contrary to the dipole gluon~\cite{Kutak:2012rf, vanHameren:2016ftb}.  Secondly,
we see that the Sudakov factor suppresses the gluon distribution at low $|\vec{k}_T|$
and enhances it at higher $|\vec{k}_T|$. Because the Sudakov form factor is derived in
the regime $\mu \propto |\vec{p}_T| \gg |\vec{k}_T|$, we apply it only to that part of the gluon
density where $\mu>|\vec{k}_T|$. In the remaining domain, we use the gluon without
Sudakov, given in Eq.~(\ref{eq:WWfromDip}). This is visible in
Fig.~\ref{fig:Fgg3} as a kink of the curve corresponding to $\mu = 17\, \GeV$.
(A similar kink exits also for the $\mu = 67\, \GeV$ curve but it is located at
larger values of $\log |\vec{k}_T|^2$.)

All the ITMD gluons discussed in this above are available publicly and can be
downloaded from \url{http://nz42.ifj.edu.pl/~sapeta/itmd-KS.tar.gz}.

%%%%%%%%%%%%%%%%%%%%%%%%%%%%%%%%%%%%%%%%%%%%%%%%%%%%%%%%%%%%%%%%%%%%%%%
\subsection{Forward dijets at the LHC}

In the following section we review some existing predictions for forward dijet production in proton-proton and proton-lead collisions at LHC energies, obtained within the ITMD framework. 
We also include some new computations, not published before.

Before we proceed, it is important to mention, that the first predictions for forward dijets in saturation formalism using the ITMD, that -- as discussed in Section~\ref{sec:ITMDsec} -- approximates the CGC for sufficiently large transverse momenta, appeared in \cite{vanHameren:2016ftb}.
This computation significantly improved the predictions of \cite{vanHameren:2014lna} obtained with $k_T$-factorization, which is not entirely accurate at small dijet imbalance.
Later, in \cite{Fujii:2020bkl} a full computation within CGC was compared to ITMD, confirming, that at larger transverse momenta ITMD is adequate. 

\begin{figure}
  \begin{center}
    \includegraphics[width=0.99\textwidth]{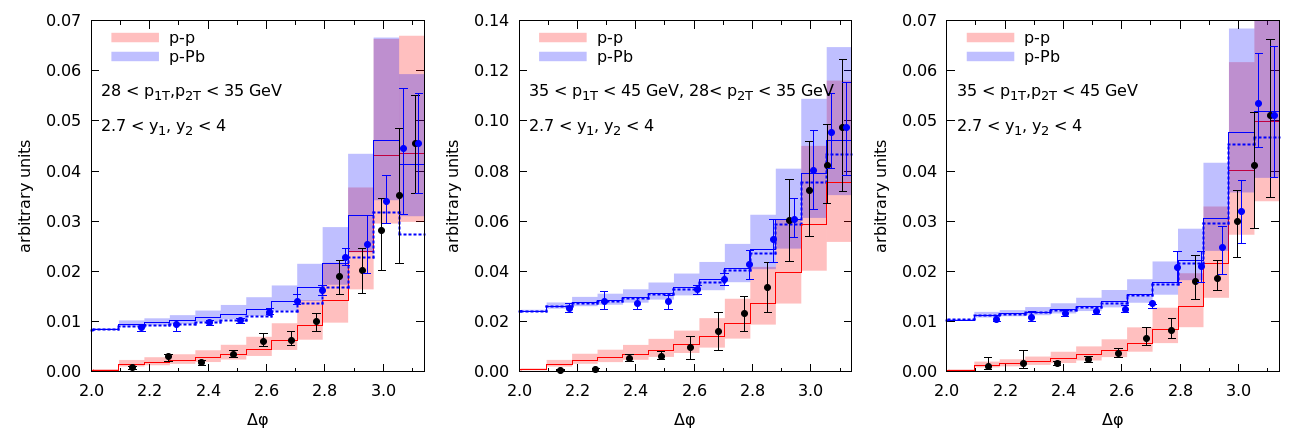}
  \end{center}
  \caption{\small 
    Broadening of  azimuthal correlations in p-Pb collisions vs p-p collisions
    for different sets of cuts imposed on the jets' transverse momenta.
    The blue and red bands show the normalized differential cross sections in azimuthal angle $\Delta\phi$, respectively for for p-Pb and p-p, shifted so that they match in the first bin.  
    The points show the experimental data \cite{ATLAS:2019jgo} for p-p and p-Pb, where the p-Pb data were shifted by a pedestal, so that the values in the bin $\Delta\phi\sim \pi$ are the same.
    Theoretical calculations are represented by the histograms with uncertainty bands coming from varying the scale by factors 1/2 and 2.
  }
  \label{fig:broadening}
\end{figure}

None of the above computations included the Sudakov resummation, however. In \cite{vanHameren:2019ysa} a calculation has been performed that included both the full ITMD and the Sudakov resummation ``model'' \cite{vanHameren:2014ala}, based on the reweighting the events with the DGLAP Sudakov form factor. The Authors compared the shape of the obtained azimuthal dijet correlations for proton-proton and proton-lead with those obtained by ATLAS collaboration \cite{ATLAS:2019jgo} (no dijet cross section was actually measured, only the conditional yields). Interestingly, when the shapes of the azimuthal correlations for p-p and p-Pb are stacked together so that they match in the first bin, one can clearly see broadening of the p-Pb cross section. It turns out, that similar broadening is obtained within the ITMD, if both the saturation as well as the Sudakov resummation are present. We show this result in Fig.~\ref{fig:broadening}. 

In the following, we shall report on further advances in dijet computations within ITMD, with the full Sudakov resummation \cite{almashad2022dijet}.
In particular, we will present the numerical results for 
the differential cross sections in terms of the azimuthal angle $\Delta\phi$ between the leading and the sub-leading jets for the proton-proton and the proton-lead collisions at LHC energies. We shall also discuss
 the nuclear modification ratio $R_{\mathrm{p-Pb}}$
 \begin{equation}
    R_{\mathrm{p-Pb}}=\frac{\frac{d \sigma^{p+P b}}{d \mathcal{O}}}{A \frac{d \sigma^{p+p}}{d \mathcal{O}}}\,,
\end{equation}
where $\mathcal{O}$ is a differential related to an observable. Finally, we shall investigate differential cross sections in the rapidity of the leading and sub-leading jet.

The cross sections were computed using the \KaTie\ Monte Carlo program~\cite{vanHameren:2016kkz} within the ITMD factorization, both described in the preceding Sections. We considered the proton-proton and proton-lead collisions at $\sqrt{s}=5.02\,\mathrm{TeV}$, $8.16\,\mathrm{TeV}$ and $8.8\, \mathrm{TeV}$ per nucleon. For proton-proton collisions, we also computed the proton-proton cross section for $\sqrt{s}=14\,\mathrm{TeV}$. In order to define the leading and the sub-leading jets, mentioned above, we used the anti-$k_T$ jet clustering algorithm \cite{Cacciari:2008ab} with a radius of $R = 0.4$. Since our computation is leading order, the jet algorithm is actually equivalent to a simple cut in rapidity-azimuthal plane.
Motivated by the current and planned LHC experiments, we applied the following cuts to the transverse momentum of these jets:
\begin{enumerate}[label={\it\roman*}$\,$)]%,leftmargin=35pt]
    \item $28\, \mathrm{GeV} < p_{T1},p_{T2} < 35\, \mathrm{GeV}$,
    \item $35\, \mathrm{GeV} < p_{T1},p_{T2} < 45\, \mathrm{GeV}$,
    \item $35\, \mathrm{GeV} < p_{T1} < 45\, \mathrm{GeV}$ and  $28\, \mathrm{GeV} < p_{T2} < 35\, \mathrm{GeV}$,
    \item $p_{T1}, p_{T2} > 10\, \mathrm{GeV}$,
    \item $p_{T1}, p_{T2} > 20\, \mathrm{GeV}$.
\end{enumerate}
%Note, however, not all of these cuts were considered for each of the collision energies mentioned above. 
Specifically, we used the first three cuts $i) - iii)$ for the transverse momentum of the jets in the rapidity range $2.7<y^{\star}_{1},y^{\star}_{2}<4.0$, both for proton-proton and proton-lead collisions at $\sqrt{s}=8.16\,\mathrm{TeV}$. These  cuts correspond to the FCal calorimeter of the ATLAS detector and are motivated by the measurement \cite{ATLAS:2019jgo}. %For the same energy, we also considered the fourth cut $iv)$ for the jets in the rapidity range $3.8<y^{\star}_{1},y^{\star}_{2}<5.1$, both for proton-proton and proton-lead collisions. These correspond to the planned  FoCal extension of the ALICE detector \cite{ALICECollaboration:2719928}. 
%The results for  these kinematics were published in \cite{almashad2022dijet}. 
The last two cuts $iv)- v)$ were applied in the rapidity range $3.8<y^{\star}_{1},y^{\star}_{2}<5.1$, both for proton-proton and proton-lead collisions at $\sqrt{s}=5.02\,\mathrm{TeV}$, and $8.8\, \mathrm{TeV}$ energies per nucleon. These correspond to the planned  FoCal extension of the ALICE detector \cite{ALICECollaboration:2719928}.  For the same kinematic domain (rapidity and transverse momentum cuts for the jets), we also considered protons collisions at $\sqrt{s}=14\,\mathrm{TeV}$ (proton-lead collisions are not feasible at this energy). The result for $5.02\,\mathrm{TeV}$ and $14\,\mathrm{TeV}$  were not published in \cite{almashad2022dijet} and are thus new. 

The factorization and renormalization scales were set using the transverse momentum of the leading and the sub-leading jets  $\mu=(p_{T1}+p_{T2})/2$. For the TMD gluon distributions, we used the ones calculated in \cite{vanHameren:2016ftb} based on the Kutak-Sapeta (KS) fit of the dipole gluon density \cite{Kutak:2012rf}, as described in the preceding subsections. For the collinear PDFs we used the CTEQ10NLO PDF set \cite{Lai_2010} from LHAPDF6 \cite{Buckley_2015}. For the computation of the cross sections, we included the channels
    $qg^*\to qg$, with five quark flavours, and $gg^* \to gg$.
The channel $ gg^*\to \overline q q$ was neglected because its contribution, for the considered kinematic domain, is quite small \cite{Kutak:2012rf,vanHameren:2016ftb}. 

\begin{figure} 
  \centering
    \includegraphics[width=0.65\linewidth]{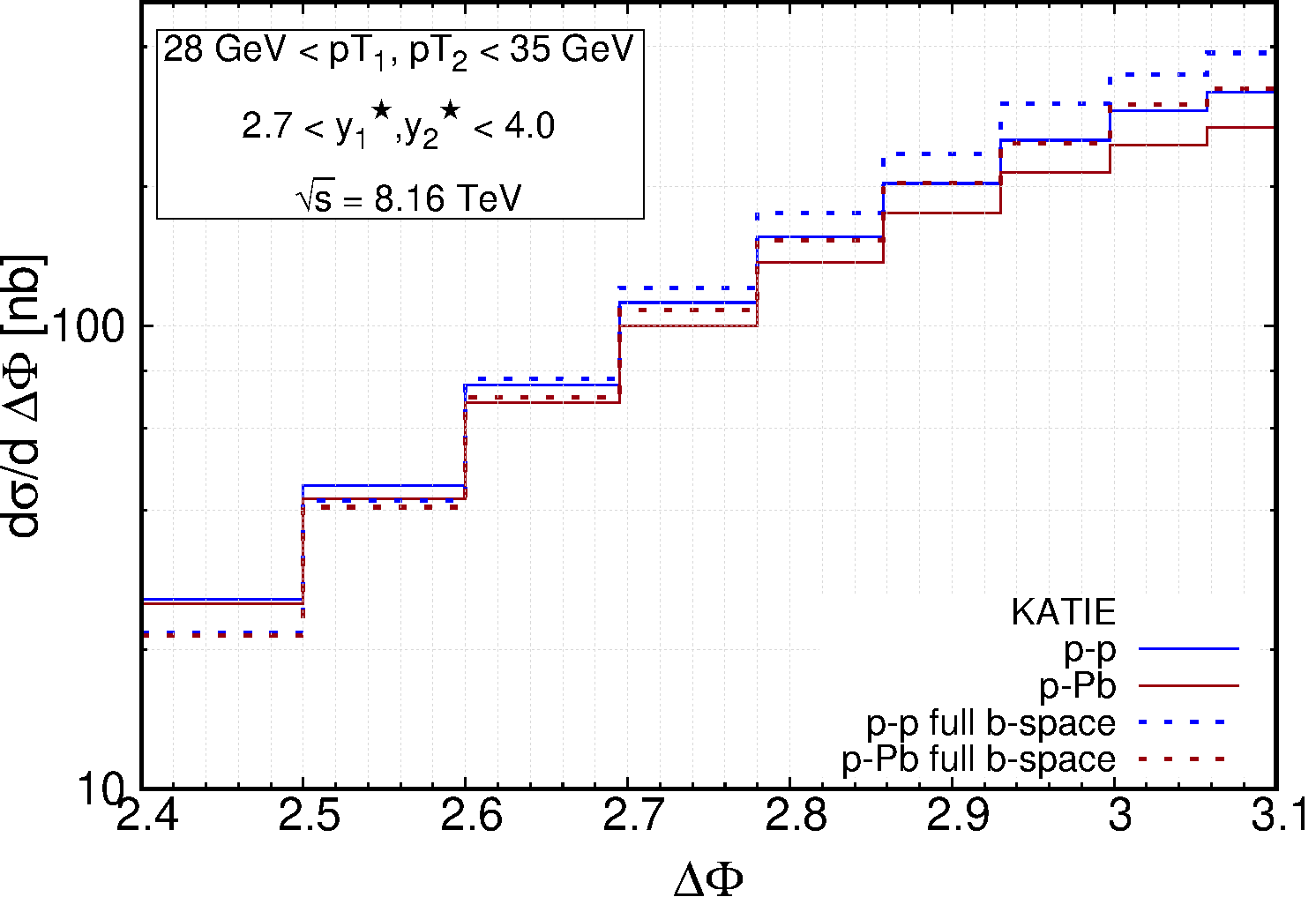}
    \includegraphics[width=0.65\linewidth]{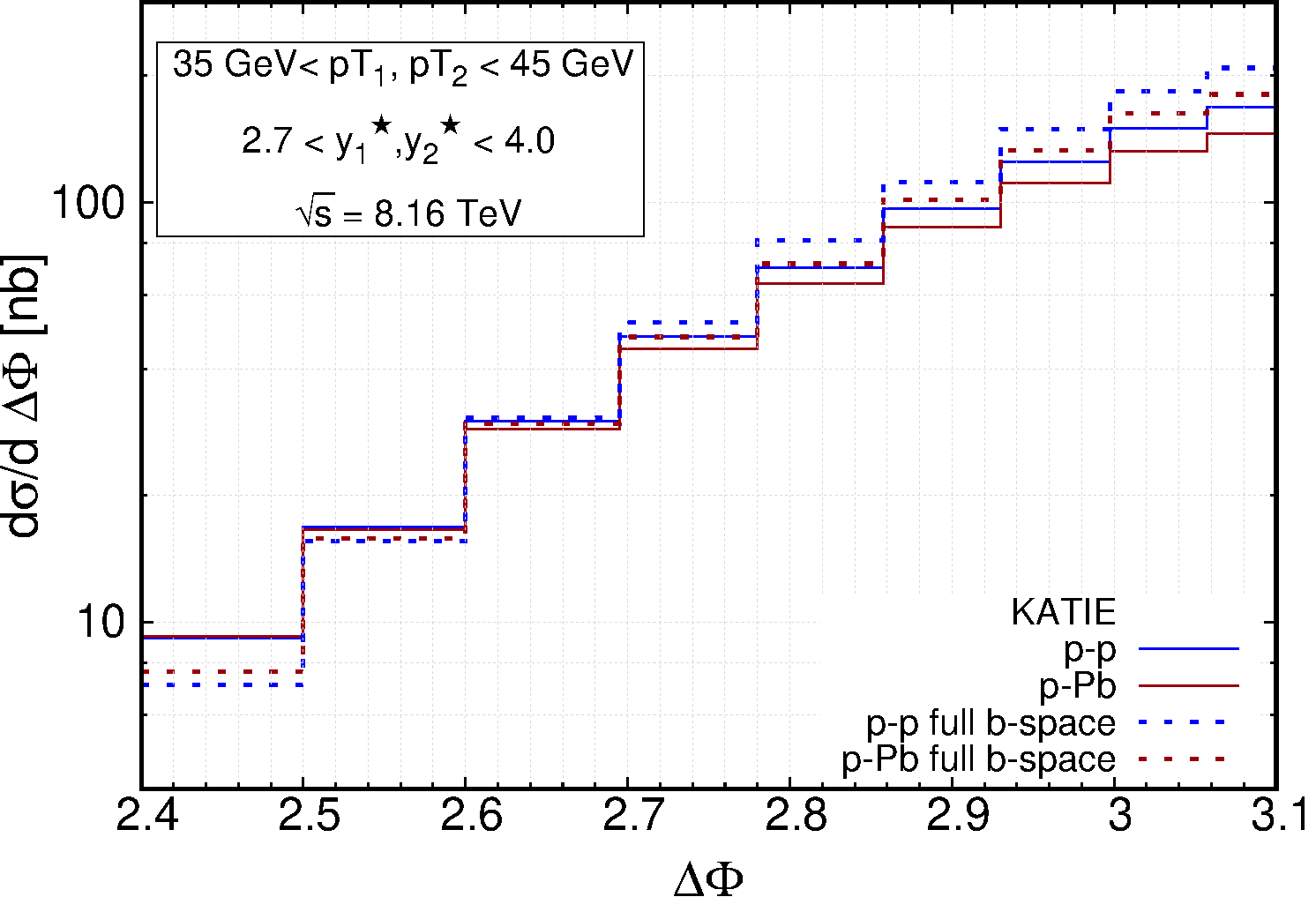}
    \includegraphics[width=0.65\linewidth]{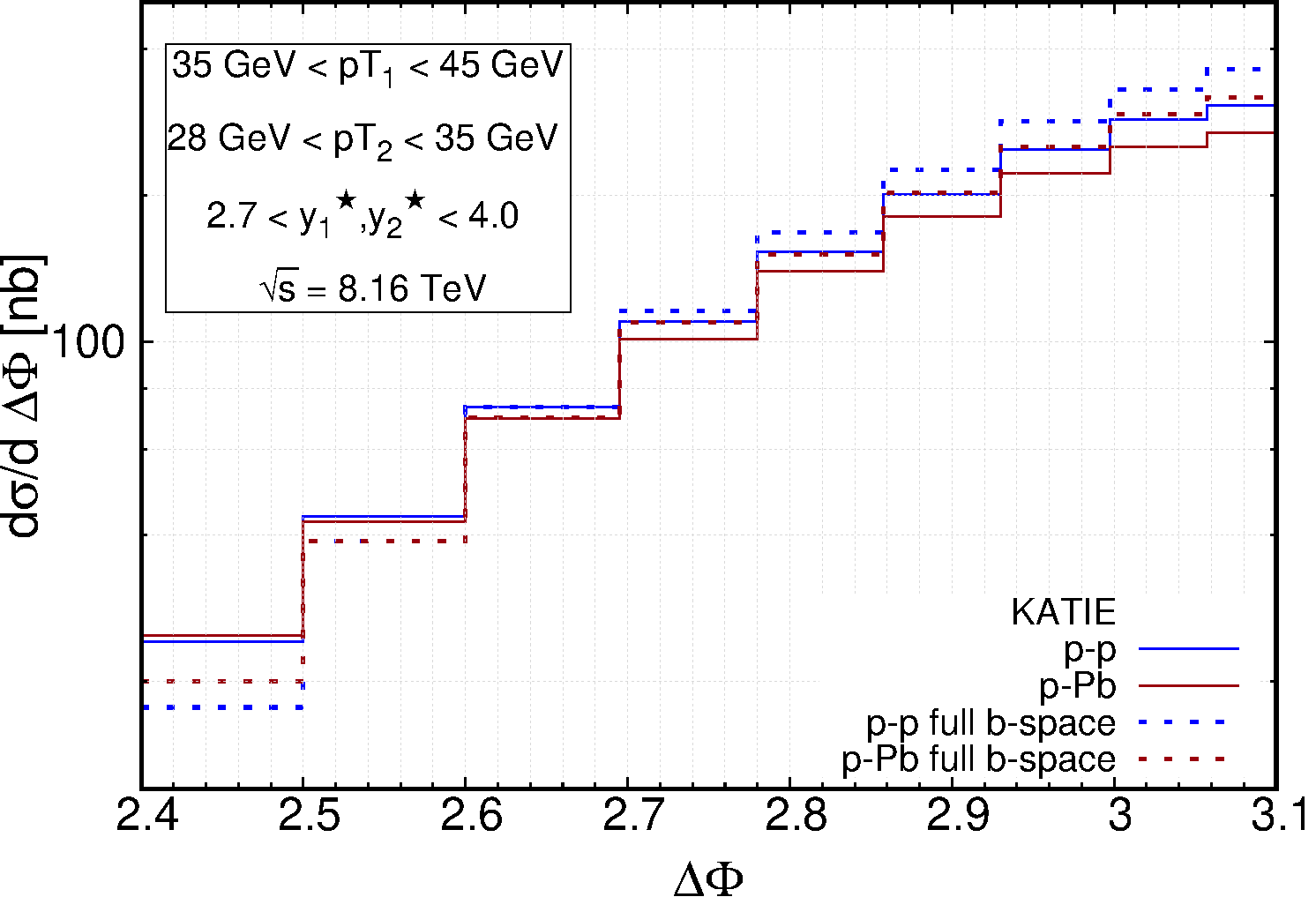}
  \caption{\small The differential cross sections in terms of the azimuthal angle $\Delta\phi$ between the leading and the sub-leading jets for the proton-proton and the proton-lead collisions at $\sqrt{s}=8.16\,\mathrm{TeV}$ in the FCal ATLAS kinematics. These were computed using \KaTie\ within the ITMD factorization formula with: the simplified Sudakov resummation Eq.~(\ref{eq:ITMD_factorization}) (solid lines), the full $b$-space resummation Eq.~(\ref{eq:itmd_Sud}) (dotted lines). The plots were taken from \cite{almashad2022dijet}.}
  \label{fig:only_katie} 
\end{figure}

\begin{figure} 
  \centering
    \includegraphics[width=0.65\linewidth]{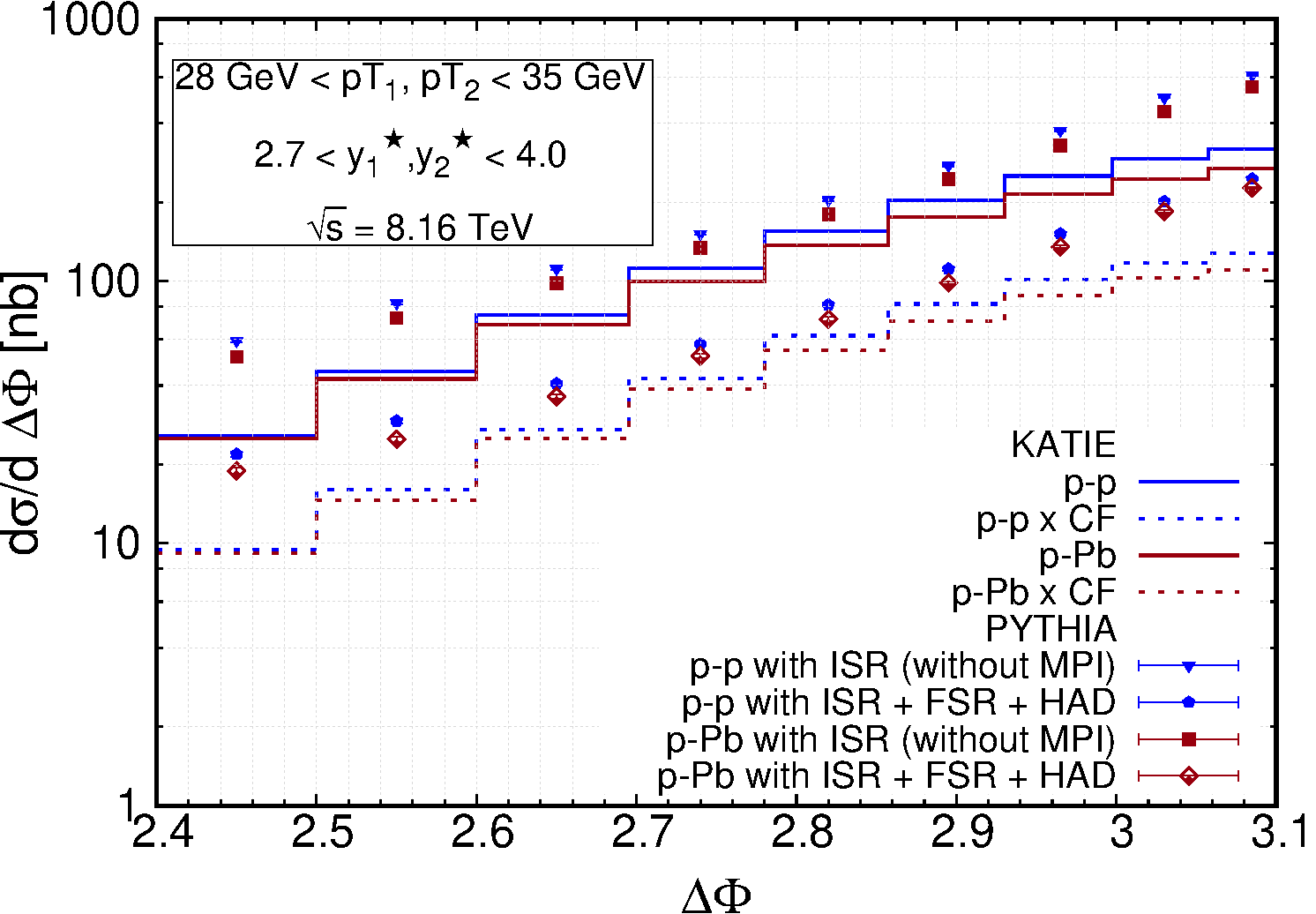}
    \includegraphics[width=0.65\linewidth]{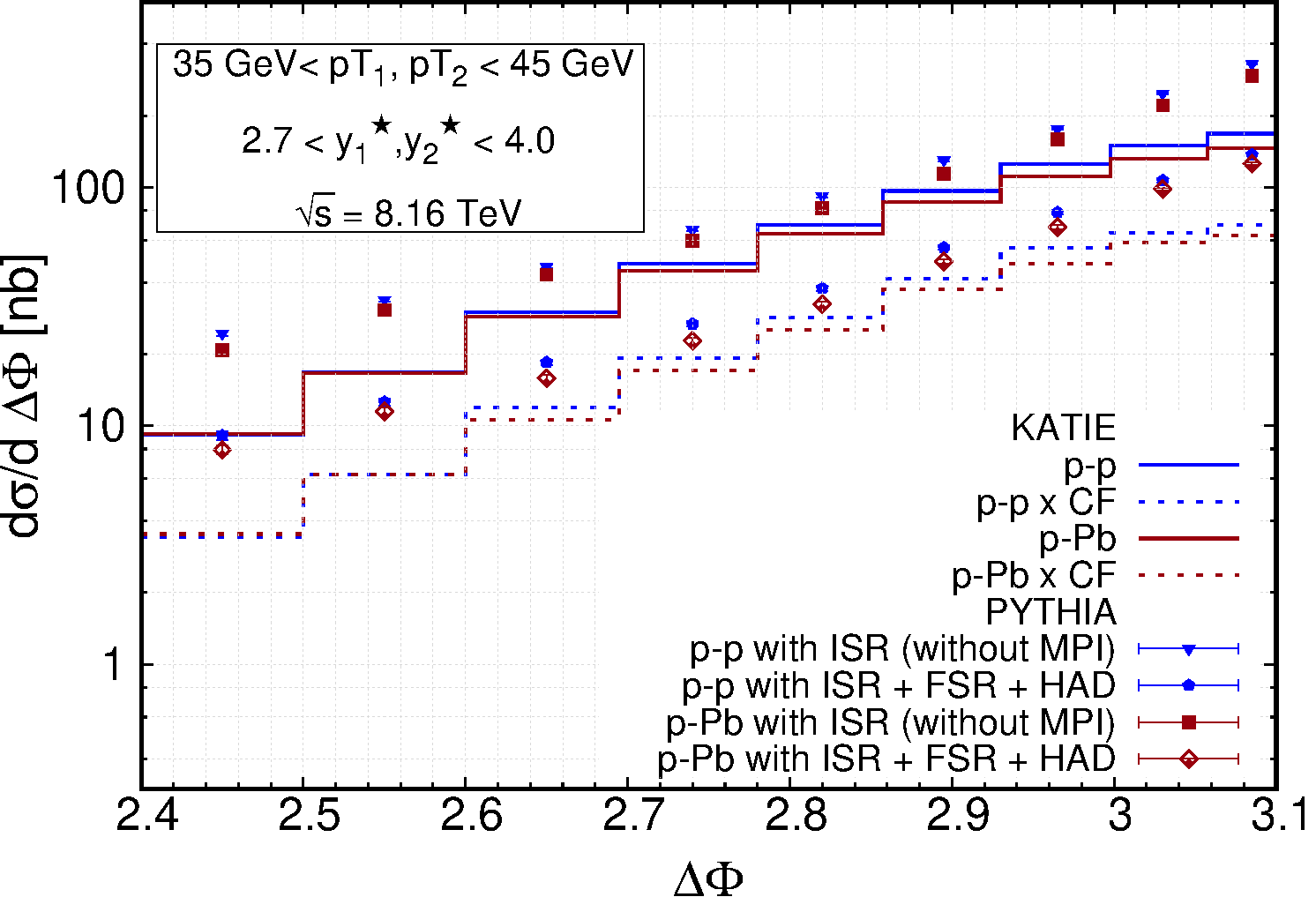}
    \includegraphics[width=0.65\linewidth]{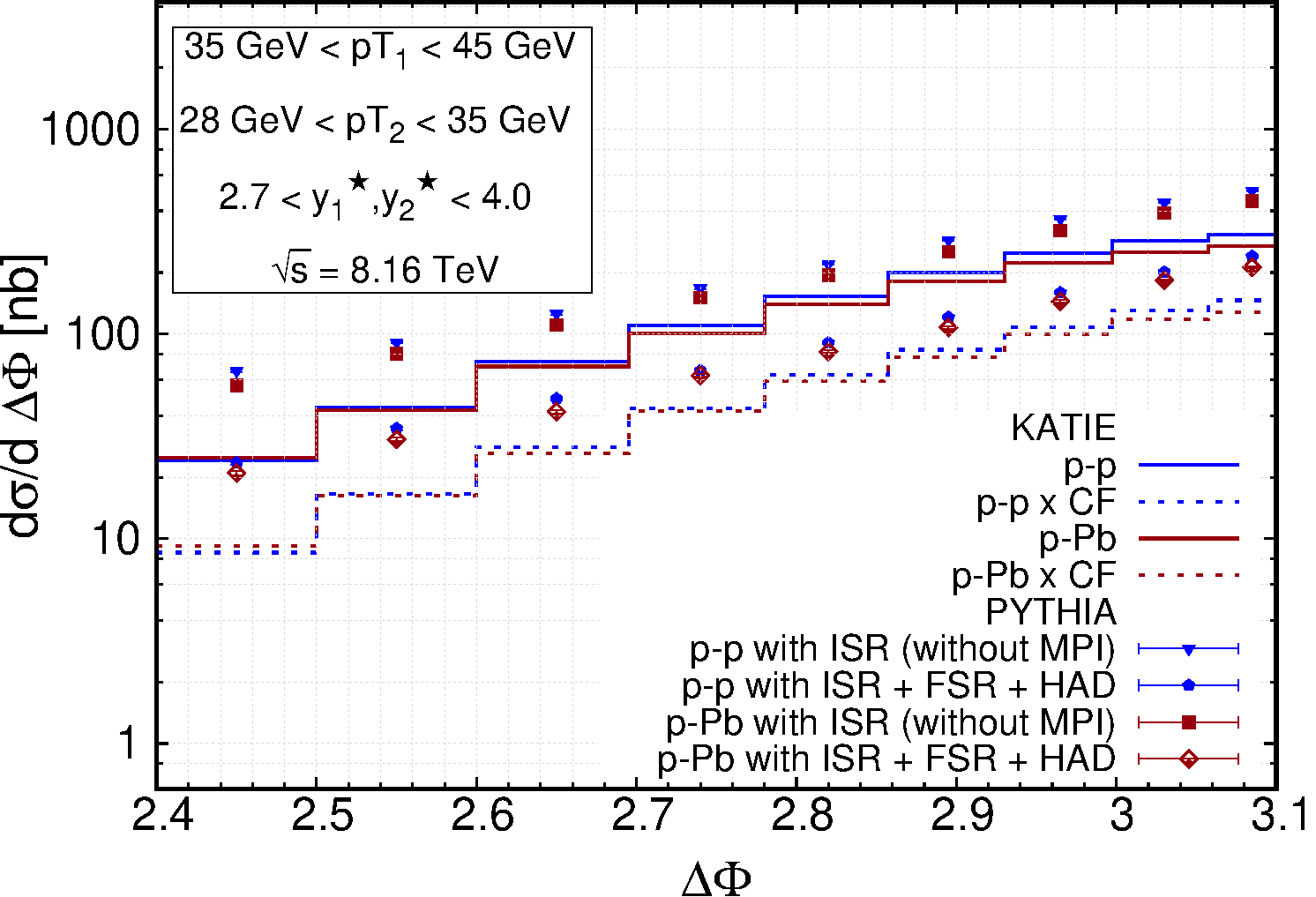}
  \caption{\small The differential cross sections in terms of the azimuthal angle $\Delta\phi$ between the leading and the sub-leading jets for the proton-proton and the proton-lead collisions at $\sqrt{s}=8.16\,\mathrm{TeV}$ in the FCal ATLAS kinematics. The solid lines represent the results from \KaTie\ computed within the ITMD approach, the points represent the results from {\Pythia} with different components, and the dotted lines represent the \KaTie\ results corrected with the non-perturbative correction factor extracted from {\Pythia}. The plots were taken from \cite{almashad2022dijet}.}
  \label{fig:pythia_katie} 
\end{figure}

\begin{figure} 
  \centering
    \includegraphics[width=0.65\linewidth]{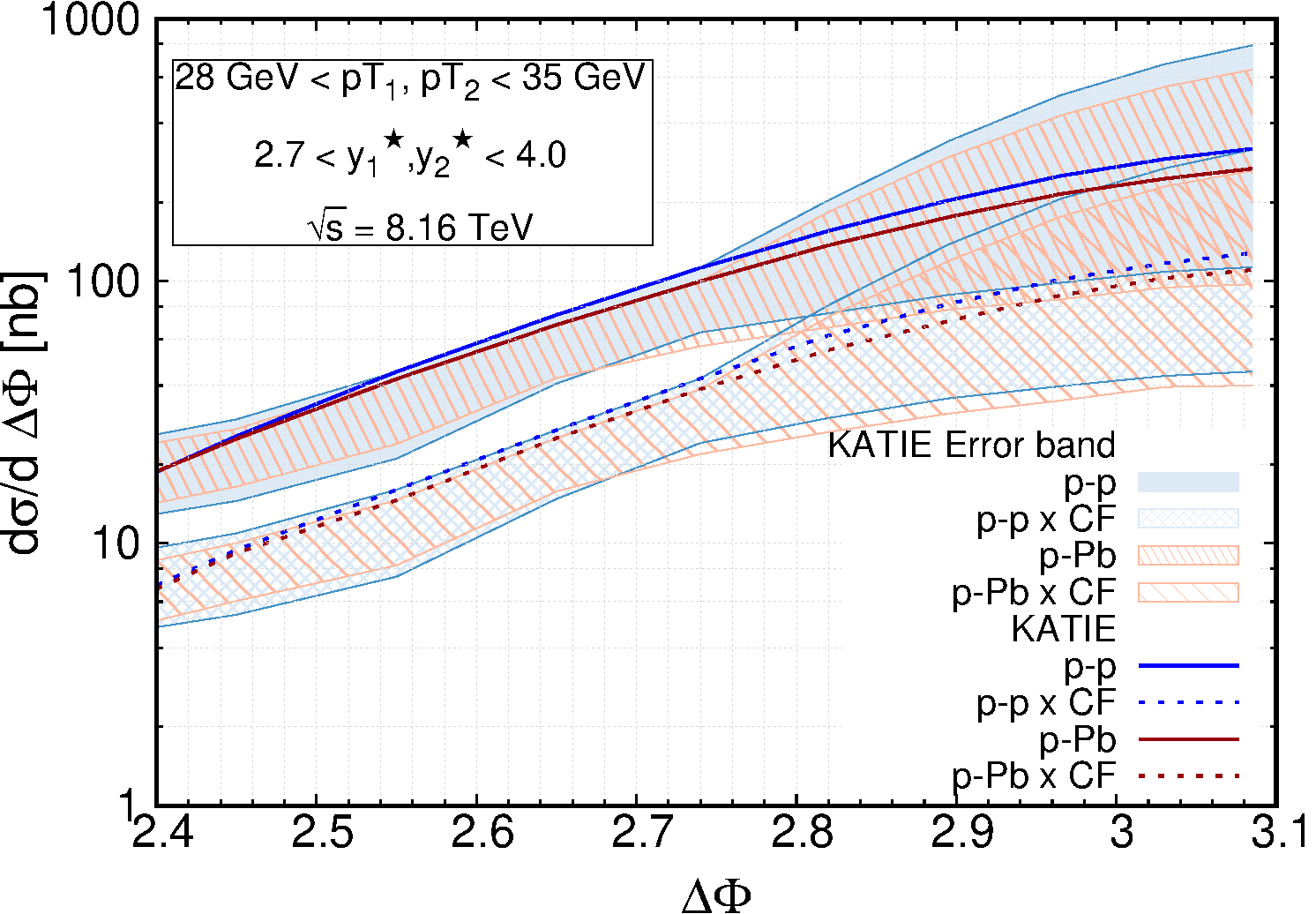}
    \includegraphics[width=0.65\linewidth]{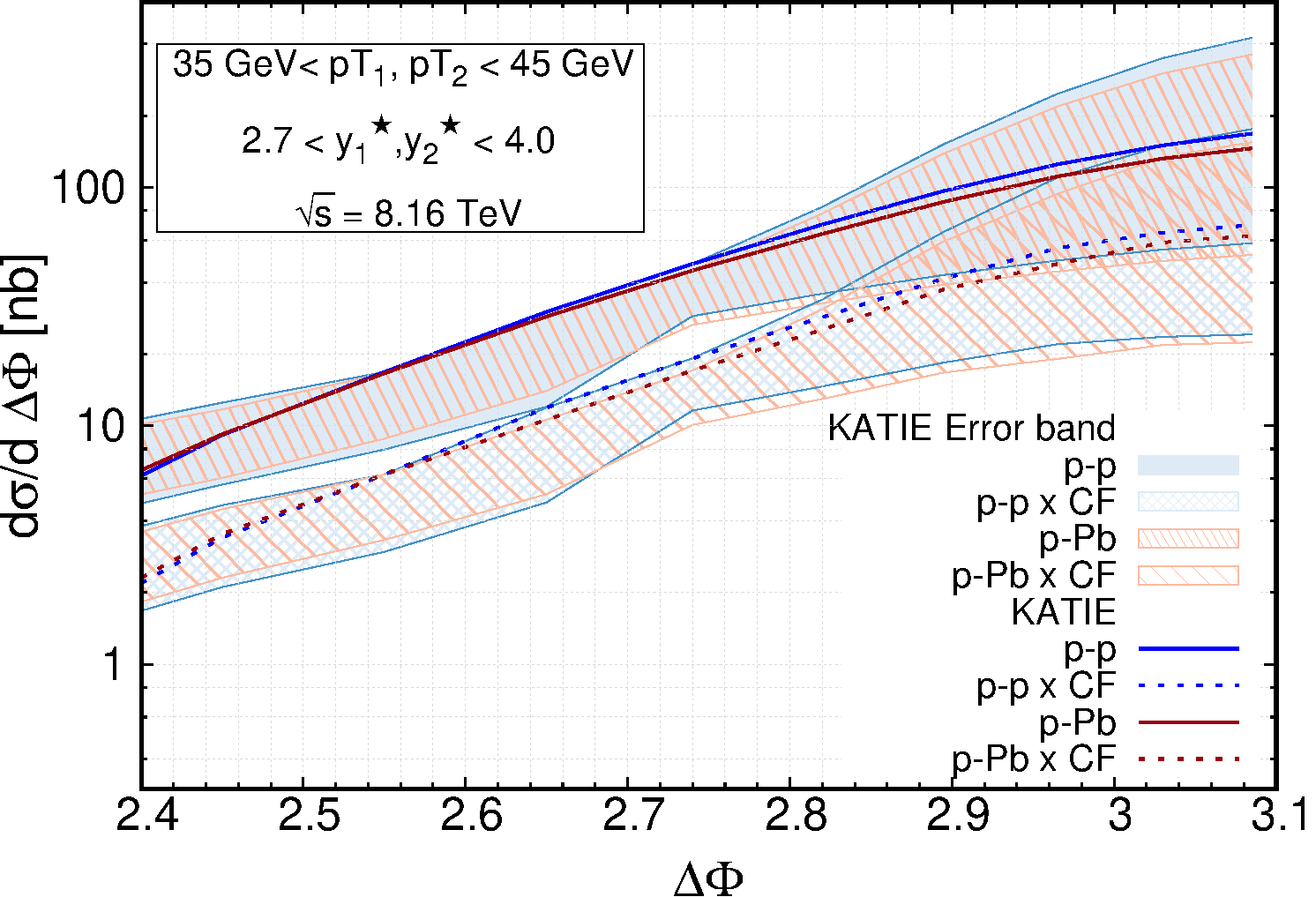}
    \includegraphics[width=0.65\linewidth]{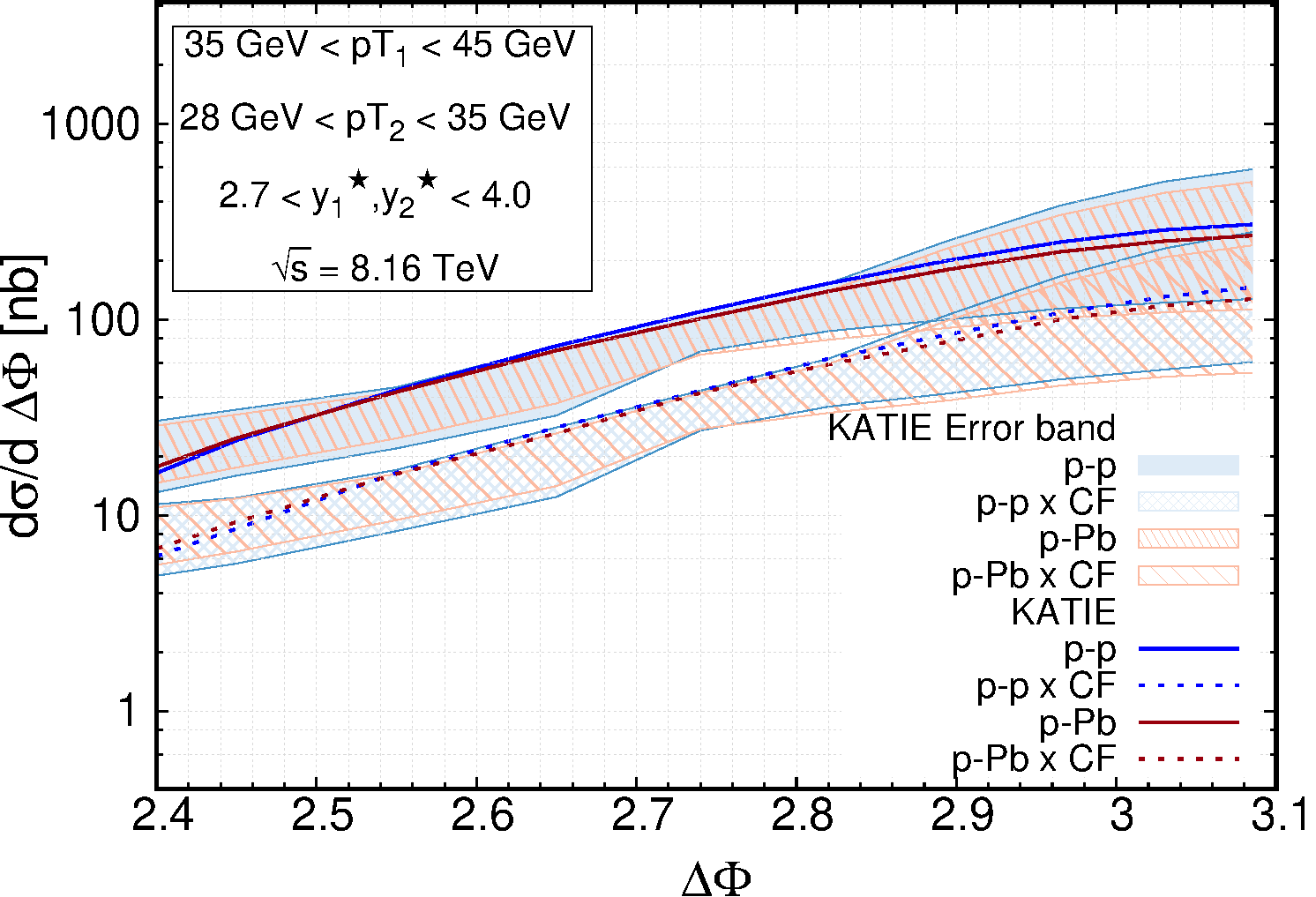}
  \caption{\small The differential cross sections in terms of the azimuthal angle $\Delta\phi$ between the leading and the sub-leading jets for the proton-proton and the proton-lead collisions at $\sqrt{s}=8.16\,\mathrm{TeV}$ in the FCal ATLAS kinematics. The solid lines represent the results from \KaTie\ computed within the ITMD approach, and the dotted lines represent the \KaTie\ results corrected with the non-perturbative correction factor extracted from {\Pythia}. The bands represent the error due to the variation of the factorization/renormalization scales from a value of $(p_{T1}+p_{T2})/2$ by a factor of 1/2 and 2. The plots were taken from  \cite{almashad2022dijet}.}
  \label{fig:katie_EB} 
\end{figure}

\begin{figure} 
  \centering
    \includegraphics[width=0.65\linewidth]{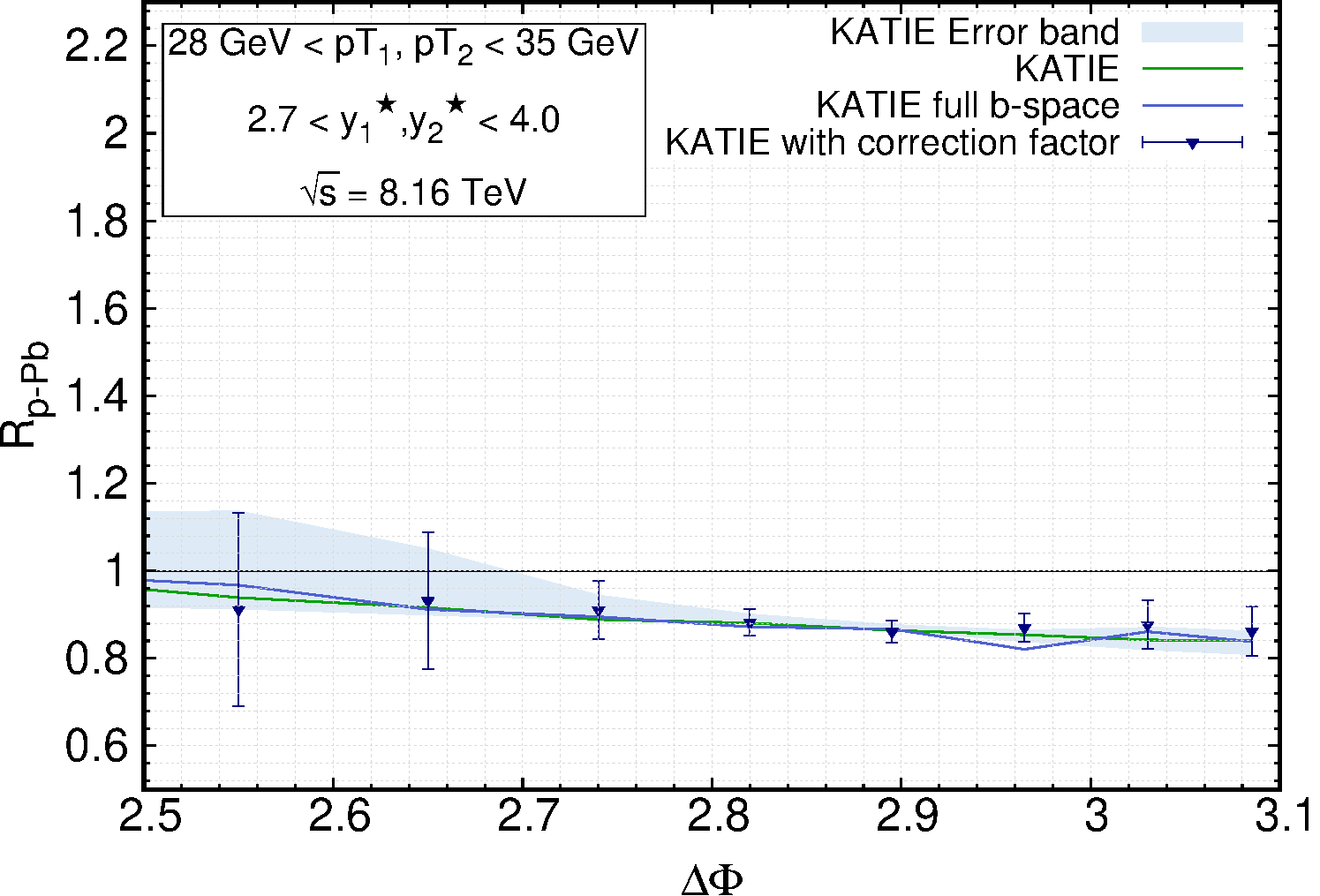}
    \includegraphics[width=0.65\linewidth]{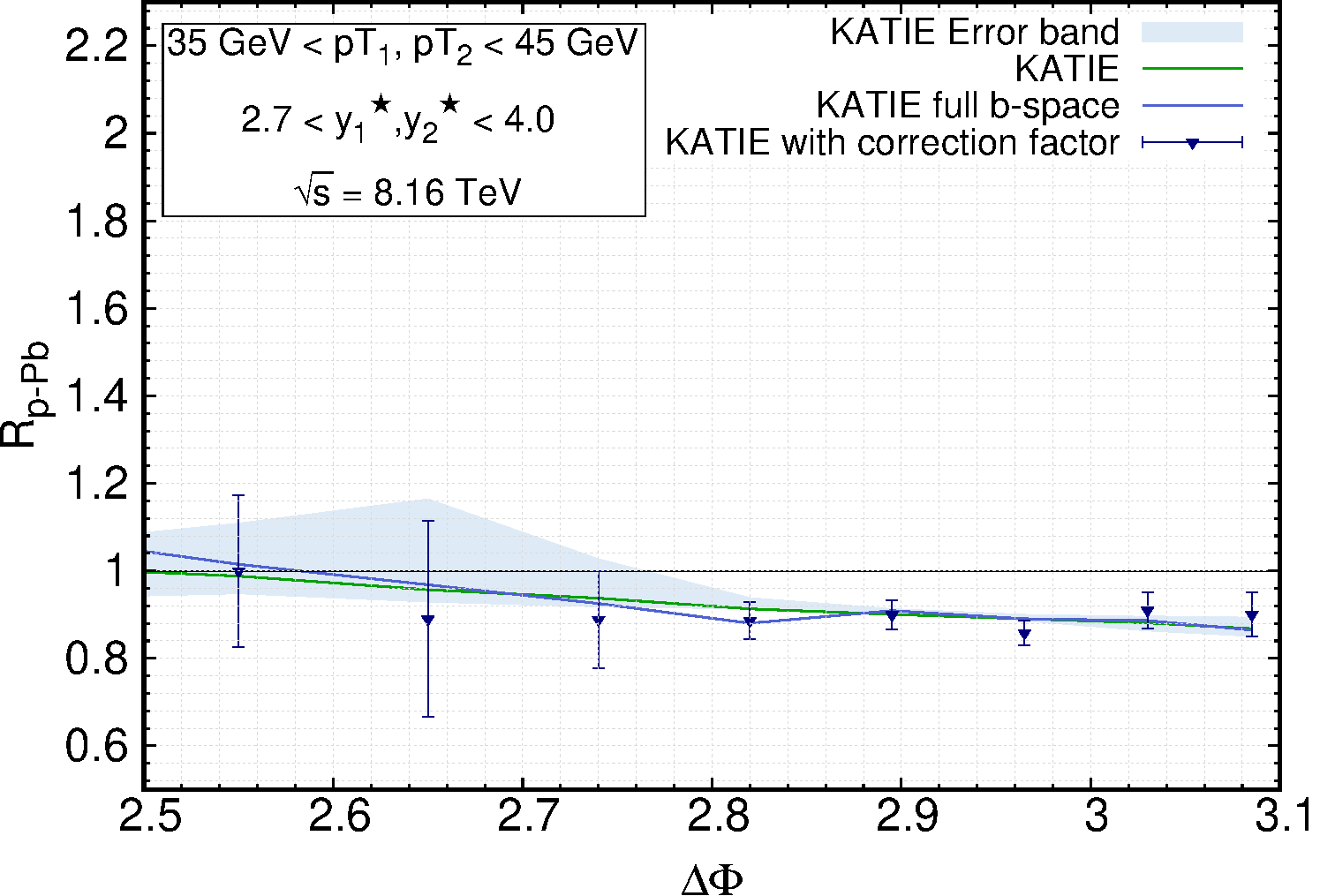}
    \includegraphics[width=0.65\linewidth]{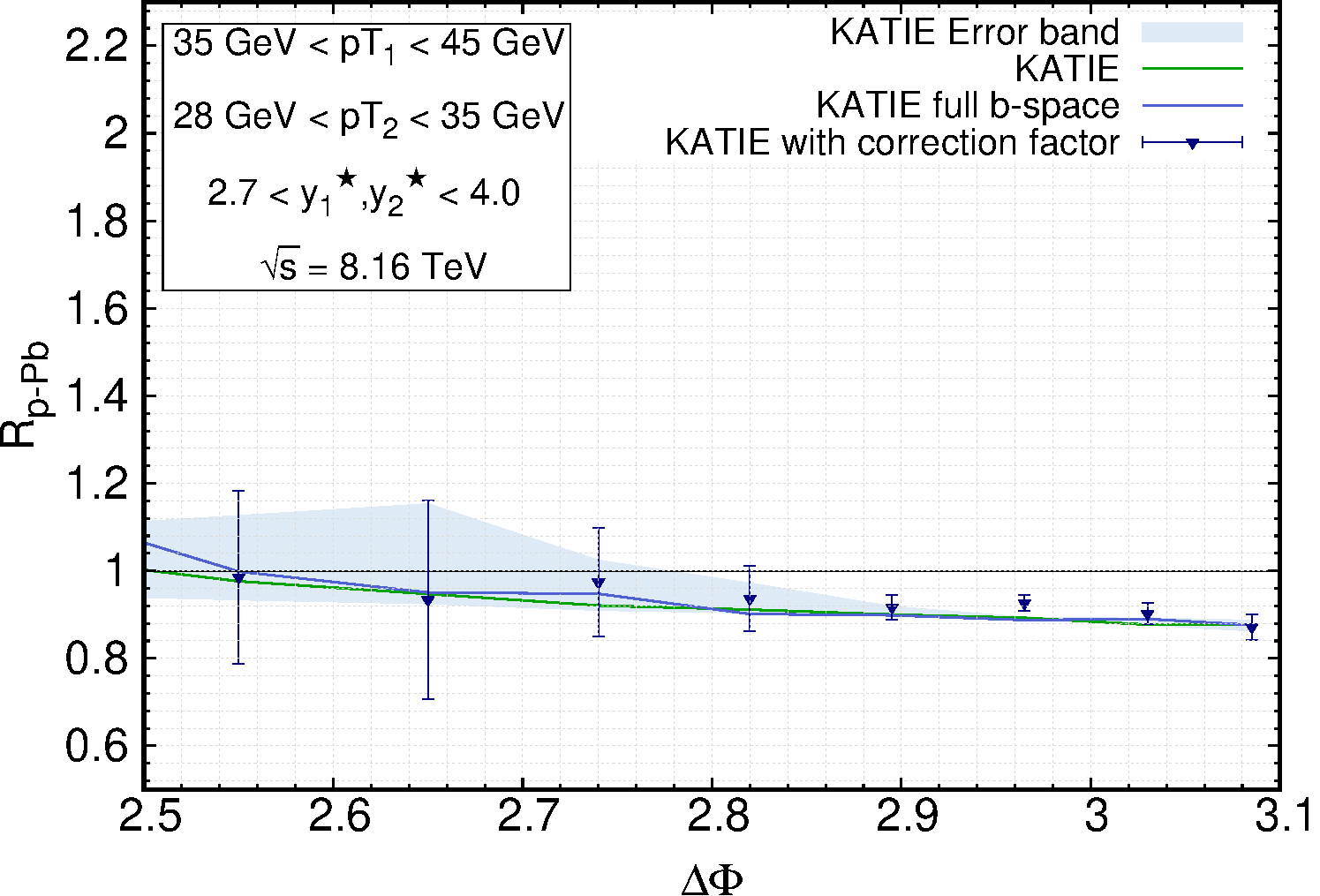}
  \caption{\small Nuclear modification ration $R_{\mathrm{p-Pb}}$ in terms of the azimuthal angle $\Delta\phi$ between the leading and the sub-leading jets for the proton-proton and the proton-lead collisions at $\sqrt{s}=8.16\,\mathrm{TeV}$ in the FCal ATLAS kinematics. The solid lines represent the results from \KaTie\ computed within the ITMD approach with: the simplified Sudakov resummation Eq.~(\ref{eq:ITMD_factorization}) (red line), the full $b$-space resummation Eq.~(\ref{eq:itmd_Sud}) (blue line). The band represents the error due to the variation of the factorization/renormalization scales from a value of $(p_{T1}+p_{T2})/2$ by a factor of 1/2 and 2. The points $\triangledown$  represent the \KaTie\ results corrected with the non-perturbative correction factor extracted from {\Pythia}. The error bars associated with these points account for both errors due to variation of scale in  \KaTie\ and the statistical uncertainties in the correction factor from {\Pythia}. The plots were taken from \cite{almashad2022dijet}.}
  \label{fig:R_at} 
\end{figure}

Let us now discuss the results. In Fig.~\ref{fig:only_katie} we show the results for the azimuthal correlations for p-p and p-Pb collisions in the ATLAS kinematics at $\sqrt{s}=8.16\,\mathrm{TeV}$. We compare the Sudakov resummation obtained via two methods, the full-$b$ space resummation (dotted) and the simplified approach, where the collinear PDF is not affected by the Sudakov form factor. As can be seen, they differ a bit, but as we shall see below, the difference cancels in the nuclear modification ratio, thus does not affect the saturation signal. Moreover, as we also show below, there are very large scale uncertainties, within which both methods are qualitatively similar. In our computations we used the ITMD factorization alone, without parton shower or hadronisation corrections. In order to asses this effect, we computed cross section with PYTHIA \cite{pythia:2006ab,pythia:2015ac} with all correction turned on, and then just with initial-state parton shower. The latter roughly corresponds to the TMD framework (as discussed eg. in \cite{Bury:2017jxo}), therefore both calculations allow for extraction of a ``correction factor''. We repeated the same procedure using the nucler PDFs in PYTHIA. In Fig.~\ref{fig:pythia_katie} we applied that correction to \KaTie\ results and compared with PYTHIA computations. Although the correction factor is rather large, as we shall see below it does not affect the saturation signal in the nuclear modification ratio. Next, in Fig.~\ref{fig:katie_EB} we show the error bands due to both scale variation. Finally, in Fig.~\ref{fig:R_at} we compute the nuclear modification ratio. For all three $p_T$ cuts, we show the ITMD result with simiplified Sudakov resummation, with the full-$b$ space resummation, and the result where the non-perturbative correction factor from PYTHIA was applied. We also compute the error band due to scale variation and due to the correction factor. We see, that the saturation effect due to the nonlinear evolution is visible, but not very significant for the considered kinematics. For the lower $p_T$ cut we observe suppression of about 15\%, but the uncertainty due to the correction factor is not much less. Interestingly (but understandably) this error decreases for larger $p_T$ bins.

\begin{figure} 
  \centering
    \includegraphics[width=0.6\linewidth]{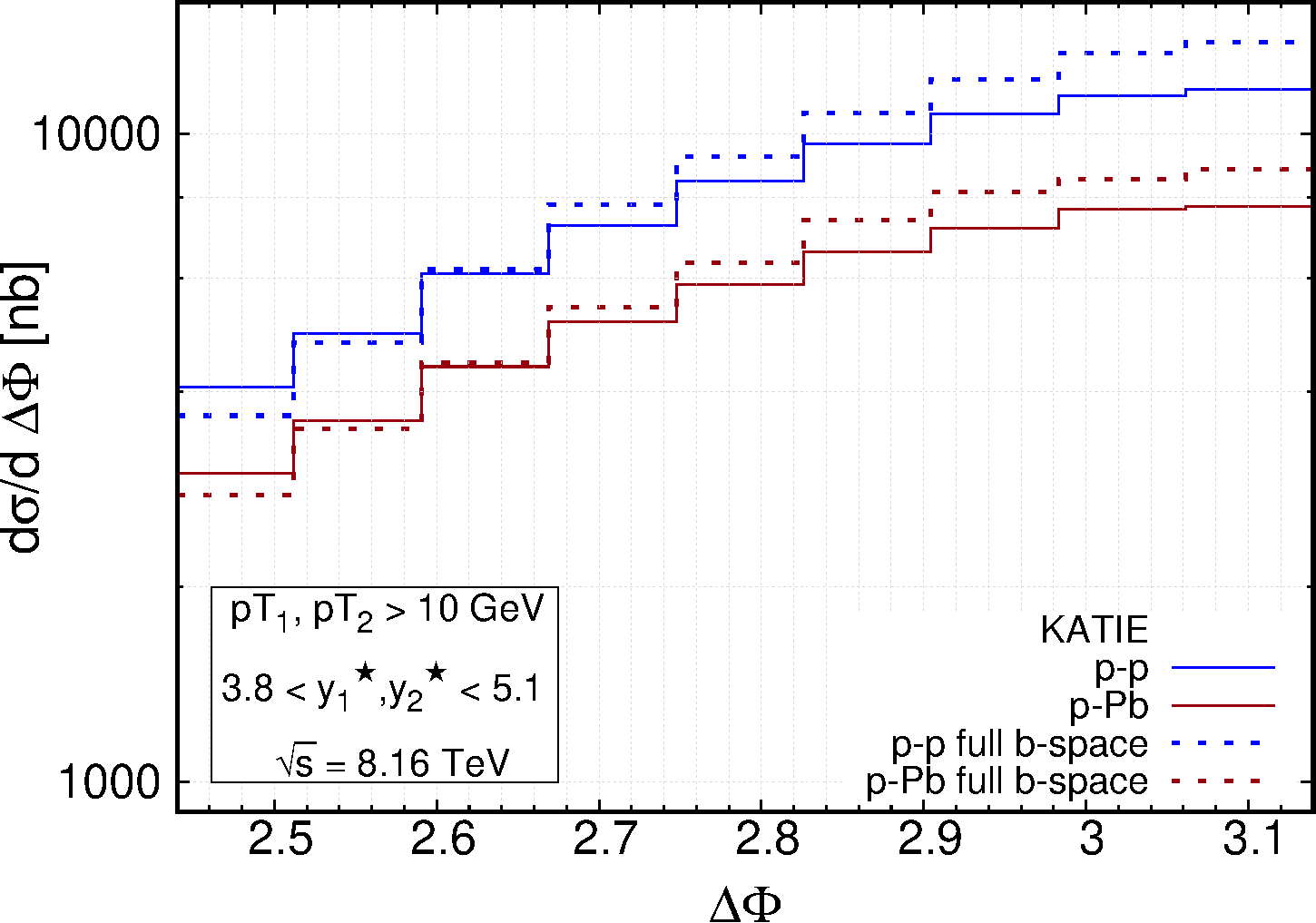}
    \includegraphics[width=0.6\linewidth]{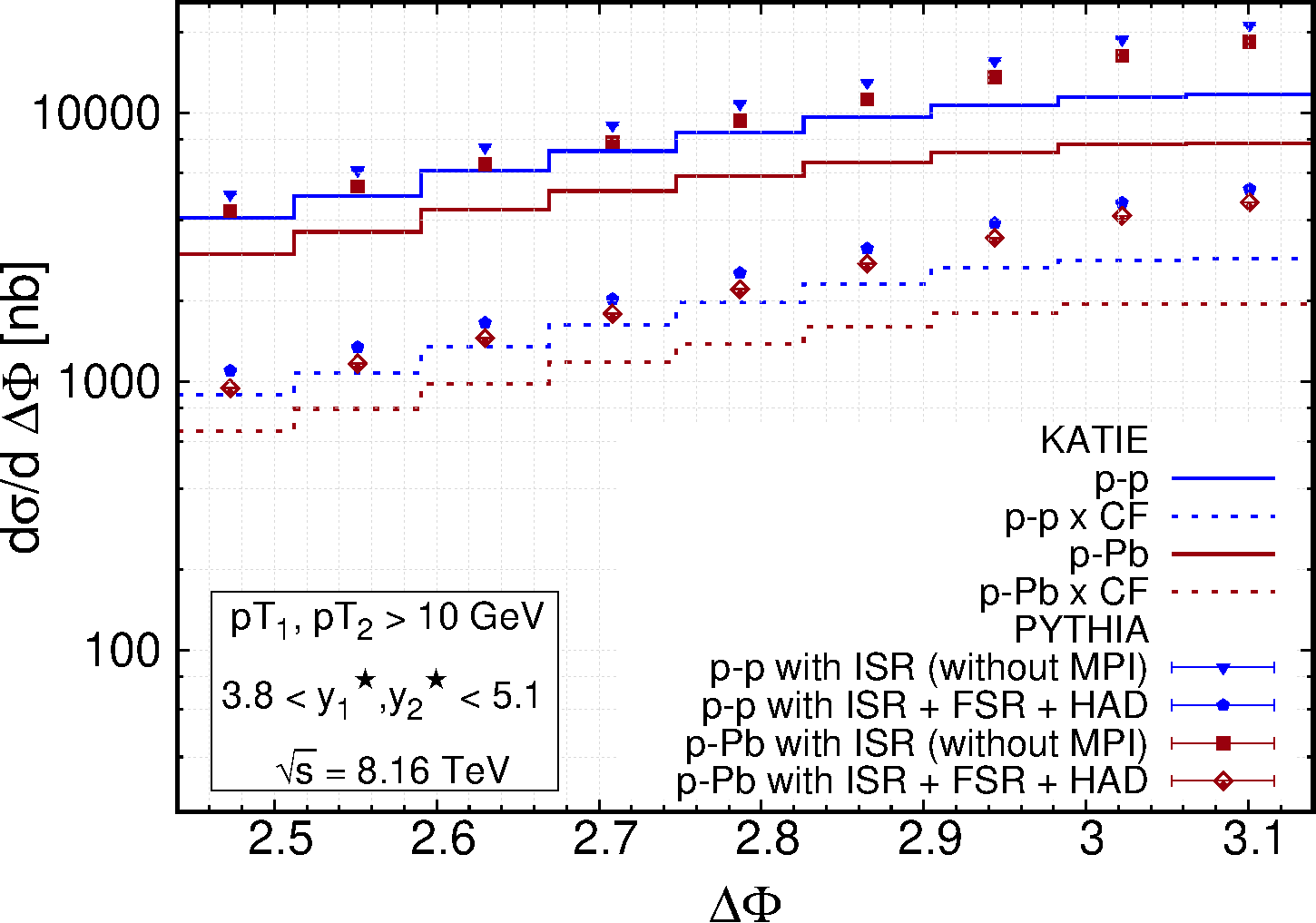}
    \includegraphics[width=0.6\linewidth]{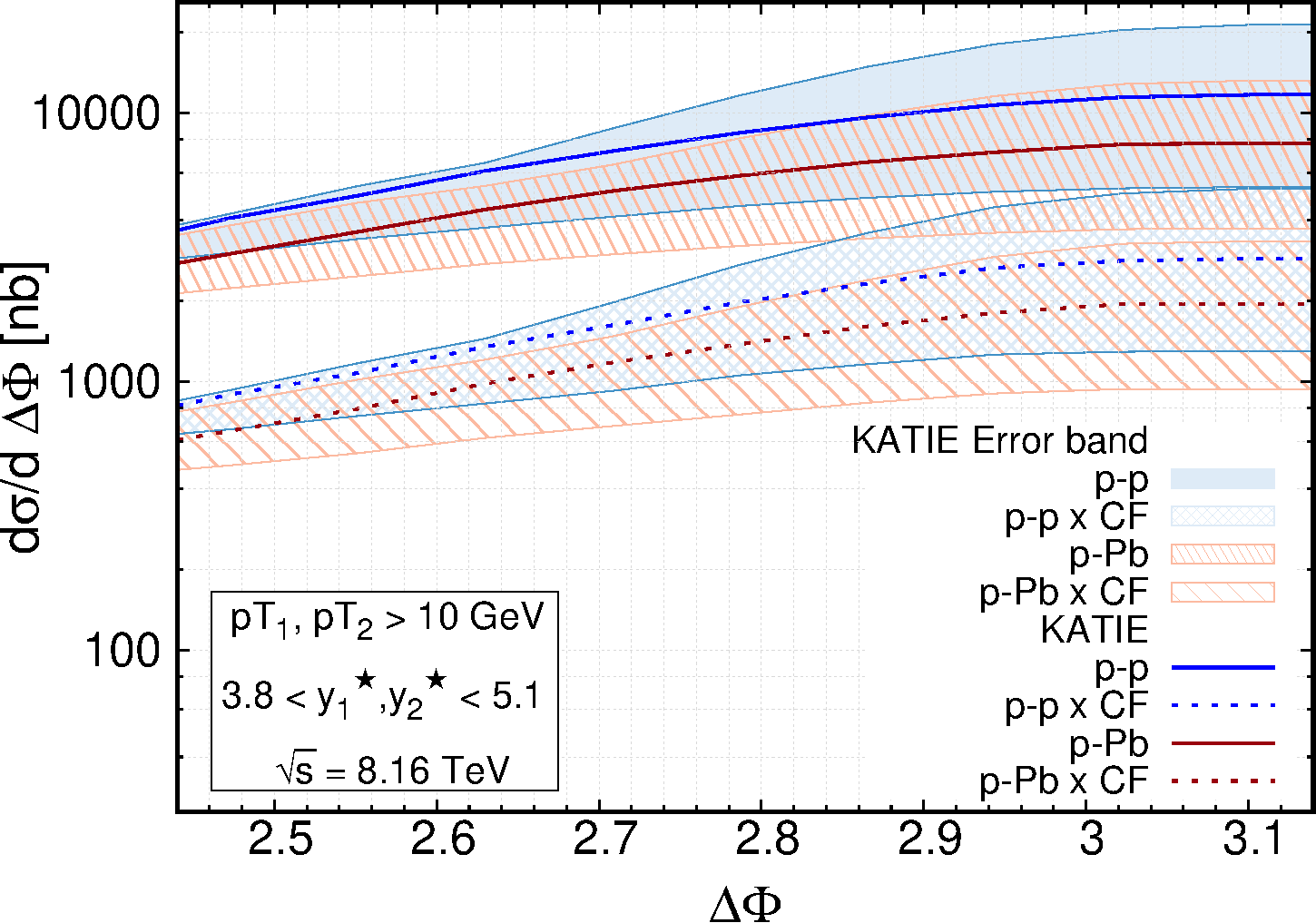}
  \caption{\small The differential cross sections in terms of the azimuthal angle $\Delta\phi$ between the leading and the sub-leading jets for the proton-proton and the proton-lead collisions at $\sqrt{s}=8.16\,\mathrm{TeV}$ in the ALICE FoCal kinematics. The first plot represents the corss sections computed using \KaTie\ within the ITMD factorization formula with: the simplified Sudakov resummation Eq.~(\ref{eq:ITMD_factorization}) (solid lines), the full $b$-space resummation Eq.~(\ref{eq:itmd_Sud}) (dotted lines).  In the second plot, the solid lines represent the cross sections from \KaTie\ computed within the ITMD approach, the points represent the results from {\Pythia} with different components, and the dotted lines represent the \KaTie\ results corrected with the non-perturbative correction factor extracted from {\Pythia}. In the third plot, the solid lines represent the cross sections from \KaTie\ computed within the ITMD approach, and the dotted lines represent the \KaTie\ results corrected with the non-perturbative correction factor extracted from {\Pythia}. The bands represent the error due to the variation of the factorization/renormalization scales in \KaTie\  from a value of $(p_{T1}+p_{T2})/2$ by a factor of 1/2 and 2.
  The plots were taken from \cite{almashad2022dijet}.}
  \label{fig:Focal_old} 
\end{figure}

\begin{figure} 
  \centering
    \includegraphics[width=0.65\linewidth]{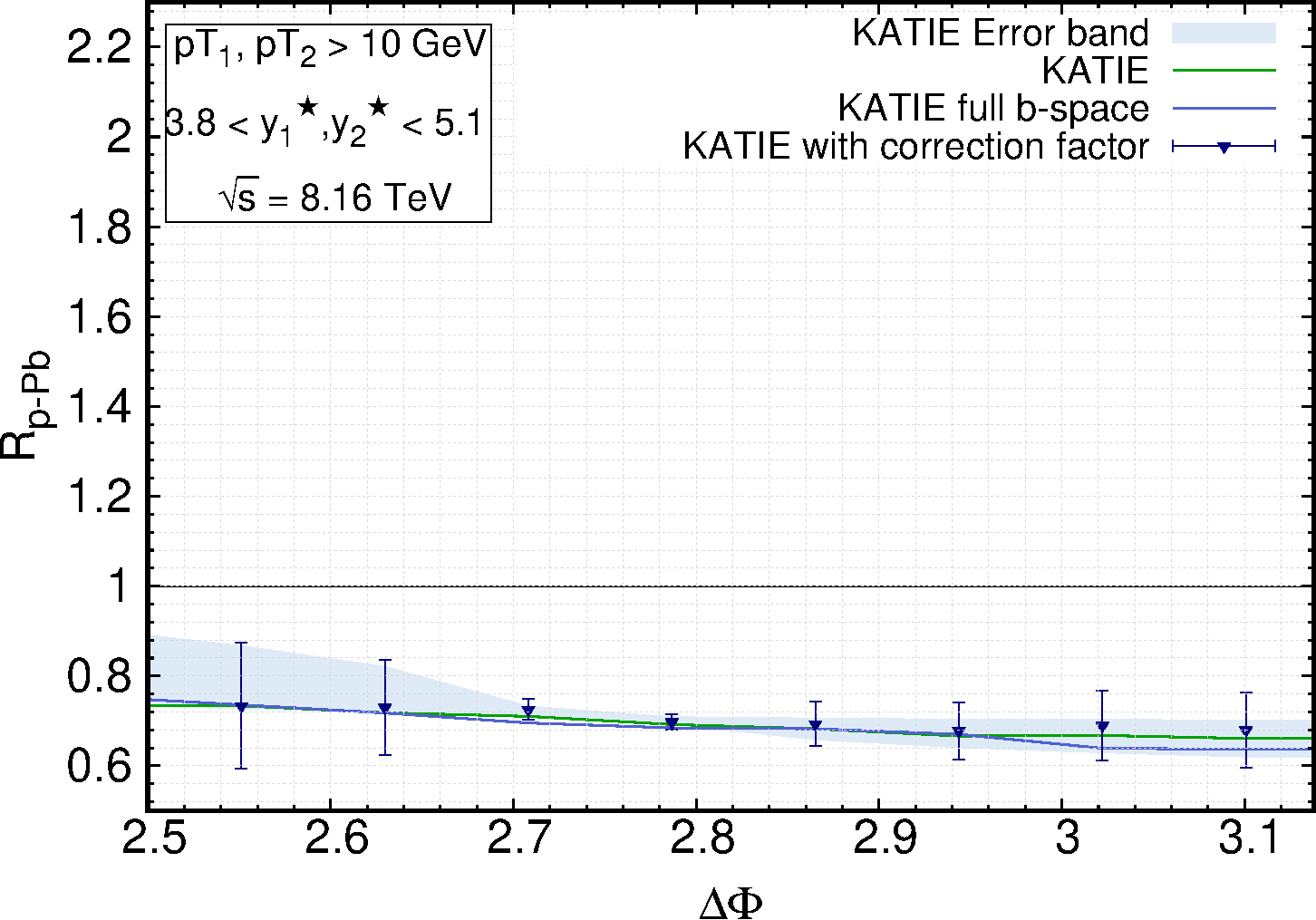}
\caption{\small Nuclear modification ration $R_{\mathrm{p-Pb}}$ in terms of the azimuthal angle $\Delta\phi$ between the leading and the sub-leading jets for the proton-proton and the proton-lead collisions at $\sqrt{s}=8.16\,\mathrm{TeV}$ in the ALICE FoCal kinematics. The solid lines represent the results from \KaTie\ computed within the ITMD approach with: the simplified Sudakov resummation Eq.~(\ref{eq:ITMD_factorization}) (red line), the full $b$-space resummation Eq.~(\ref{eq:itmd_Sud}) (blue line). The band represents the error due to the variation of the factorization/renormalization scales from a value of $(p_{T1}+p_{T2})/2$ by a factor of 1/2 and 2. The points $\triangledown$  represent the \KaTie\ results corrected with the non-perturbative correction factor extracted from {\Pythia}. The error bars associated with these points account for both errors due to variation of scale in  \KaTie\ and the statistical uncertainties in the correction factor from {\Pythia}. The plots were taken from \cite{almashad2022dijet}.}
  \label{fig:focal_Rold} 
\end{figure}

\begin{figure} 
  \centering
    \includegraphics[width=0.65\linewidth]{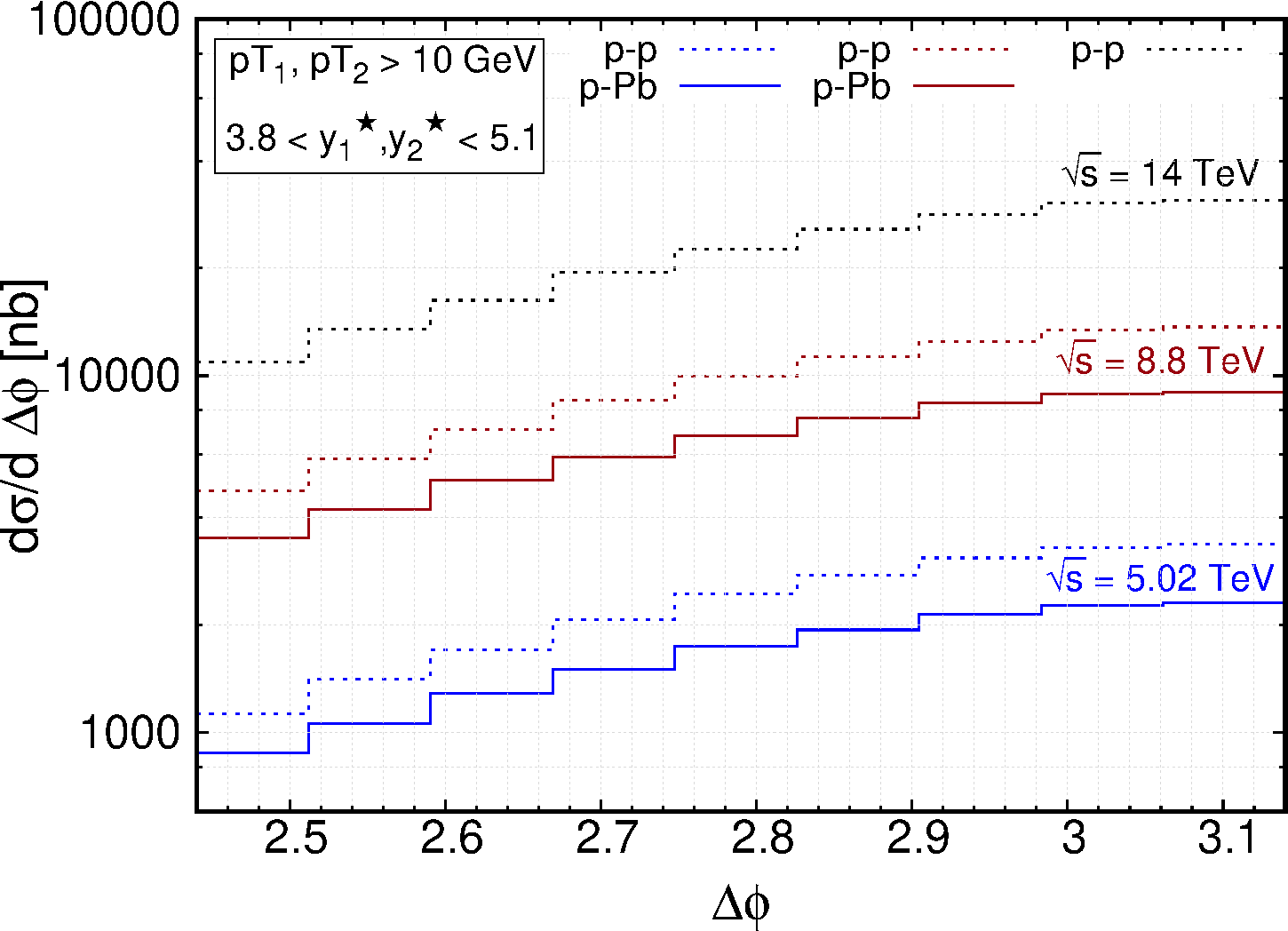}
    \includegraphics[width=0.65\linewidth]{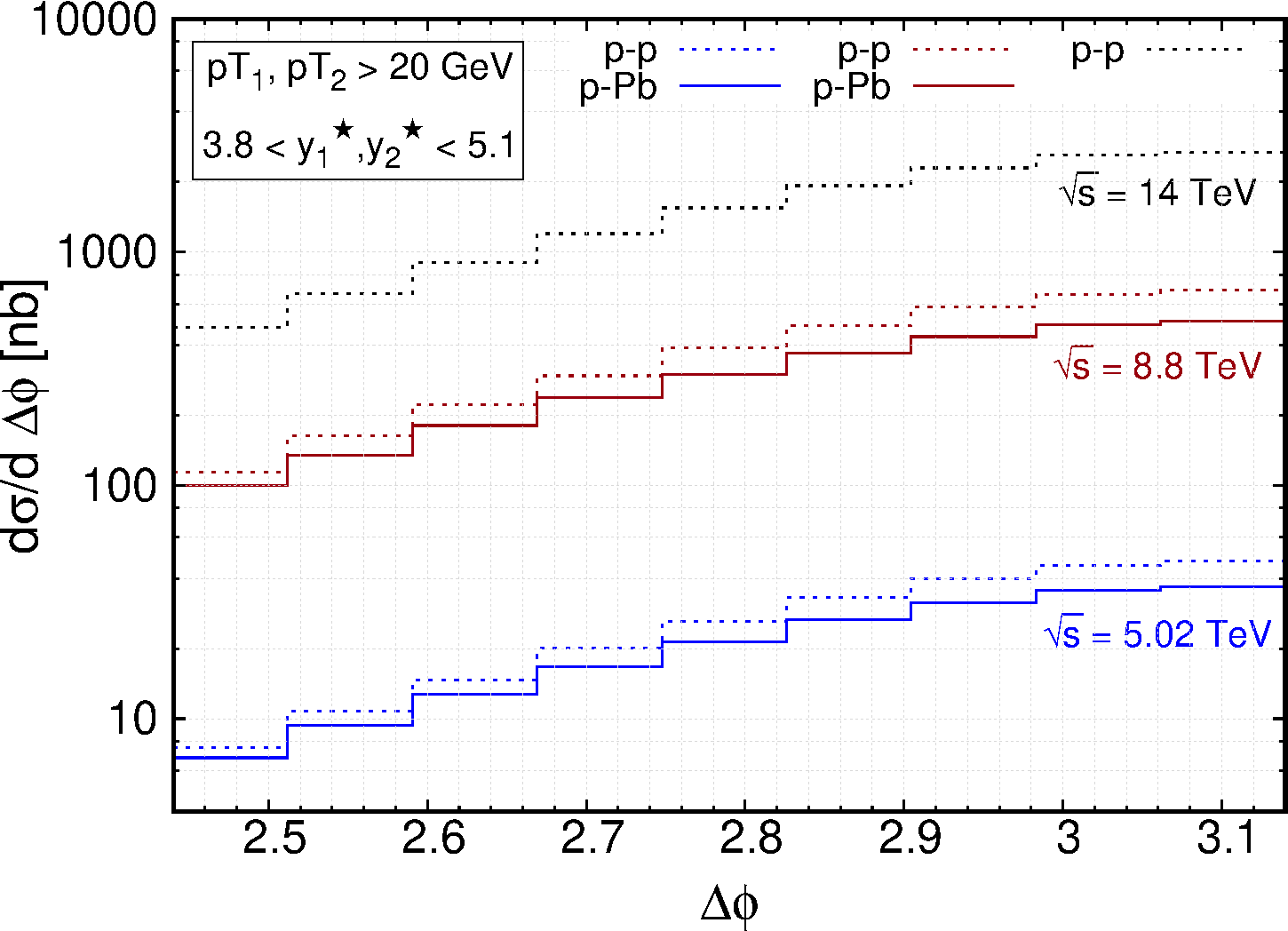}
\caption{\small The differential cross sections in terms of the azimuthal angle $\Delta\phi$ between the leading and the sub-leading jets for the proton-proton and the proton-lead collisions at $\sqrt{s}=5.02\,\mathrm{TeV}$ (blue) and $8.8\,\mathrm{TeV}$ (red) in the ALICE FoCal kinematics. For the same kinematics, each plot also represents the differential cross section for the proton-proton collision at $\sqrt{s}=5.02\,\mathrm{TeV}$ (black). Since proton-lead collisions are not feasible at this energy, we didn't compute those.  All of these cross sections were computed using \KaTie\ within the ITMD factorization formula with the simplified Sudakov resummation Eq.~(\ref{eq:ITMD_factorization}) The solid lines represent the results for proton-lead collisions and the dotted lines represent the results for the proton-proton collisions.}
  \label{fig:focal_new_DCS} 
\end{figure}

\begin{figure} 
  \centering
    \includegraphics[width=0.65\linewidth]{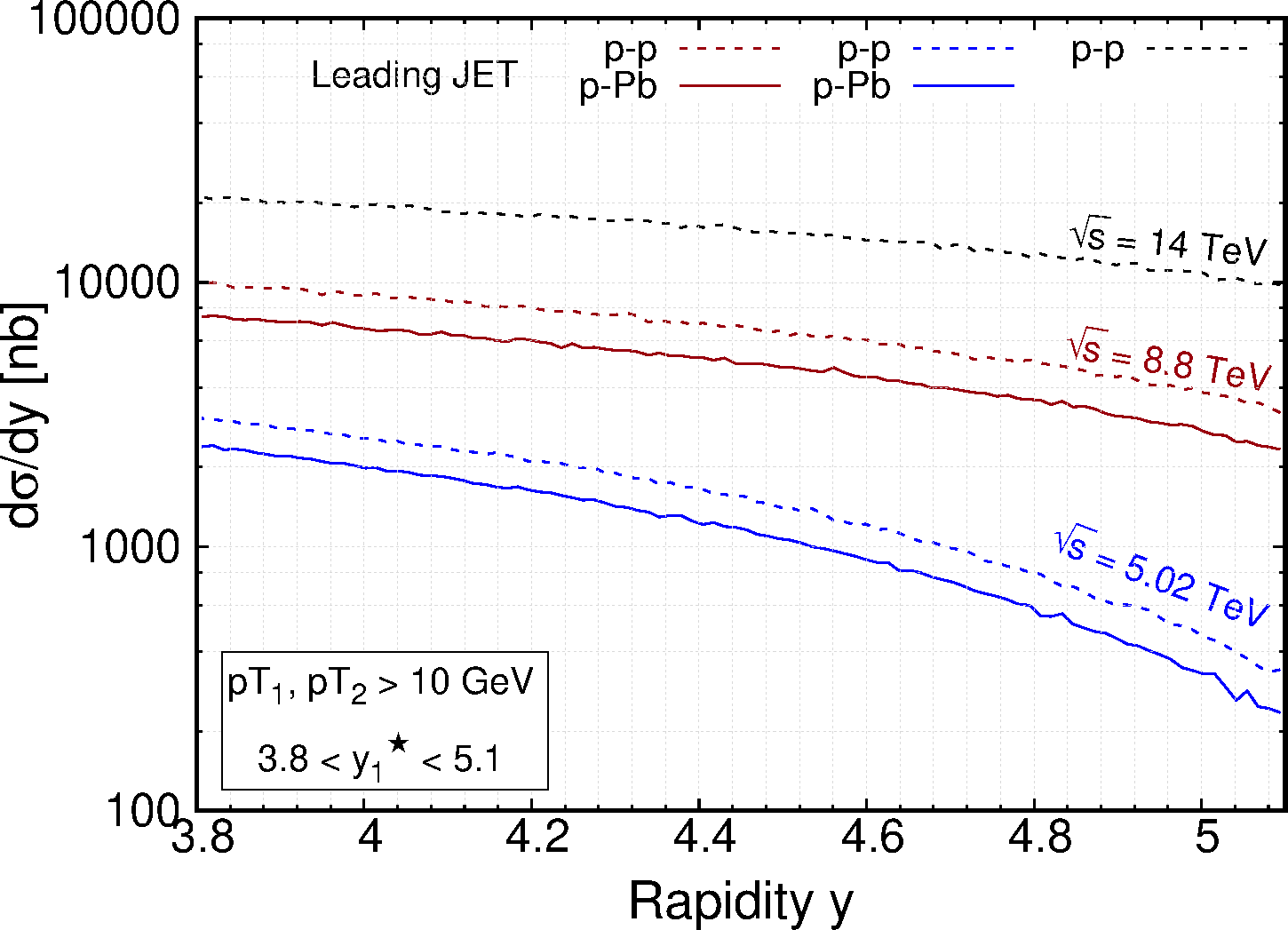}
    \includegraphics[width=0.65\linewidth]{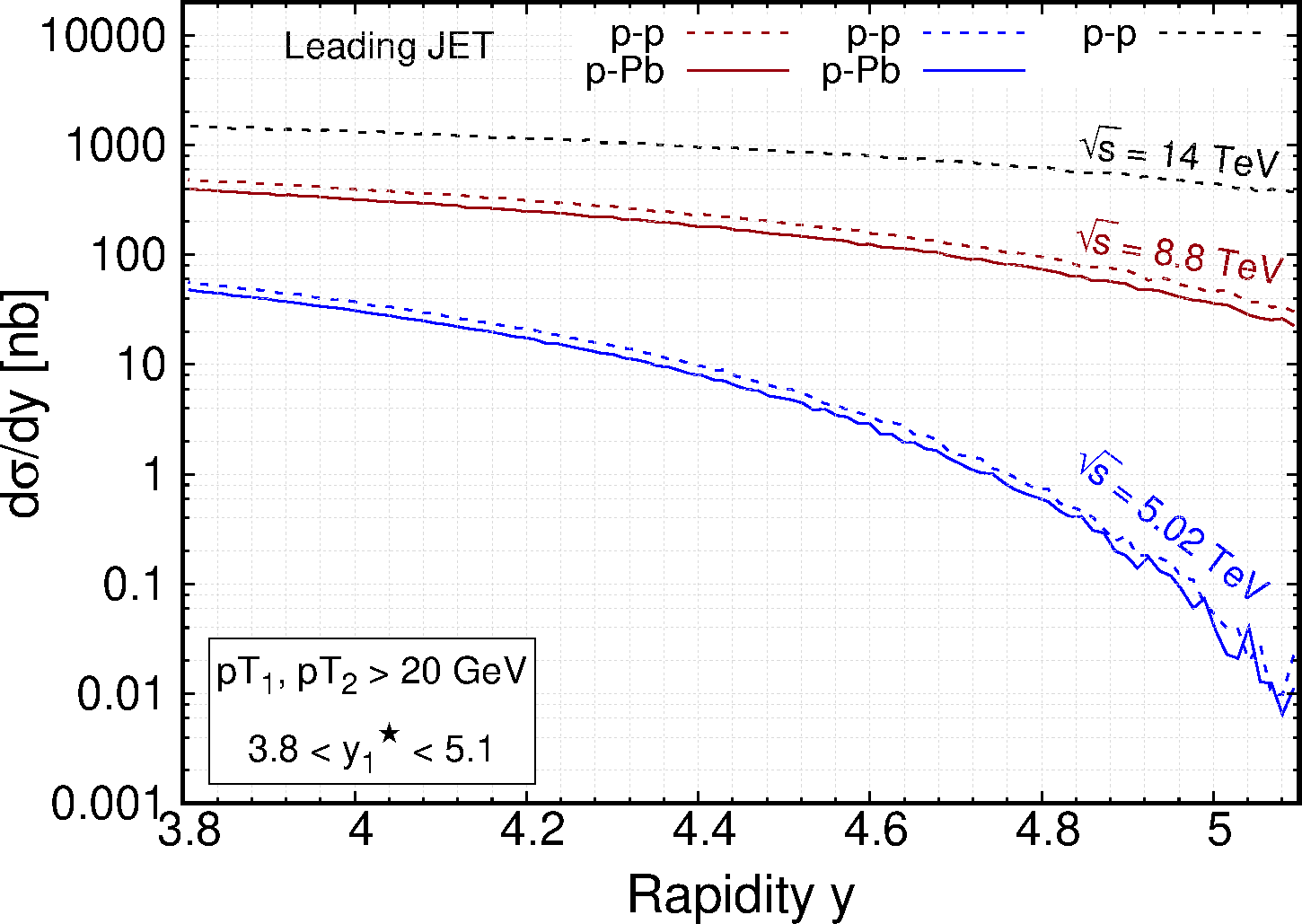}
\caption{\small The differential cross sections in terms of the rapidity of the leading jet $y_1^{\star}$ for the proton-proton and the proton-lead collisions at $\sqrt{s}=5.02\,\mathrm{TeV}$ (blue) and $8.8\,\mathrm{TeV}$ (red) in the ALICE FoCal kinematics. For the same kinematics, each plot also represents the differential cross section for the proton-proton collision at $\sqrt{s}=5.02\,\mathrm{TeV}$ (black). Since proton-lead collisions are not feasible at this energy, we didn't compute those.  All of these cross sections were computed using \KaTie\ within the ITMD factorization formula with the simplified Sudakov resummation Eq.~(\ref{eq:ITMD_factorization}) The solid lines represent the results for proton-lead collisions and the dotted lines represent the results for the proton-proton collisions.}
  \label{fig:focal_lead_jets} 
\end{figure}

\begin{figure} 
  \centering
    \includegraphics[width=0.65\linewidth]{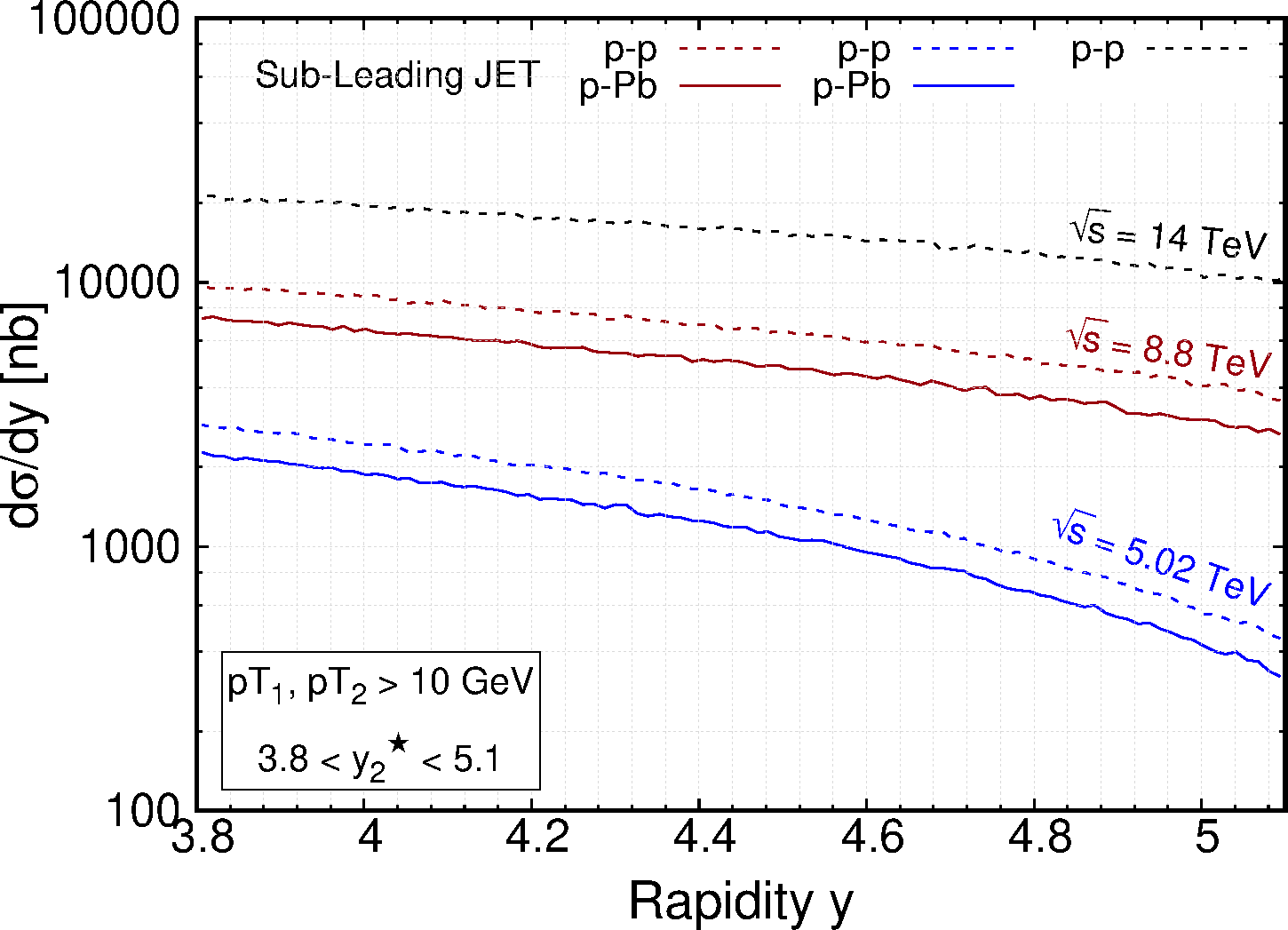}
    \includegraphics[width=0.65\linewidth]{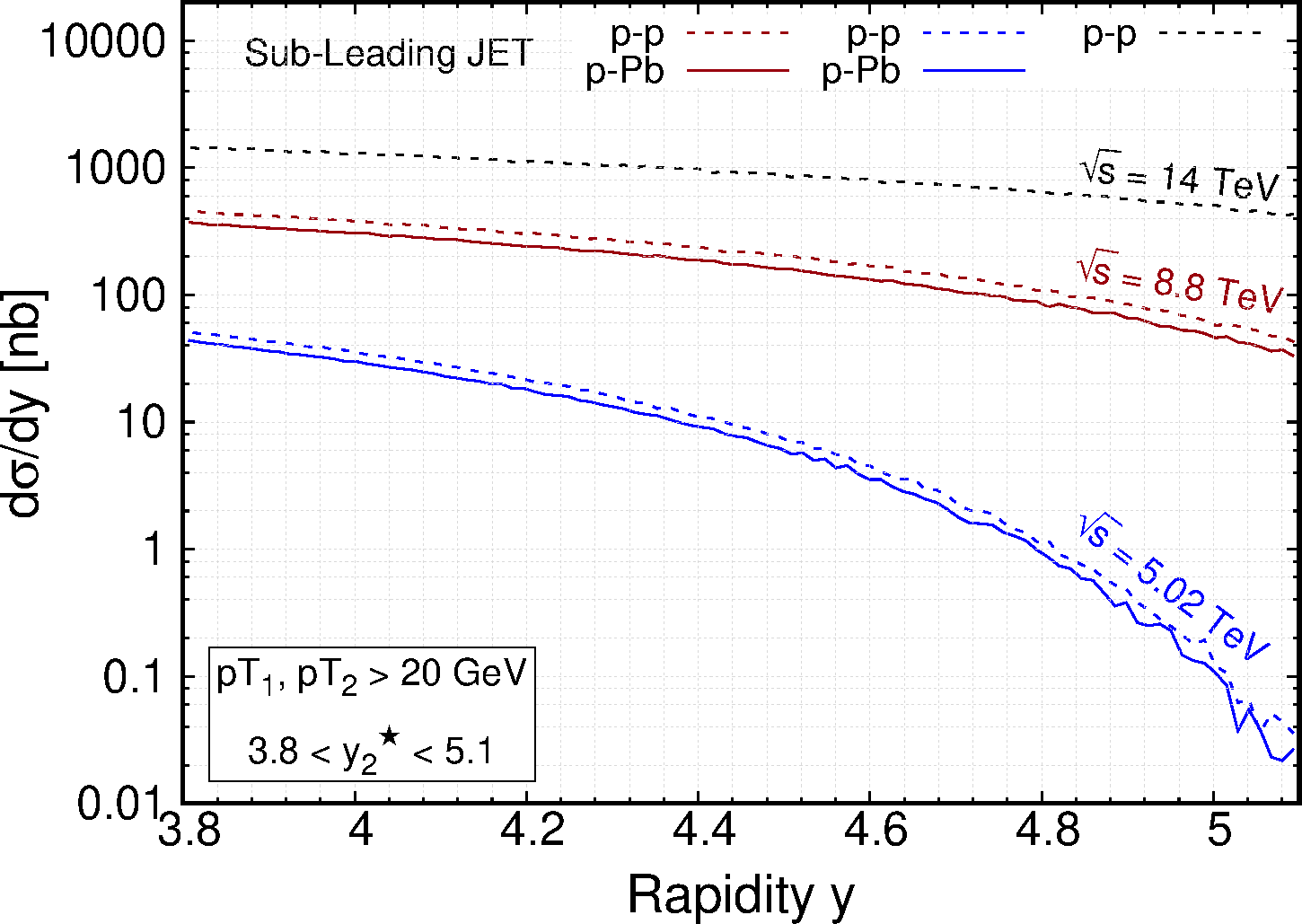}
\caption{\small The differential cross sections in terms of the rapidity of the sub-leading jet $y_2^{\star}$ for the proton-proton and the proton-lead collisions at $\sqrt{s}=5.02\,\mathrm{TeV}$ (blue) and $8.8\,\mathrm{TeV}$ (red) in the ALICE FoCal kinematics. For the same kinematics, each plot also represents the differential cross section for the proton-proton collision at $\sqrt{s}=5.02\,\mathrm{TeV}$ (black). Since proton-lead collisions are not feasible at this energy, we didn't compute those.  All of these cross sections were computed using \KaTie\ within the ITMD factorization formula with the simplified Sudakov resummation Eq.~(\ref{eq:ITMD_factorization}) The solid lines represent the results for proton-lead collisions and the dotted lines represent the results for the proton-proton collisions.}
  \label{fig:focal_sublead_jets} 
\end{figure}

\begin{figure}
    \centering
    \includegraphics[width=0.65\linewidth]{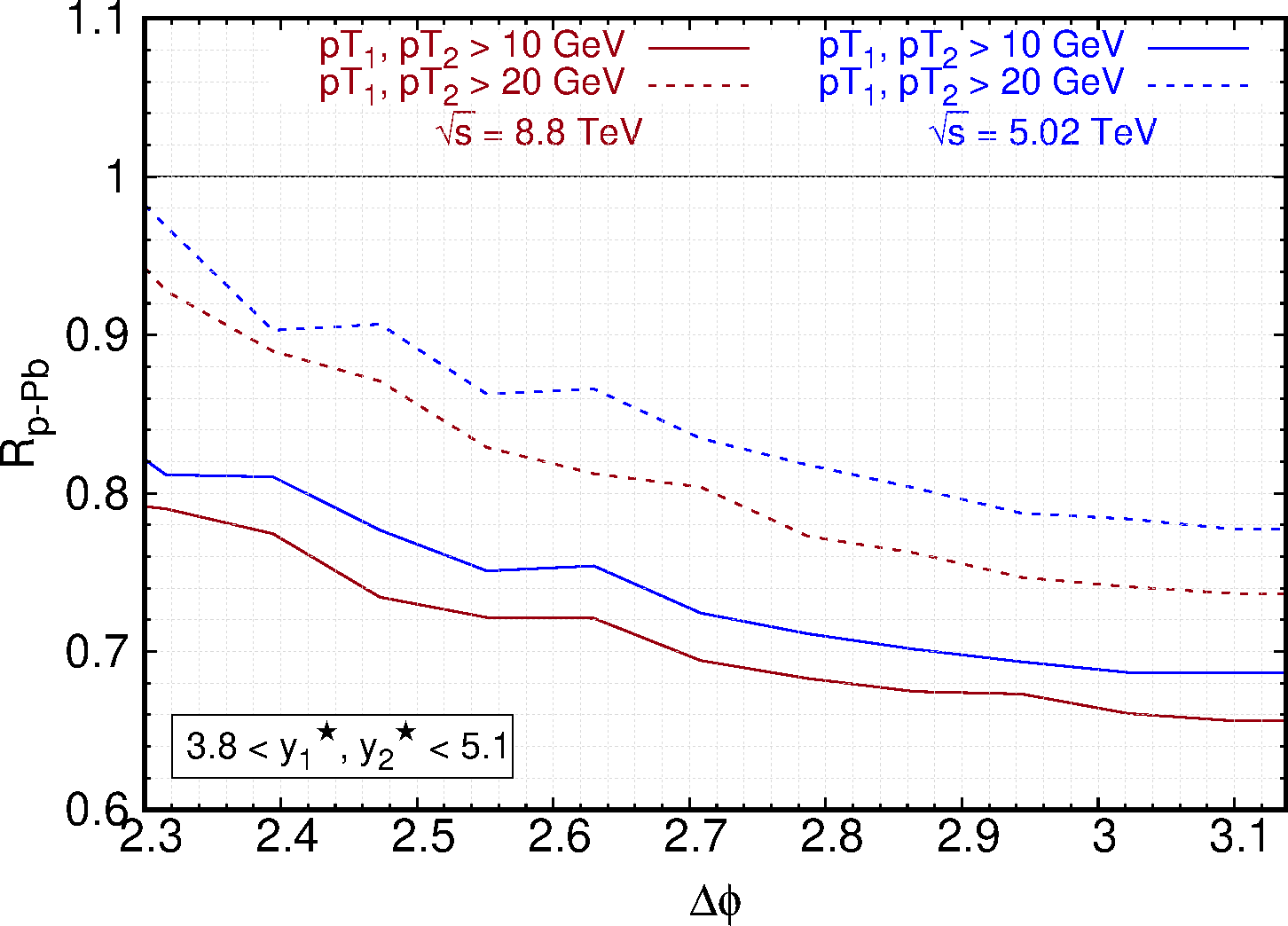}
    \caption{\small Nuclear modification ration $R_{\mathrm{p-Pb}}$ in terms of the azimuthal angle $\Delta\phi$ between the leading and the sub-leading jets for the proton-proton and the proton-lead collisions at $\sqrt{s}=5.02\,\mathrm{TeV}$ (blue) and $8.8\,\mathrm{TeV}$ (red) in the ALICE FoCal kinematics. The solid lines represent the results from \KaTie\ computed within the ITMD approach with the simplified Sudakov resummation Eq.~(\ref{eq:ITMD_factorization}) for $p_{T1}, p_{T2} > 10\, \mathrm{GeV}$. And the dotted lines represent the results for $p_{T1}, p_{T2} > 20\, \mathrm{GeV}$.}
    \label{fig:Focal_Rnew}
\end{figure}

Next, we move on to the more forward region with FoCal kinematics. Here we focus first on the $p_T>10$~GeV cut. The results for azimuthal correlations are shown in Fig.~\ref{fig:Focal_old}. In one figure we show ITMD result with Sudakov obtained via both models, as well as PYTHIA corrections and errorbands, as discussed before. In Fig.~\ref{fig:focal_Rold} we compute the nucler modification ration for that case. As can be seen from the plot, the suppression due to saturation is very large, about 25-30\%, and is not destroyed by the errors, both due to the scale variationn and due to the non-perturbative effects.
In Fig.~\ref{fig:focal_new_DCS} we show latest results, where we compare p-p and p-Pb cross sections at different energies, computed also with larger $p_T>20$~GeV cut. As expected, the difference between p-p and p-Pb is smaller in that case. Next, in Fig.~\ref{fig:focal_lead_jets}-\ref{fig:focal_sublead_jets} we present latest results for rapidity distributions for the two $p_T$ cuts, for the leading jet, and the sub-leading jet.
As the shape of the rapidity distribution is correlated with the $x$ dependence, and thus with the evolution in the energy, measurement of the rapidity distribution may provide a valuable discriminatory tool for the evolution equations.
Finally, in Fig.~\ref{fig:Focal_Rnew} we show the nuclear modification ratio for the two energies and both $p_T$ cuts. As can be seen, with forward FoCal kinematic, even with the $p_T$ cut of 20~GeV the suppression is about 20\%.

%-----------------------------------------------------
%-----------------------------------------------------
\subsection{Forward  dijets at Electron Ion Collider}

\begin{figure}
  \begin{center}
   \includegraphics[width=0.4\textwidth]{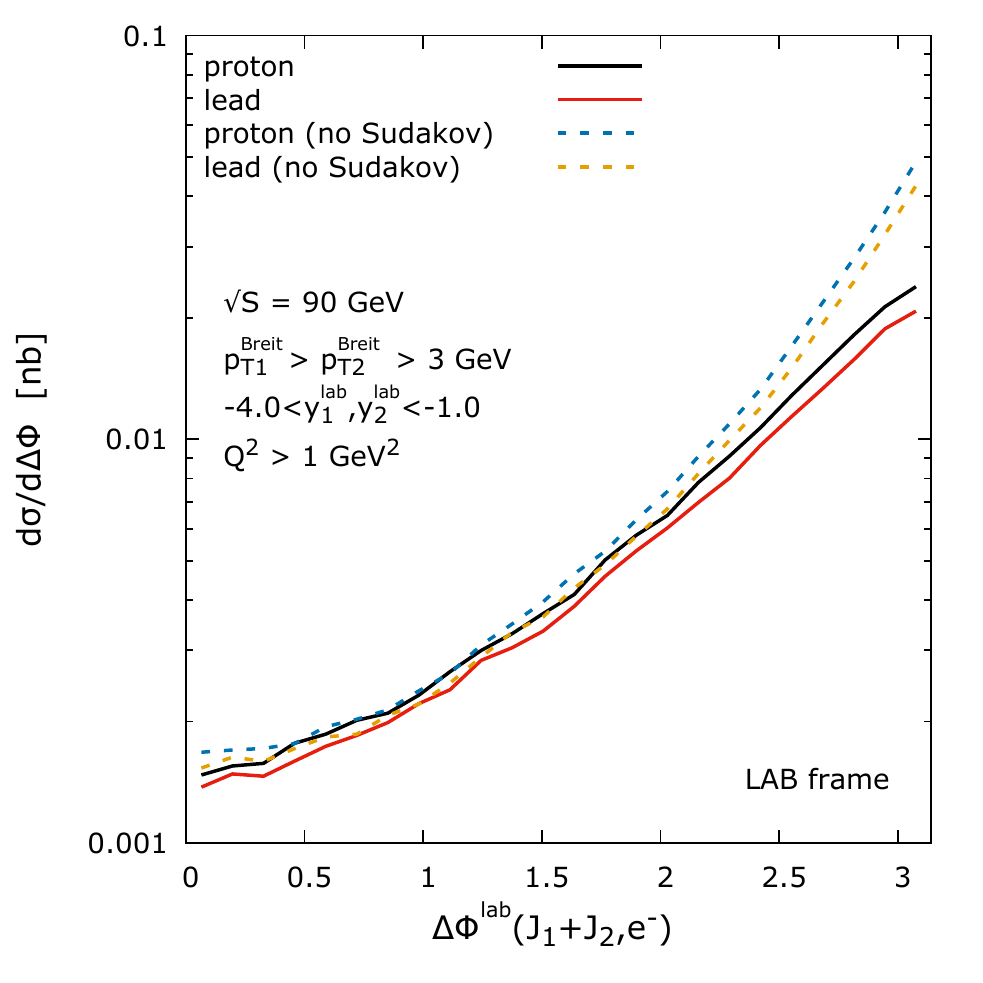}
   \includegraphics[width=0.4\textwidth]{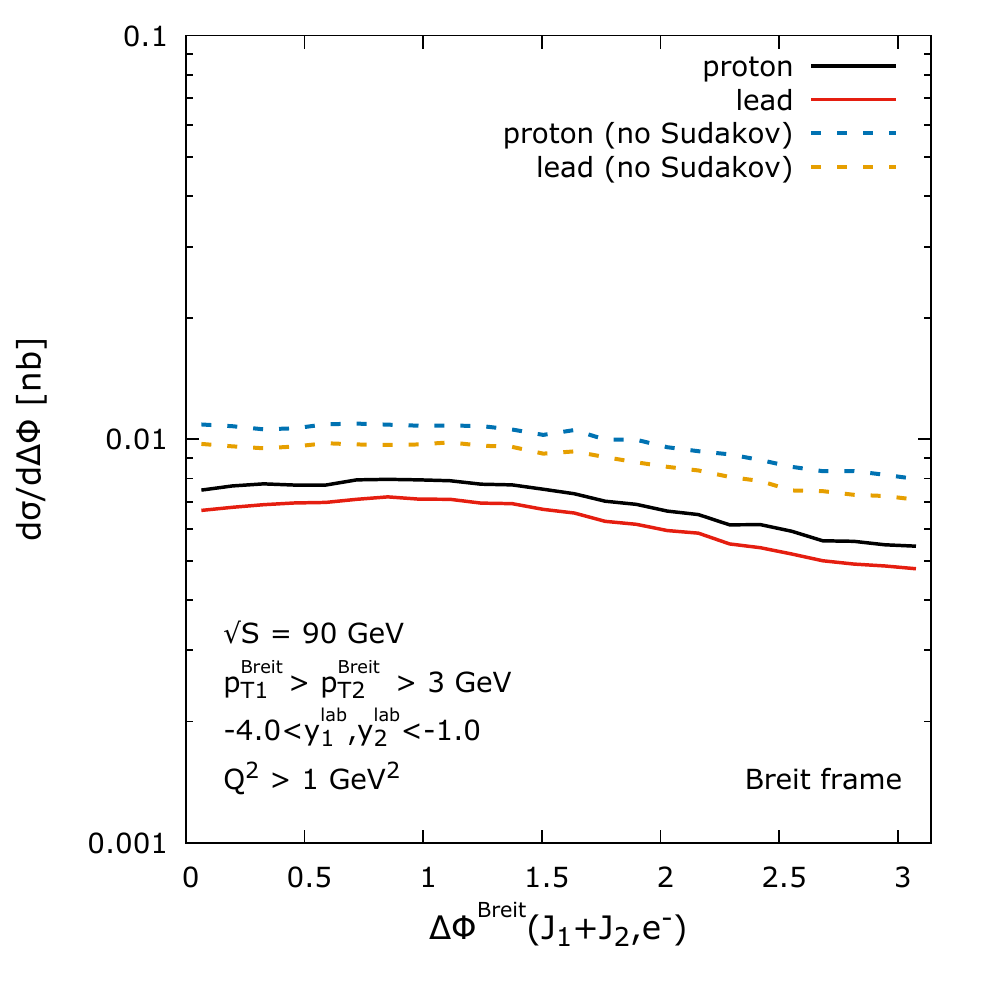}
  \end{center}
  \caption{\small
  Azimuthal correlations between the total transverse momentum of the
  dijet system and the transverse momentum of the scattered electron at EIC in
  two frames: the LAB frame (left), the Breit frame (right). The calculation has
  been done within the ITMD* framework with the Weizs\"acker-Williams gluon distribution obtained from the Kutak-Sapeta fit to HERA data.
  }
  \label{fig:angleplot}
\end{figure}

While the focus of this review is on forward jets at LHC, the new accelerator Electron Ion Collider (EIC) is  just around the  corner \cite{AbdulKhalek:2021gbh}. A  vital part of  its  research program is to look for  gluon saturation. 
However, since the involved hard scales are lower than at the LHC, in fact, hadrons constitute better probes of the saturation physics than jets. Nevertheless, there has been a large interest in jets at EIC, see for example \cite{Arratia:2019vju,Li:2020rqj,Boussarie:2021ybe}.
Moreover, at the partonic level, both hadron and jet production need classes of the same diagrams. Therefore, the upcoming EIC experiment triggered tremendous progress in computing NLO corrections for the DIS process within the CGC formalism, see for example \cite{Altinoluk:2020oyd,Taels:2022tza,Altinoluk:2022jkk,Bergabo:2022zhe,Bergabo:2023wed,Caucal:2023nci}.

In the context of the ITMD factorization, dijet production at EIC is actually a very interesting and important process. As discussed eg. in \cite{Dominguez:2011wm} the only TMD gluon distribution that is relevant for dijets in DIS, in the back-to-back limit, is the Weizsacker-Williams (WW) gluon distribution $F_{gg}^{(3)}$ (called also $xG^{(1)}$)\footnote{This distribution is also probed in forward Ultra Peripheral Collisions at LHC. We however do not discuss this process here. First application of ITMD to UPC processes can be found in \cite{Kotko:2017oxg}}.
For the ITMD factorization, this is only true for sufficiently small photon virtuality $Q^2\ll p_T$, otherwise longitudinaly polarized gluons contribute, and they are not formally accounted for in the ITMD approach. Due to this fact, for processes where such contributions formally enter, the ITMD factorization is called ITMD$^*$ (see for example \cite{Bury:2020ndc} for the trijet production case).

The dijet production in the DIS process gives actually more interesting observables than dijets in hadron-hadron scattering. This is because there is also an identified electron in the final state. Thus, in addition to dijet azimuthal correlations, one can also study correlations with the outgoing electron.

The ITMD$^*$ factorization formula for the process
\begin{equation}
    A(P_A)+e^{-} (P_B) \rightarrow J(p_1) + J(p_2)+e^{-} (P'_B)+X \,,
\end{equation}
reads
\begin{equation}
  d\sigma_{eA \to e' + 2j + X} =
  \int \frac{d x_A}{x_A} d^2 k_T\, 
    \frac{1}{4\pi x_A P_A\!\cdot\!P_B}\,
  \calF^{(3)}_{gg} \big(x_A, |\vec{k}_T|, \mu\big)\,
  |\overline{M}_{eg^* \to e'+2j}|^2 \, d\Gamma_3\, .
  \label{eq:ITMD}
\end{equation}
As mentioned above, there is only WW gluon distribution in the above formula.

In \cite{vanHameren:2021sqc} we computed several observables using the above framework, not only to quantify possible saturation effects in dijet observable at EIC but also to demonstrate the relevance of interplay of the saturation, the Sudakov effect, and the exact kinematics. 

First, we found that in the context of saturation physics, it is best to consider forward dijets. This provides a very good focusing of the $x_A$ distribution around the sufficiently small values so that the use of the small-$x$ formalism is well justified. For the WW gluon distribution, we used the Kutak-Sapeta fit of the dipole TMD to HERA data, and then the WW TMD distribution was calculated in the mean field approximation (see Section~\ref{subsec:ITMD_distrib}). The calculation was supplemented with the Sudakov form factor, see Fig.~\ref{fig:Fgg3} and the corresponding discussion in the text.
In Fig.~\ref{fig:angleplot} we  show prediction for the cross section dependence on azimuthal angle between the dijet system and the scattered electron for CM energy per nucleon $\sqrt{s}=90$~GeV. The calculations are done both in the lab frame and the Breit frame. One clearly sees effects coming from the  Sudakov form factor while the saturation is rather mild, despite the cut on the jet transverse momentum (in the Breit frame) is very low. Primarily, it is a consequence of the saturation pattern visible in the WW gluon distribution. Unlike the dipole gluon distribution, it does not have a peak as a function of the transverse momentum, \cf\ Fig.~\ref{fig:Fgg3} and~\ref{fig:KSgluons}. In our computation, we put quite a low cut on the photon virtuality $Q^2>1$~GeV$^2$ in order to suppress the longitudinal gluons. For a complete discussion see \cite{vanHameren:2021sqc}  and \cite{Goda:2023jie}.

%---------------------------------------------------------
\section{Summary and outlook}
\label{sec:Summary}

The text discusses the challenges and selected ongoing research related to searches of gluon saturation in QCD using dijet observables. Although there are indications of saturation in the experimental data, achieving a complete consensus on its existence is still a challenge due to demanding kinematics and the complex collision environment. While there are simpler observables than dijets, the latter have the advantage of providing the azimuthal correlations, which in the back-to-back kinematics are sensitive to saturation even for relatively hard jets. 

The study of forward dijet production, using the Improved TMD Factorization (ITMD) framework, is a relatively novel proposal in the search for gluon saturation. The framework has been already used in addressing many dijet related observables that were reviewed in this work. 
The text highlights the need for more measurements and the construction of the FoCal detector by the ALICE collaboration to shed more light on the saturation phenomenon. Furthermore, it discusses complementary dijet measurements in photon nucleus scattering that can be done at the Electron Ion Collider.

While in the review we focused our attention on dijets, there are other forward physics final states that can be used to search for saturation effects \cite{Hentschinski:2022xnd}. 
Those include di-hadrons \cite{Stasto:2018rci,Albacete:2018ruq,Jalilian-Marian:2011tvq}, $J/\psi$ production \cite{Ducloue:2015gfa}, and trijets \cite{Bury:2020ndc,Altinoluk:2020qet}. The processes  are complementary to dijets, however, in some aspects more challenging. The di-hadrons require knowledge of fragmentation function and are sensitive to longitudinal polarization effects while trijets are just more difficult to measure and interpret. 

The future research will be focused on increasing the precision of the calculations.
The ITMD framework is well suited to extrapolate methods known in collinear factorization to compute higher order corrections to off-shell gauge invariant matrix elements, and in principle to automatize NLO computations.
First steps toward that goal have already been undertaken \cite{Hentschinski:2014dra,Blanco:2020akb,vanHameren:2022mtk,Blanco:2022iai}. Also, it is important to increase the accuracy of the evolution equations and there is also a substantial progress in this direction, see for example \cite{Hentschinski:2012kr,Iancu:2015joa,Lappi:2016fmu,Hentschinski:2016wya,Hentschinski:2017ayz,Shi:2021hwx,Boussarie:2021wkn,Li:2022avs,Hautmann:2022xuc}. Last but not least, the ITMD framework should account for the longitudinally polarized gluons, which contribute to some processes (see eg. \cite{Marquet:2017xwy}). The corresponding amplitudes should be obtainable  automatically, similarly to to off-shell gauge invariant amplitudes coupled eikonally that are used currently in the ITMD. Research in that direction is also ongoing.

\section{Acknowledgements}
\label{sec:acknowledgements}
We would like to thank Jacek Otwinowski for interesting and stimulating discussions. \\
HK is supported by the National Science Centre, Poland grant no.\ DEC-2021/41/N/ST2/02956.
PK is partially supported by the National Science Centre, Poland grant no.\ DEC-2020/39/O/ST2/03011.
KK acknowledges the Polish National Science Centre grant no.\ DEC-2017/27/B/ST2/01985.\\
AvH is supported by the National Science Centre, Poland grant no.\ DEC-2019/35/B/ST2/03531

\bibliographystyle{JHEP} 
\bibliography{sudakov,references}

\end{document}